\documentclass[a4paper,onecolumn, superscriptaddress,10pt,accepted=2024-10-12,issue=1, volume=7, shorttitle=Information\space Decomposition\space Diagrams\space beyond\space Shannon\space Entropy]{compositionalityarticle}
\pdfoutput=1
\usepackage[utf8]{inputenc}
\usepackage[english]{babel}
\usepackage[T1]{fontenc}
\usepackage{amsmath}
\usepackage{hyperref}
\hypersetup{colorlinks, linkcolor = blue, citecolor=magenta, linktocpage=true}  % Potentially remove in published version?

\usepackage{tikz}
\usepackage{lipsum}
\usepackage[numbers, sort&compress]{natbib}

\usepackage{amsmath,cleveref,amsfonts,bm, amsthm, thmtools, dsfont, mathtools}

\newcounter{mycounter}

\crefname{proof}{proof}{proofs}
\newenvironment{Prf}[2][]{%
    \par\bigskip\noindent%
    \refstepcounter{mycounter}%
    \label[proof]{prf:#2}%  creates new label starting with prf:<label_of_theorem>
    \textbf{Proof \themycounter\ for \Cref{#2}}%
    \ifx&#1&\else\space(#1)\fi%
    \textbf{.}%
}{\hfill\qed\par\bigskip}%

\newcommand{\N}{\mathds{N}}

\newcommand{\R}{\mathds{R}}

\declaretheorem[numberwithin=section]{theorem}

\declaretheorem[style=plain, name=Proposition, sibling=theorem]{proposition}
\declaretheorem[style=plain, name=Lemma, sibling=theorem]{lemma}
\declaretheorem[style=plain, name=Corollary, sibling=theorem]{corollary}
\declaretheorem[style=plain, name=Thesis, sibling=theorem]{thesis}
\declaretheorem[style=plain, name=Definition, sibling=theorem]{definition}
\declaretheorem[style=plain, name=Example, sibling=theorem]{example}
\declaretheorem[style=plain, name=Summary, sibling=theorem]{summary}
\declaretheorem[style=plain, name=Remark, sibling=theorem]{remark}
\newcommand{\vs}[1]{E_{#1}}
\newcommand{\ec}[1]{\left[#1\right]}
\newcommand{\one}{\boldsymbol{1}}
\newcommand{\zero}{\boldsymbol{0}}

\newcommand{\Asterisk}{\mathop{\scalebox{1.5}{\raisebox{-0.2ex}{$\ast$}}}}%
\newcommand{\Kc}{Kc}
\newcommand{\Exp}{\operatorname{E}}
\newcommand{\ple}{\overset{+}{=}}
\newcommand{\plle}{\overset{+}{<}}
\newcommand{\plge}{\overset{+}{>}}
\newcommand{\indep}{\perp\!\!\!\perp}

\DeclareMathOperator{\Maps}{Maps}
\DeclareMathOperator{\Meas}{Meas}

\DeclareMathOperator{\Ad}{Ad}

\DeclareMathOperator{\im}{im}
\DeclareMathOperator{\id}{id}

\DeclareMathOperator{\M}{M}
\DeclareMathOperator{\ev}{ev}
\DeclareMathOperator{\con}{con}

\newcommand{\bins}{\{0, 1\}^*}
\newcommand{\q}{q}
\DeclareMathOperator{\Concat}{Concat}

\usepackage{url}
\usepackage{amsmath}
\usepackage{float} % Allows to place figures ``here'' with H

\usepackage{graphicx}
\usepackage{amssymb}
\usepackage{silence}
\usepackage{verbatim}
\usepackage{tikz-cd}
\usetikzlibrary{graphs,decorations.pathmorphing,decorations.markings}
\usepackage{enumerate}
\allowdisplaybreaks

\begin{document}

\title{Information Decomposition Diagrams Applied beyond Shannon Entropy: a Generalization of Hu's Theorem}
\date{}
\author{Leon Lang} 
\email{l.lang@uva.nl}
\thanks{Main contributing author.}
\affiliation{Informatics Institute, University of Amsterdam}
\orcid{0000-0002-1950-2831}
\author{Pierre Baudot}
\email{pierre.baudot@mediantechnologies.com}
\affiliation{Median Technologies, France}
\orcid{0000-0002-5574-6809}
\author{Rick Quax}
\email{r.quax@uva.nl}
\affiliation{Informatics Institute, University of Amsterdam}
\orcid{0000-0002-0299-0074}
  \author{Patrick Forré}
\email{p.d.forre@uva.nl}
\affiliation{Informatics Institute, University of Amsterdam}
\orcid{0000-0003-4663-3842}

%%%%%%%%%%%%%%%%%%%%%%%%%%%%%%%%%%%%%%%%%%%%%%%%%%%%%%%%%%%%%%%%%%%%%%
\newcommand{\authorsforheader}{Lang, Baudot, Quax, and Forré}
\newcommand{\paperdoi}{https://doi.org/10.46298/compositionality-7-1}
\newcommand{\receiveddate}{2024-03-01}
\newcommand{\accepteddate}{2024-10-12}
%\titlerunning{Information Decomposition Diagrams Applied beyond Shannon Entropy}
%%%%%%%%%%%%%%%%%%%%%%%

\maketitle

\begin{abstract}
In information theory, one major goal is to find useful functions that summarize the amount of information contained in the interaction of several random variables.
  Specifically, one can ask how the classical Shannon entropy, mutual information, and higher interaction information relate to each other.
  This is answered by Hu's theorem, which is widely known in the form of information diagrams: it relates shapes in a Venn diagram to information functions, thus establishing a bridge from set theory to information theory.
  In this work, we view random variables together with the joint operation as a monoid that acts by conditioning on information functions, and entropy as a function satisfying the chain rule of information.
  This abstract viewpoint allows to prove a \emph{generalization of Hu's theorem}.
  It applies to Shannon and Tsallis entropy, (Tsallis) Kullback-Leibler Divergence, cross-entropy, Kolmogorov complexity, submodular information functions, and the generalization error in machine learning.
  Our result implies for Chaitin's Kolmogorov complexity that the \emph{interaction complexities} of all degrees are in expectation close to Shannon interaction information.
  For well-behaved probability distributions on increasing sequence lengths, this shows that the per-bit expected interaction complexity and information asymptotically coincide, thus showing a strong bridge between algorithmic and classical information theory.
\end{abstract}

\tableofcontents

\newpage

\section{Introduction}

Information diagrams, most often drawn for two or three random variables (see Figures~\ref{fig:hu_kuo_ting_simple_visualization} and~\ref{fig:hu_kuo_ting_triple_visualization}), provide a concise way to visualize information functions.
Not only do they show (conditional) Shannon entropy~\citep{Shannon1948}, mutual information, and interaction information --- also called co-information~\citep{bell2003} --- of several random variables in one overview, they also provide an intuitive account of the \emph{relations} between these functions. 

This well-known fact goes beyond just three variables:
diagrams with four (see Figure~\ref{fig:hu_kuo_ting_quadruple_visualization}) and more variables exist as well.
Hu's theorem~\citep{Hu1962,Yeung1991,Yeung2002} renders all this mathematically precise by connecting the set-theoretic operations of union, intersection, and set difference to joint information, interaction information, and conditioning of information functions, respectively.
The map from sets to information functions is then a \emph{measure} and turns disjoint unions into sums.
Certain summation rules of information functions then follow visually from disjoint unions in the diagrams.

Our work is concerned with the question of whether Hu's theorem can be generalized to other information functions than entropy, such as Kullback-Leibler divergence and cross-entropy.
Such functions are important in the context of statistical modeling of multivariate data, in which one aims to find a probabilistic model able to reproduce the information structure of the data.
For instance, an information diagram for cross-entropy would then allow to visualize how the cross-entropy between a model probability distribution and the data distribution is decomposed into higher-order terms.  
\cite{Cocco2012} used these higher-order terms (which they called cluster (cross)-entropies) in their adaptive cluster expansion approach to statistical modeling of data with Ising models. 

Kullback-Leibler divergence has been studied in the context of decompositions of joint entropy and information~\citep{Amari2001} and is often minimized in machine learning and deep learning~\citep{Bishop2007,Bishop2023}. 
This becomes especially interesting for graphical methods, including diffusion models~\citep{Dickstein2015}, which form the basis for widespread text-to-image generation methods like Dalle~\citep{Ramesh2021}, Imagen~\citep{Chitwan2022}, and stable diffusion~\citep{Stable_Diffusion2021}.
Diffusion models involve a decomposition of a joint Kullback-Leibler divergence over a Markov chain.
Once information diagrams are established in a generalized context, this might facilitate to study decompositions of loss functions for more general graphical models.

Our claim is that the language employed in the foundations of information cohomology~\cite{baudot2015a} gives the perfect starting point for generalizing Hu's theorem.
Namely, by replacing discrete random variables with partitions on a sample space, they give random variables the structure of a \emph{monoid} that is commutative and idempotent.
Furthermore, conditional information functions are formally described by a \emph{monoid action}.
And finally, the most basic information function that generates all others, Shannon entropy, is fully characterized as the unique function that satisfies the \emph{chain rule of information}.
We substantially generalize Hu's theorem by giving a proof only based on the properties just mentioned, leading to new applications to Kolmogorov complexity, Kullback-Leibler divergence, and beyond.

To clarify, the main contribution of this work is not to provide major previously unknown ideas --- indeed, our proof is very similar to the original one given in~\cite{Yeung1991} --- but instead, to place and prove this result in its proper abstract context.
This then reveals information diagrams for new information measures.

Section~\ref{sec:technical_introduction} summarizes classical definitions and results for Shannon information theory, generalized to \emph{countable} discrete random variables to be later applied to Kolmogorov complexity.
Section~\ref{sec:generalized_hu_kuo_ting} --- which can be read independently of the preceding section --- contains our main result, the generalized Hu theorem.
In Section~\ref{sec:kolmogorov_complexity}, we prove a Hu theorem for Kolmogorov complexity.
We also combine Hu's theorems for Shannon entropy and Kolmogorov complexity to generalize the well-known result that ``expected Kolmogorov complexity is close to entropy''~\citep{grunwald2008}:
general \emph{interaction complexity} is close to interaction information.
For the case of well-behaved sequences of probability measures on binary strings with increasing length, this leads to an asymptotic result:
in the limit of infinite sequence length, the \emph{per-bit} interaction complexity and interaction information coincide.
In Section~\ref{sec:example_applications}, we consider further examples of Hu's theorem, including Kullback-Leibler divergence and the generalization error in machine learning.
We conclude with a discussion in Section~\ref{sec:discussion}, followed by proofs in the appendices.

\subsection*{Preliminaries and Notation}

We mainly assume the reader to be familiar with the basics of measure theory and probability theory.
They can be learned from any book on the topic, for example~\cite{Schilling2017} or~\cite{Tao2013}.
The main concepts we assume to be known are $\sigma$-algebras, the Borel $\sigma$-algebra on $\R^n$, measurable spaces, measures, measure spaces, probability measures, probability spaces, and random variables.
We assume some very basic familiarity with abelian groups, (commutative, idempotent) monoids, and additive monoid actions.
In contrast, we carefully define all basic notions from (algorithmic) information theory from scratch.

On notation:
to aid familiarity, we will start writing the Shannon entropy with the symbol $H$, but then switch to the notation $I_1$ once we embed Shannon entropy in the concept of interaction information, Definition~\ref{def:interaction_information}.
Instead of the typical notation $H(Y \mid X)$ for the conditional entropy, we will use $X.H(Y) = X.I_1(Y)$.
This is the general notation of monoid actions and is thus preferable in our abstract context.
Furthermore, for two disjoint sets $A$ and $B$, we write their union as $A \dot\cup B$.
The number of elements in $A$ is written as $|A|$.
The power set of $A$, i.e., the set of its subsets, is denoted $2^A$.
And finally, the natural and binary logarithms  of $x$ are denoted $\ln(x)$ and $\log(x)$, respectively.

 \section{Preliminaries on Shannon Entropy of Countable Discrete Random Variables}\label{sec:technical_introduction}

In this technical introduction, we explain preliminaries on discrete random variables, entropy, mutual information, and interaction information.
Our treatment will also emphasize abstract structures that lead us to the generalizations in Section~\ref{sec:generalized_hu_kuo_ting}.
The goal is to arrive at Summary~\ref{sum:summary}, which summarizes the properties of classical information functions in an abstract way suitable for our generalizations.
We will omit many proofs of elementary and well-known results.
When we say a set is \emph{countable}, then we mean it is finite or countably infinite.
Whenever we talk about \emph{discrete measurable spaces}, we mean countable measurable spaces in which all subsets are measurable.
Some technical considerations related to the measurability of certain functions in the infinite, discrete case are found in Appendix~\ref{sec:measure_theory_technicalities}.

\subsection{Entropy, Mutual Information, and Interaction Information}\label{sec:information_functions}

We fix in this section a discrete sample space $\Omega$.
We define 
\begin{equation*}
  \Delta(\Omega) \coloneqq \bigg\lbrace P: \Omega \to [0, 1] \ \ \Big| \ \   \sum_{\omega \in \Omega}P(\omega) = 1\bigg\rbrace = \bigg\lbrace (p_{\omega})_{\omega \in \Omega} \in [0, 1]^{\Omega} \ \ \Big|   \ \ \sum_{\omega \in \Omega} p_\omega = 1\bigg\rbrace.
\end{equation*}
If $\Omega$ is finite, we view it as a measurable space with the $\sigma$-algebra of Borel measurable sets.
When $\Omega$ is infinite and discrete, we equip $\Delta(\Omega)$ with the smallest $\sigma$-algebra that makes all evaluation maps
\begin{equation*}
  \ev_A: \Delta(\Omega) \to \R, \ \ \ P \mapsto \ev_A(P) \coloneqq P(A)
\end{equation*}
for all subsets $A \subseteq \Omega$ measurable.
In the finite case, this definition is equivalent to the one given before.
We remark that we do not distinguish between probability measures and their mass functions in the notation or terminology:
for a subset $A \subseteq \Omega$ and a probability measure $P: \Omega \to [0,1]$, we simply set $P(A) \coloneqq \sum_{\omega \in A}P(\omega)$.

Our aim is the study of discrete random variables $X: \Omega \to \vs{X}$.
Here, being discrete means that $\vs{X}$ --- next to $\Omega$ --- is discrete.
For any probability measure $P$ on $\Omega$ and any random variable $X: \Omega \to \vs{X}$, we define the \emph{distributional law} $P_X: \vs{X} \to [0,1]$ as the unique probability measure with
\begin{equation*}
  P_X(x) \coloneqq P\big(X^{-1}(x)\big) = \sum_{\omega \in X^{-1}(x)} P(\omega)
\end{equation*}
for all $x \in X$.
Clearly, $P_X \in \Delta(\vs{X})$.

For the following definition of Shannon entropy, introduced in~\cite{Shannon1948, Shannon1964}, we employ the convention $0 \cdot \infty = 0 \cdot (- \infty) = 0$ and $\ln(0) = - \infty$.
Furthermore, set $\overline{\R} \coloneqq \R \cup \{+\infty\}$.

\begin{definition}[Shannon Entropy]\label{def:shannon_entropy}
  Let $P \in \Delta(\Omega)$ be a probability measure.
  Then the \emph{Shannon entropy of $P$} is given by
  \begin{equation*}
    H(P) \coloneqq - \sum_{\omega \in \Omega} P(\omega) \ln P(\omega) \in \overline{\R}.
  \end{equation*}
  Here, $\ln: [0, \infty) \to \R \cup \{- \infty\}$ is the natural logarithm. 
  Now, let $X: \Omega \to \vs{X}$ be a discrete random variable. 
  The \emph{Shannon entropy of $X$ with respect to $P \in \Delta(\Omega)$} is given by
  \begin{equation*}
    H(X; P) \coloneqq H(P_X) = - \sum_{x \in E_X} P_X(x) \ln P_X(x) \in \overline{\R}.
  \end{equation*}
\end{definition}

For the identity $\id_{\Omega}: \Omega \to \Omega$, $\omega \mapsto \omega$ we then have $P_{\id_{\Omega}} = P$ and therefore $H(\id_{\Omega}; P) = H(P)$. 
For discrete probability distributions with infinite Shannon entropy, see~\cite{baccetti2013}.

Now, set 
\begin{equation*}
  \Delta_f(\Omega) \coloneqq \Delta(\Omega) \setminus \{ P \in \Delta(\Omega) \mid H(P) = \infty\}.
\end{equation*}
$\Delta_f(\Omega)$ is the measurable space of probability measures with finite entropy.
We restrict entropy functions to this space for algebraic reasons:

\begin{definition}
  [Entropy Function of a Random Variable]
  \label{def:entropy_function}
  Let $X: \Omega \to \vs{X}$ be a discrete random variable.
  Then its \emph{entropy function} or \emph{Shannon entropy} is the measurable function
  \begin{equation*}
    H(X): \Delta_f(\Omega) \to \R, \ \ \ P \mapsto H(X; P)
  \end{equation*}
  defined on probability measures with finite entropy. 
  Its measurability is proven in Corollary~\ref{cor:measurability_of_entropy_function}. 
\end{definition}

Let $P: \Omega \to \R$ be a probability measure and $X: \Omega \to \vs{X}$ a discrete random variable.
Then we define the conditional probability measure $P|_{X = x}: \Omega \to \R$ by
\begin{equation}\label{eq:conditional_distribution_def}
  P|_{X = x}(\omega) \coloneqq 
  \begin{cases}
    \frac{P\big(\{\omega\} \cap X^{-1}(x)\big)}{P_X(x)}, \ \ \ & P_X(x) \neq 0; \\
    P(\omega), & P_X(x) = 0.\footnotemark
  \end{cases}
\end{equation}
\footnotetext{ 
  Note that the precise definition for the case $P_X(x) = 0$ does not matter since it almost surely does not appear. 
However, defining the conditional also in this case makes many formulas simpler since we do not need to restrict sums involving $P|_{X = x}$ to the case $P_X(x) \neq 0$.}
For all $A \subseteq \Omega$, we then have 
\begin{equation*}
  P|_{X = x}(A) =
  \begin{cases} 
    \frac{P\big(A \cap X^{-1}(x)\big)}{P(X^{-1}(x))}, \ \ \ & P_X(x) \neq 0; \\
    P(A), & P_X(x) = 0.
  \end{cases}
\end{equation*}

For the following definition, recall that a series of real numbers converges absolutely if the series of its absolute values converges.
It converges unconditionally if every reordering of the original series still converges with the same limit.
According to the Riemann series theorem~\citep{macrobert1926}, these two properties are equivalent.

\begin{definition}[Conditionable Functions, Averaged Conditioning]\label{def:averaged_conditioning}
  Let $F: \Delta_f(\Omega) \to \R$ be a measurable function.
  $F$ is called \emph{conditionable} if for all discrete random variables $X: \Omega \to \vs{X}$ and all $P \in \Delta_f(\Omega)$, the sum
  \begin{equation}
    (X.F)(P) \coloneqq \sum_{x \in \vs{X}} P_X(x) F(P|_{X = x})
    \label{eq:sum_converges_unconditionally}
  \end{equation}
  converges unconditionally.
  Note that $P|_{X = x} \in \Delta_f(\Omega)$, which makes $F(P|_{X = x})$ in Equation~\eqref{eq:sum_converges_unconditionally} well-defined.

  For all conditionable measurable functions $F: \Delta_f(\Omega) \to \R$ and all discrete random variables $X: \Omega \to \vs{X}$, the function $X.F: \Delta_f(\Omega) \to \R$ is a measurable function by Corollary~\ref{cor:result_after_pliability_measurable}, which we call the \emph{averaged conditioning} of $F$ by $X$.
  The space of all conditionable measurable functions $F: \Delta_f(\Omega) \to \R$ is denoted by $\Meas_{\con}(\Delta_f(\Omega), \R)$.
\end{definition}

If $X: \Omega \to \vs{X}$ and $Y: \Omega \to \vs{Y}$ are two (not necessarily discrete) random variables, then their (Cartesian) product, or joint variable, $XY: \Omega \to \vs{X} \times \vs{Y}$ is defined by
\begin{equation}\label{eq:def_product_random_variable}
  (XY)(\omega) \coloneqq \big(X(\omega), Y(\omega)\big) \in \vs{X} \times \vs{Y}.\footnotemark
\end{equation}
\footnotetext{In the case that $\vs{X} = \vs{Y} = \R$, there is some ambiguity of notation, as the reader could understand $XY$ to be given by $(XY)(\omega) = X(\omega) \cdot Y(\omega)$. 
  This definition plays a role in the \emph{algebra of random variables}~\citep{springer1979}.
In our work, we instead \emph{always} mean the Cartesian product.}If we have two discrete random variables $X$ and $Y$ and a probability measure $P \in \Delta(\Omega)$, then this allows to consider $(P|_{X = x})_Y(y)$ for $(x, y) \in \vs{X} \times \vs{Y}$.
In order to not overload notation, we will write this often as $P(y \mid x)$.
Similarly, we will often write $P(x) \coloneqq P_X(x)$ and $P(\omega \mid x) \coloneqq P|_{X = x}(\omega)$.
We obtain the following elementary lemma and corollary whose proofs are left to the reader:

\begin{lemma}
\label{lem:converges_unconditionally}
  Let $Y$ be a discrete random variable on $\Omega$.
  Then $H(Y)$ is conditionable.
  More precisely, for another discrete random variable $X$ on $\Omega$ and $P \in \Delta_f(\Omega)$, $H(X; P)$ and $H(XY; P)$ are finite and we have
  \begin{equation*}
    \big[X.H(Y)\big](P) = H(XY; P) - H(X; P),
  \end{equation*}
  which results in $\big[ X.H(Y)\big](P)$ converging unconditionally.
\end{lemma}

\begin{corollary}
\label{cor:cocycle_condition_entropy}
  The following chain rule
  \begin{equation*}
    H(XY) = H(X) + X.H(Y) 
  \end{equation*}
  holds for arbitrary discrete random variables $X: \Omega \to \vs{X}$ and $Y: \Omega \to \vs{Y}$.
\end{corollary}

We will also write $Y.F(P) \coloneqq (Y.F)(P)$.
For example, if $F = H(X)$ is the Shannon entropy of the discrete random variable $X$, we write 
\begin{equation*}
  Y.H(X;P) = Y.H(X)(P) = [Y.H(X)](P) = \sum_{y \in \vs{Y}} P_Y(y) H(X; P|_{Y = y}).
\end{equation*}
We emphasize explicitly that $Y$ can not act on $H(X;P)$ since this is only a number, and not a measurable function. 
Nevertheless, we find the notation $Y.H(X;P)$ for $[Y.H(X)](P)$ convenient.
We obtain the following properties resembling those of an additive monoid action:
\begin{proposition}\label{pro:conditioning_rules}
  Let $X, Y$ be two discrete random variables on $\Omega$, $\one: \Omega \to \Asterisk \coloneqq \{\ast\}$ a trivial random variable, and $F, G: \Delta_f(\Omega) \to \R$ two conditionable measurable functions.
  Then the following hold:
  \begin{enumerate}
    \item $\one.F = F$;
    \item $Y.F$ is also conditionable, and we have $X.(Y.F) = (XY).F$;
    \item $F + G$ is also conditionable, and we have $X.(F + G) = X.F + X.G$.
  \end{enumerate}
\end{proposition}

\begin{proof}
  Properties 1 and 3 are elementary and left to the reader to prove.
  $2$ follows from $P(x, y) = P(x) \cdot P(x \mid y)$ and $(P|_{X = x})|_{Y=y} = P|_{XY = (x, y)}$.
\end{proof}

Next, we define mutual information and, more generally, interaction information --- also called co-information~\citep{bell2003}.
As we want to view interaction information as a ``higher degree generalization'' of entropy and treat both on an equal footing in Hu's theorem, we now change the notation:
for any discrete random variables $X$, we set $I_1(X) \coloneqq H(X)$.

\begin{definition}[Mutual Information, Interaction Information]\label{def:interaction_information}
    Let $q \in \N$ and assume that $I_{q-1}$ is already defined.
    Assume also that $Y_1, \dots, Y_q$ are $q$ discrete random variables on $\Omega$.
    Then we define $I_q(Y_1;\dots;Y_q): \Delta_f(\Omega) \to \R$, the \emph{interaction information of degree $q$}, as the function
    \begin{equation*}
      I_q(Y_1; \dots ; Y_q ) \coloneqq I_{q-1}(Y_1; \dots; Y_{q-1}) - Y_q.I_{q-1}(Y_1; \dots ; Y_{q-1}).
    \end{equation*}
    $I_2$ is also called mutual information.
\end{definition}

\begin{remark}
  \label{rem:interaction_vs_mutual}
  What we call \emph{interaction information} is in the literature sometimes called \emph{(higher / multivariate) mutual information}.
  In that case, the term $J_q(Y_1; \dots ; Y_q) \coloneqq (-1)^{q+1}I_q(Y_1; \dots ; Y_q)$ is called interaction information, see for example~\cite{baudot2021a}.
\end{remark}

\begin{proposition}
    \label{pro:interaction_information_pliable}
    For all $q \geq 1$ and all discrete random variables $Y_1, \dots, Y_q$, $I_q(Y_1; \dots ; Y_q): \Delta_f(\Omega) \to \R$ is a well-defined conditionable measurable function.
\end{proposition}

\begin{proof}
  $I_1(Y_1)$ is conditionable by Lemma~\ref{lem:converges_unconditionally}.
  Assuming by induction that $I_{q-1}(Y_1; \dots ; Y_{q-1})$ is well-defined and conditionable, we obtain the following:
  $Y_q.I_{q-1}(Y_1; \dots ; Y_{q-1})$ is well-defined and conditionable by Proposition~\ref{pro:conditioning_rules}, part 2, and $I_q(Y_1; \dots ; Y_q)$ is well-defined and conditionable by Proposition~\ref{pro:conditioning_rules}, part 3.
\end{proof}

\subsection{Equivalence Classes of Random Variables}\label{sec:equivalence_classes}

Assume all random variables are discrete.
For two random variables $X$ and $Y$ on $\Omega$, we write $X \precsim Y$ if there is a function $f_{XY}: \vs{Y} \to \vs{X}$ such that $f_{XY} \circ Y = X$. 
  The definition of $\precsim$ is equivalent to a preorder put forward in the context of conditional independence relations~\citep{dawid1979,dawid1980,Dawid2001}.
  The latter work defines in their Section 6.2: $X \precsim Y$ if for all $\omega, \omega' \in \Omega$, the following implication holds true:
  \begin{equation*}
    Y\big(\omega\big) = Y\big(\omega'\big) \ \ \Longrightarrow \ \ X\big(\omega\big) = X\big(\omega'\big).
  \end{equation*}
  It is straightforward to show that this coincides with our own definition.

  Clearly, our relation is reflexive and transitive and thus a \emph{preorder}. 
  We define the equivalence relation $\sim$ by $X \sim Y$ iff $X \precsim Y$ and $Y \precsim X$.
  We denote by $\ec{X}$ the equivalence class of $X$.

\begin{proposition}[See Proof~\ref{prf:pro:inequality_of_rvs_preserved}]\label{pro:inequality_of_rvs_preserved}
  Let $Y \precsim X$ be two discrete random variables on $\Omega$. Then we have $I_1(Y) \leq I_1(X)$ as functions on $\Delta_f(\Omega)$, meaning that $I_1(Y; P) \leq I_1(X; P)$ for all $P \in \Delta_f(\Omega)$.
  In particular, if $X$ and $Y$ are equivalent (i.e., $X \precsim Y$ and $Y \precsim X$), then $I_1(X) = I_1(Y)$.
\end{proposition}

\begin{proposition}[See Proof~\ref{prf:pro:averaged_conditioning_equivalence}]\label{pro:averaged_conditioning_equivalence}
  Let $X \sim Y$ be two equivalent discrete random variables on $\Omega$.
  Then for all conditionable measurable functions $F: \Delta_f(\Omega) \to \R$ we have $X.F = Y.F$.
\end{proposition}

  \begin{proposition}\label{pro:well-definednedd_of_interaction_information}
    Let $q \geq 1$ and $Y_1, \dots, Y_q$ and $Z_1, \dots, Z_q$ be two collections of discrete random variables on $\Omega$ such that $Y_k \sim Z_k$ for all $k = 1, \dots, q$.
    Then $I_q(Y_1; \dots; Y_q) = I_q(Z_1; \dots; Z_q)$.
  \end{proposition}

  \begin{proof}
    For $q = 1$, this was shown in Proposition~\ref{pro:inequality_of_rvs_preserved}.
    The case $q > 1$ can be shown by induction using Definition~\ref{def:interaction_information} and Proposition~\ref{pro:averaged_conditioning_equivalence}.
  \end{proof}

  This proposition shows that interaction information is naturally defined for collections of \emph{equivalence classes of random variables}, instead of the random variables themselves.

\subsection{Monoids of Random Variables}\label{sec:abstract_properties_interaction_information}

  Again, assume all random variables to be discrete.

\begin{lemma}\label{lem:joint_and_equivalence}
    Let $X, Y, Z, X'$, and $Y'$ be random variables on $\Omega$. 
    Let $\one: \Omega \to \Asterisk$ be a trivial random variable, with $\Asterisk = \{\ast\}$ a measurable space with one element.
    Then the following properties hold:
    \begin{enumerate}\addtocounter{enumi}{-1}
      \item If $X \sim X'$ and $Y \sim Y'$, then $XY \sim X'Y'$;
      \item $\one X \sim X \sim X \one$; 
      \item $(XY)Z \sim X(YZ)$; 
      \item $XY \sim YX$; 
      \item $XX \sim X$. 
      \end{enumerate}
      Additionally, we have $X \precsim Y$ if and only if $XY \sim Y$.
\end{lemma}

\begin{proof}
  All of these statements are elementary and left to the reader to prove.
\end{proof}

Recall that a monoid is a tuple $(M, \cdot, \one)$ with $M$ a set, $\cdot$ a multiplication, and $\one \in M$, such that $\one$ is neutral and the multiplication is associative. 
A monoid is commutative and idempotent if $m \cdot n = n \cdot m$ and $m \cdot m = m$ for all $m, n \in M$.
Notice that rules 1 to 4 in the lemma resemble the properties of a commutative, idempotent monoid.
 
We remark that a commutative, idempotent monoid is algebraically the same as a join-semilattice (sometimes also called \emph{bounded} join-semilattice), i.e., a partially ordered set which has a bottom element (corresponding to $\one \in M$) and binary joins (corresponding to the multiplication in a monoid).
      The partial order can be reconstructed from a commutative, idempotent monoid $M$ by writing $m \leq n$ if $m \cdot n = n$, which corresponds to the last statement in Lemma~\ref{lem:joint_and_equivalence}.
      The language of join-semilattices is, for example, used in the development of the theory of conditional independence~\citep{Dawid2001}.

\begin{proposition}[See Proof~\ref{prf:pro:monoid_of_rvs}]\label{pro:monoid_of_rvs}
	Let $\widehat{M} = \{X: \Omega \to \vs{X}\}_X$ be a collection of random variables with the following two properties:
	\begin{enumerate}[a)]
	\item There is a random variable $\one: \Omega \to \Asterisk$ in $\widehat{M}$ which has a one-point set $\Asterisk = \{\ast\}$ as the target;
	\item For every two $X, Y \in \widehat{M}$ there exists a $Z \in \widehat{M}$ such that $XY \sim Z$. 
	  \end{enumerate}
	  Let $\ec{X}$ denote the equivalence class of $X$ under the relation $\sim$.
	  Define $M \coloneqq \widehat{M}/\sim$ as the collection of equivalence classes of elements in $\widehat{M}$.
	  Define $\ec{X} \cdot \ec{Y} \coloneqq \ec{Z}$ for any $Z \in \widehat{M}$ with $XY \sim Z$. 
      Then the triple $(M, \cdot, \ec{\one})$ is a commutative, idempotent monoid.
\end{proposition}

We note that the monoid of equivalence classes of discrete random variables is isomorphic to the monoid of partitions on $\Omega$, which is the formalization used in~\cite{baudot2015a}.

We can now study finite monoids of random variables as instances of the construction in Proposition~\ref{pro:monoid_of_rvs}.
      Let $n \geq 0$ be a natural number. 
      Let $X_1, \dots, X_n$ be fixed random variables on $\Omega$.
      Define $[n] \coloneqq \{1, \dots, n\}$.
      For arbitrary $I \subseteq [n]$, define $X_I \coloneqq \prod_{i \in I} X_i$, the joint of the variables $X_i$ for $i \in I$.
      For $X_J$ and $X_I$, we have the equivalence $X_JX_I \sim X_{J \cup I}$.
      Note that $X_{\emptyset}: \Omega \to \Asterisk = \{\ast\}$ is a trivial random variable.

      \begin{definition}[Monoid of $X_1, \dots, X_n$]\label{def:simplicial_monoid}
	The monoid $\M(X_1, \dots, X_n)$ of the variables $X_1, \dots, X_n$ consists of the following data:
	\begin{enumerate}
	  \item The elements are equivalence classes $\ec{X_I}$ for $I \subseteq [n]$.
	  \item The multiplication is given by $\ec{X_J} \cdot \ec{X_I} = \ec{X_{J \cup I}}$.
	  \item $\one \coloneqq \ec{X_{\emptyset}}$ is the neutral element with respect to multiplication.
	  \end{enumerate}
	  This is a well-defined commutative, idempotent monoid by Proposition~\ref{pro:monoid_of_rvs}.
      \end{definition}

      Recall that an additive monoid action is a triple $(M, G, .)$, where $M$ is a monoid, $G$ is an abelian group, and $.: M \times G \to G$ is a function such that $\one \in M$ acts neutrally, with associativity (meaning $m.(n.g) = (m \cdot n).g$), and distributivity over addition in $G$.

\begin{proposition}\label{pro:our_monoid_action}
  Let $M$ be a monoid of (equivalence classes of) discrete random variables on $\Omega$ as in Proposition~\ref{pro:monoid_of_rvs}.
  Let $G = \Meas_{\con}\big(\Delta_f(\Omega), \R\big)$ be the group of conditionable measurable functions from $\Delta_f(\Omega)$ to $\R$.
  Then the averaged conditioning $.: M \times G \to G$ given by 
  \begin{equation*}
    \big(\ec{X}.F\big)(P) \coloneqq \big( X.F\big)(P) = \sum_{x \in \vs{X}} P_X(x) F(P|_{X = x})
  \end{equation*}
  is a well-defined monoid action.
\end{proposition}

\begin{proof}
  The action is well-defined by Proposition~\ref{pro:averaged_conditioning_equivalence} and Proposition~\ref{pro:conditioning_rules}, part 2.
  It is a monoid action by Proposition~\ref{pro:conditioning_rules}.
\end{proof}

\begin{summary}\label{sum:summary}
  We now summarize the abstract properties of interaction information $I_q$. 
  Let $M$ be a commutative, idempotent monoid of discrete random variables as in Proposition~\ref{pro:monoid_of_rvs}.
  By abuse of notation, we do not distinguish between random variables and their equivalence classes, i.e., we write $Y$ instead of $\ec{Y}$.
  Denote by $G \coloneqq \Meas_{\con}\big(\Delta_f(\Omega), \R\big)$ the group of conditionable measurable functions from $\Delta_f(\Omega)$ to $\R$.
  By Proposition~\ref{pro:our_monoid_action}, averaged conditioning $.: M \times G \to G$ is a well-defined monoid action.

  By Proposition~\ref{pro:well-definednedd_of_interaction_information}, we can view $I_q$ as a function
  $I_q: M^q \to G$ that is defined on tuples of \emph{equivalence classes} of discrete random variables.
  By Proposition~\ref{cor:cocycle_condition_entropy}, entropy $I_1$ satisfies the equation
  \begin{equation*}
    I_1(XY) = I_1(X) + X.I_1(Y)
  \end{equation*}
  for all $X, Y \in M$, where $X.I_1(Y)$ is the result of the action of $X \in M$ on $I_1(Y) \in G$ via averaged conditioning.
  Finally, by Definition~\ref{def:interaction_information}, for all $q \geq 2$ and all $Y_1, \dots, Y_q \in M$, one has
  \begin{equation*}
    I_q(Y_1; \dots ; Y_q) = I_{q-1}(Y_1; \dots ; Y_{q-1}) - Y_q.I_{q-1}(Y_1; \dots ; Y_{q-1}).
  \end{equation*}
\end{summary}

\section{A Generalization of Hu's Theorem}\label{sec:generalized_hu_kuo_ting}

In this section, we formulate and prove a generalization of Hu's theorem.
Our treatment can be read mostly independently from the previous sections, but is motivated by Summary~\ref{sum:summary}.
First, in Section~\ref{sec:a_formulation}, we formulate the main result of this work, Theorem~\ref{thm:hu_kuo_ting_generalized}, together with its Corollary~\ref{cor:corollary_for_kolmogorovy} that allows it to be applied to Kolmogorov complexity in Section~\ref{sec:kolmogorov_complexity} and the generalization error in Section~\ref{sec:example_applications}.
The formulation relies on a group-valued measure whose construction we motivate visually in Section~\ref{sec:explicit_construction}.
Afterwards, in Section~\ref{sec:general_consequences}, we deduce some general consequences on how (conditional) interaction terms of different degrees can be related to each other.
The proofs can be found in Appendix~\ref{sec:proofs_section_5}.

\subsection{A Formulation of the Generalized Hu Theorem}\label{sec:a_formulation}

Let $M$ be a commutative, idempotent monoid.
We assume that $M$ is finitely generated, meaning there are elements $X_1, \dots, X_n \in M$ such that all elements in $M$ can be written as arbitrary finite products of the elements $X_1, \dots, X_n$.
Since $M$ is commutative and idempotent, all elements in $M$ are of the form $X_I = \prod_{i \in I}X_i$ for some subset $I \subseteq [n] = \{1, \dots, n\}$, and $X_IX_J \coloneqq X_I \cdot X_J = X_{I \cup J}$.
Additionally, fix an abelian group $G$ and an additive monoid action $.: M \times G \to G$.

For each $\emptyset \neq I \subseteq [n]$, we denote by $p_I$ an abstract atom.
The only property we require of them is to be pairwise different, i.e., $p_I \neq p_J$ if $I \neq J$.
Then, set $\widetilde{X}$ as the set of all these atoms:
\begin{equation}\label{eq:Sigma_definition}
  \widetilde{X} \coloneqq \big\lbrace p_{I} \mid  \emptyset \neq I \subseteq [n] \big\rbrace.
\end{equation}
The atoms $p_I$ represent all smallest parts (the intersections of sets with indices in $I$ minus the sets with indices in $[n]\setminus I$) of a general Venn diagram for $n$ sets.

For $i \in [n]$, we denote by $\widetilde{X}_i \coloneqq \big\lbrace p_I \in \widetilde{X} \ | \ i \in I \big\rbrace$  a set which we can imagine to be depicted by a ``disk'' corresponding to the variable $X_i$, and we denote by $\widetilde{X}_I \coloneqq \bigcup_{i \in I} \widetilde{X}_i$ the union of the ``disks'' corresponding to the joint variable $X_I$.
Clearly, we have $\widetilde{X} = \widetilde{X}_{[n]}$.
This is actually the simplest construction that leads to the $\widetilde{X}_i$ being in general position, as we have the following for all $\emptyset \neq I \subseteq [n]$:

\begin{equation}\label{eq:in_general_position}
  \bigcap_{i \in I} \widetilde{X}_i \ \  \setminus \ \  \bigcup_{j \in [n] \setminus I} \widetilde{X}_j = \{p_I\}.
\end{equation}

\begin{figure}
        \centering
	\includegraphics[width=\textwidth]{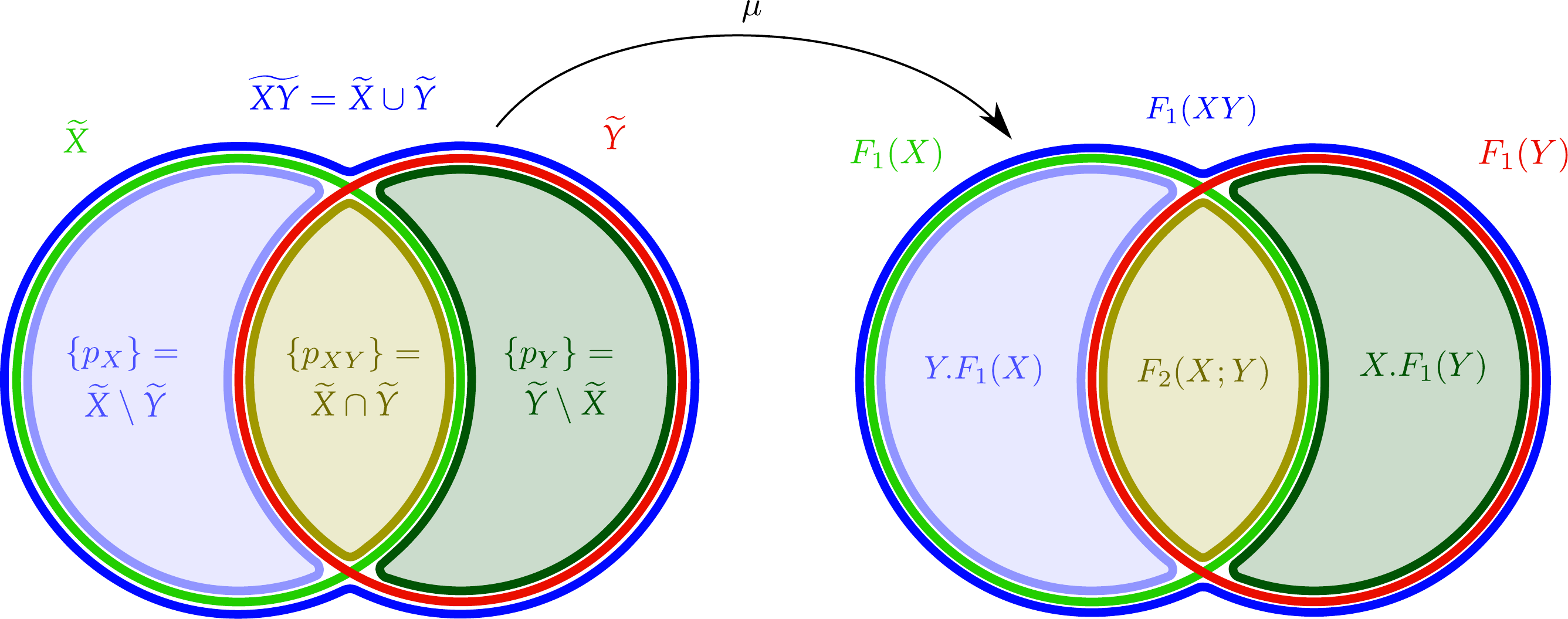}
	\caption{The generalized Hu theorem, visualized for a commutative, idempotent monoid $M$ generated by $X, Y$, and for $F_1$ and $F_2$. 
	  The measure $\mu$ turns sets into elements of the abelian group $G$ and disjoint unions into sums.
	}
	\label{fig:hu_kuo_ting_simple_visualization}
\end{figure}

\begin{figure}
        \centering
	\includegraphics[width=\textwidth]{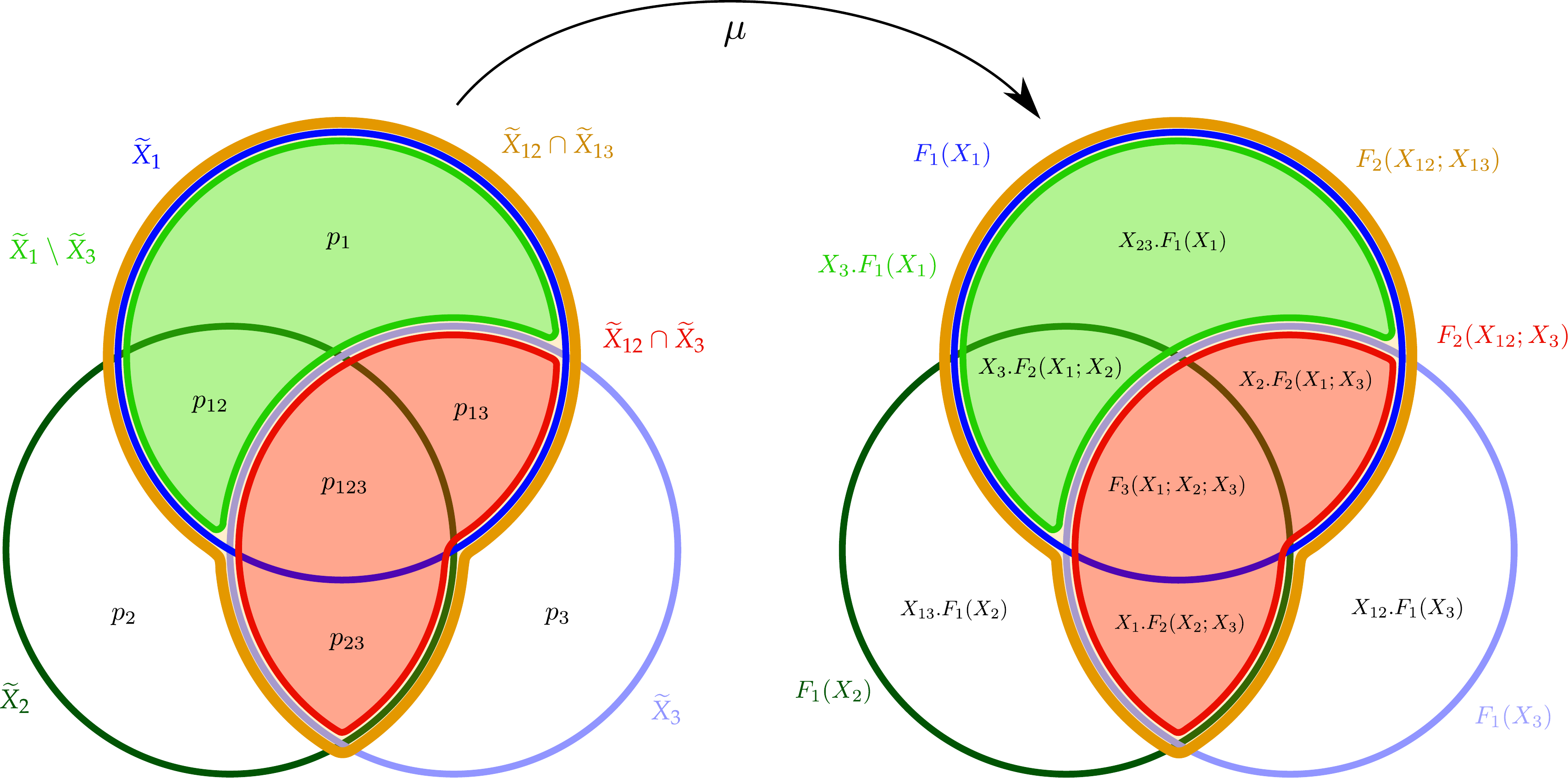}
	\caption{A visualization of the generalized Hu theorem for a commutative, idempotent monoid generated by $X_1, X_2, X_3$.
	  On the left-hand-side, three subsets of the abstract set $\widetilde{X}$ are emphasized, namely $\widetilde{X}_{12} \cap \widetilde{X}_{13}$, $\widetilde{X}_1 \setminus \widetilde{X}_3$, and $\widetilde{X}_{12} \cap \widetilde{X}_3$.
	  On the right-hand-side, Equation~\eqref{eq:hu_kuo_ting_equation} turns them into elements of the abelian group $G$, namely $F_2(X_{12}; X_{13})$, $X_3.F_1(X_1)$, and $F_2(X_{12}; X_3)$, respectively.
    Many decompositions of information functions into sums directly follow from the theorem by using that $\mu$ turns disjoint unions into sums, as exemplified by the equation $F_2(X_{12};X_{13}) =  X_3.F_1(X_1) + F_2(X_{12}; X_3)$.}
	\label{fig:hu_kuo_ting_triple_visualization}
\end{figure}

\begin{figure}
        \centering
	\includegraphics[width=\textwidth]{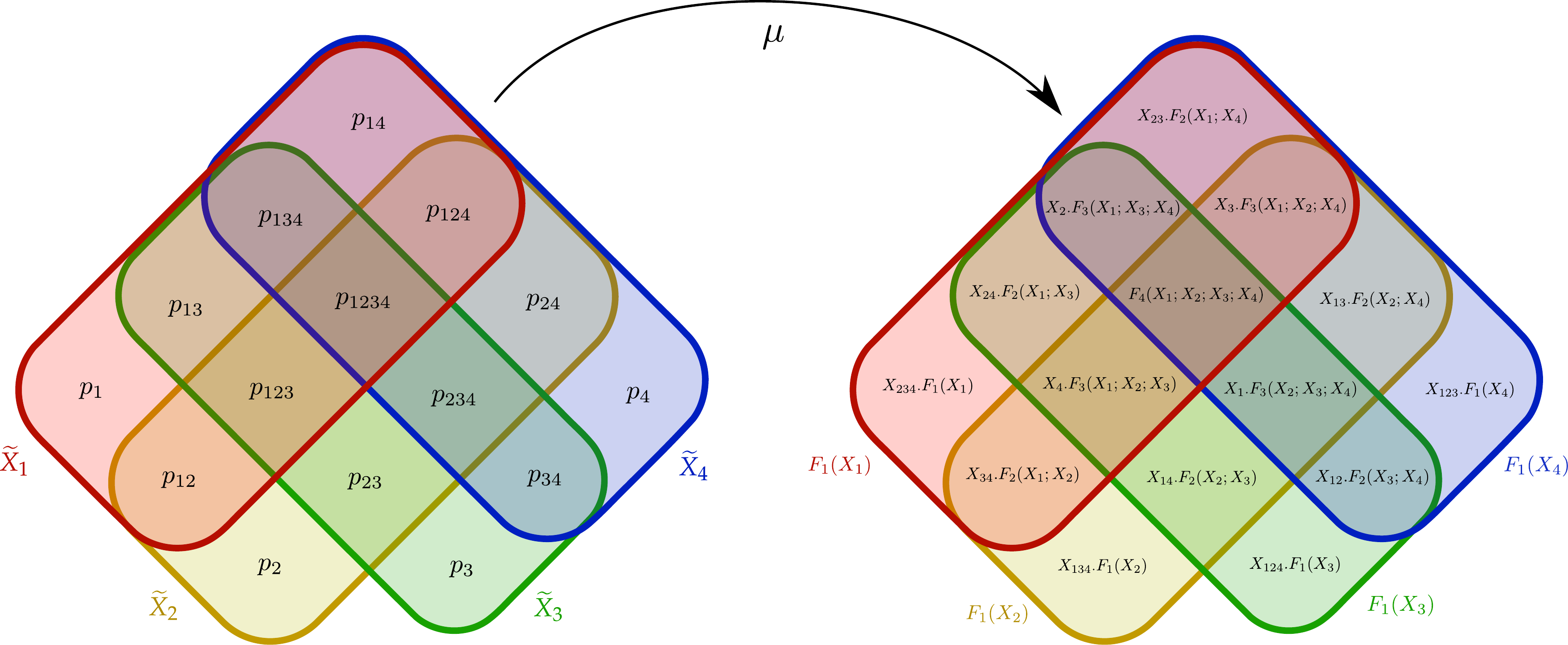}
	\caption{A visualization of the generalized Hu theorem for a commutative, idempotent monoid $M$ generated by $X_1, X_2, X_3, X_4$.
	To reduce clutter, we restrict to a visualization of the abstract sets $\widetilde{X}_i$ and the atoms $p_I$, as well as the corresponding information functions.
      On the right-hand-side, for computing $\mu(p_I)$ for the $15$ atoms $p_I$, we use Lemma~\ref{lem:p_I_charac}.}
	\label{fig:hu_kuo_ting_quadruple_visualization}
\end{figure}

We remark that $\widetilde{X}$ depends on $n$ and could therefore also be written as $\widetilde{X}(n)$.
We will in most cases abstain from this to not overload the notation.
In general, $\widetilde{X}$ has $2^n - 1$ elements.
Therefore, for $n = 2$, $n = 3$ and $n = 4$, $\widetilde{X}$ has $3$, $7$, and $15$ elements, respectively, see Figures~\ref{fig:hu_kuo_ting_simple_visualization},~\ref{fig:hu_kuo_ting_triple_visualization} and~\ref{fig:hu_kuo_ting_quadruple_visualization}.

Remember that for a set $\Sigma$, $2^{\Sigma}$ is its powerset, i.e., the set of its subsets.

\begin{definition}[($G$-Valued) Measure]\label{def:group-valued_measure_new}
  Let $G$ be an abelian group and $\Sigma$ a set.
  A \emph{$G$-valued measure} (on $\Sigma$) is a function $\mu: 2^{\Sigma} \to G$ with the property 
  \begin{equation*}
    \mu(A_1 \cup A_2) = \mu(A_1) + \mu(A_2)
  \end{equation*}
  for all disjoint $A_1, A_2 \subseteq \Sigma$.
\end{definition}

\begin{theorem}[Generalized Hu Theorem; See Section~\ref{sec:the_proof} and Proof~\ref{prf:thm:hu_kuo_ting_generalized}]\label{thm:hu_kuo_ting_generalized}
  Let $M$ be a commutative, idempotent monoid generated by $X_1, \dots, X_n$, $G$ an abelian group, $. : M \times G \to G$ an additive monoid action, and $\widetilde{X} = \widetilde{X}(n)$.
\begin{enumerate}
  \item  Assume $F_1: M \to G$ is a function that satisfies the following chain rule: for all $X, Y \in M$, one has
    \begin{equation}\label{eq:cocycle_equationn}
      F_1(XY) =  F_1(X) +  X.F_1(Y)  .
   \end{equation}
Construct $F_q: M^q \to G$ for $q \geq 2$ inductively by
\begin{equation}\label{eq:inductive_definition}
  F_q(Y_1; \dots ; Y_q) \coloneqq F_{q-1}(Y_1; \dots ; Y_{q-1}) - Y_q.F_{q-1}(Y_1; \dots ; Y_{q-1})
\end{equation}
for all $Y_1, \dots, Y_q \in M$.

Then there exists a $G$-valued measure $\mu: 2^{\widetilde{X}} \to G$ such that for all $q \geq 1$ and $J, L_1, \dots, L_q \subseteq [n]$, the following identity holds:
\begin{equation}\label{eq:hu_kuo_ting_equation}
  X_J.F_q(X_{L_1}; \dots ; X_{L_q}) = \mu\Bigg(\bigcap_{k = 1}^q \widetilde{X}_{L_k} \setminus \widetilde{X}_J\Bigg).
\end{equation}
Concretely, one can define $\mu$ as the unique $G$-valued measure that is on individual atoms $p_I \in \widetilde{X}$ defined by
\begin{equation}\label{eq:measure_pointwise_definition}
  \mu(p_I) \coloneqq \sum_{\emptyset \neq K \supseteq I^c} (-1)^{|K|+|I|+1-n} \cdot F_1(X_K),
\end{equation}
where $I^c = [n] \setminus I$ is the complement of $I$ in $[n]$.\footnote{Alternatively, noting that $F_1(X_{\emptyset}) = 0$ and writing $K = K' \cup I^c$ for some unique $K' \subseteq I$, we can also write $\mu(p_I) = \sum_{K \subseteq I} (-1)^{|K| + 1} \cdot F_1(X_K X_{I^c})$.}
\item Conversely, assume that $\mu: 2^{\widetilde{X}} \to G$ is a $G$-valued measure. 
  Assume there is a sequence of functions $F_q: M^q \to G$ that satisfy Equation~\eqref{eq:hu_kuo_ting_equation}.
  Then $F_1$ satisfies Equation~\eqref{eq:cocycle_equationn} and $F_q$ is related to $F_{q-1}$ as in Equation~\eqref{eq:inductive_definition}.
\end{enumerate}
\end{theorem}

\begin{proof}[Sketch of Proof]
  Part 1 can be shown as follows:
 When specializing Equation~\eqref{eq:hu_kuo_ting_equation} to the case $X_J = \one$ and $q = 1$, one obtains 
 \begin{equation*}
   F_1(X_K) = \mu(\widetilde{X}_K) = \sum_{I \colon I \cap K \neq \emptyset} \mu(p_I),
 \end{equation*} 
 which follows from the Möbius inversion formula on a poset~\citep[3.7.1 Proposition]{Stanley_2011} from Equation~\eqref{eq:measure_pointwise_definition}.
The general formula for $q > 1$ then follows by induction using the properties of the monoid action. 
Part 2 follows by a direct computation. 

More details can be found in Appendix~\ref{sec:the_proof}.
\end{proof}

The following corollary will be applied to Kolmogorov complexity in Section~\ref{sec:kolmogorov_complexity} and the generalization error in machine learning in Section~\ref{sec:example_applications}.

\begin{corollary}[Hu's Theorem for Two-Argument Functions; see Proof~\ref{prf:cor:corollary_for_kolmogorovy}]\label{cor:corollary_for_kolmogorovy}
  Let $M$ be a commutative, idempotent monoid generated by $X_1, \dots, X_n$, $G$ an abelian group, and $\widetilde{X} = \widetilde{X}(n)$.
  Assume that $K_1: M \times M \to G$ is a function satisfying the following chain rule:
  \begin{equation}\label{eq:new_chain_rule_two_argument}
    K_1(XY) = K_1(X) + K_1(Y \mid X),
  \end{equation}
  where we define $K_1(X) \coloneqq K_1(X \mid \one)$ for all $X \in M$.
  Construct $K_q: M^q \times M \to G$ for $q \geq 2$ inductively by
  \begin{equation}\label{eq:inductive_definition_K}
    K_q\big(Y_1; \dots ; Y_q \mid Z\big) \coloneqq K_{q-1}\big(Y_1; \dots; Y_{q-1} \mid Z\big) - K_{q-1}\big(Y_1; \dots ; Y_{q-1} \mid Y_{q}Z\big).
  \end{equation}
  Then there exists a $G$-valued measure $\mu: 2^{\widetilde{X}} \to G$ such that for all $L_1, \dots, L_q, J \subseteq [n]$, the following identity holds:
  \begin{equation}\label{eq:hkt_formula_for_K}
     K_q (X_{L_1}; \dots ; X_{L_q} \mid X_J) = \mu \Bigg( \bigcap_{k = 1}^q \widetilde{X}_{L_k} \setminus \widetilde{X}_J\Bigg).
  \end{equation}
  Concretely, one can define $\mu$ as the unique $G$-valued measure that is on individual atoms $p_I \in \widetilde{X}$ defined by
  \begin{equation}\label{eq:explicit_construction_for_K}
    \mu(p_I) \coloneqq \sum_{\emptyset \neq K \supseteq I^c} (-1)^{|K| + |I| + 1 - n} \cdot K_1(X_K),
  \end{equation}
  where $I^c = [n] \setminus I$ is the complement of $I$ in $[n]$.
\end{corollary}

We conclude by discussing how Hu's theorem can be visualized, for which we will prove one further elementary lemma.
For $I = \{i_1, \dots, i_q\} \subseteq [n]$, set 
\begin{equation}\label{eq:eta_I_def}
  \eta_I \coloneqq X_{[n] \setminus I}.F_q(X_{i_1}; \dots ; X_{i_q}).
\end{equation}
For the special case that $F_q = I_q$ is interaction information, these functions were discussed in~\cite{baudot2019} as generators of all information functions of the form $X_J.I_q(X_{L_1}; \dots ; X_{L_q})$.
The following lemma gives an explanation for this:
the functions $\eta_I$ generate the information measure (or, more generally: $G$-valued measure) $\mu$, which in turn generates all other information functions:

\begin{lemma}\label{lem:p_I_charac}
  Let $\emptyset \neq I \subseteq [n]$ be arbitrary. 
  Then $\eta_I = \mu(p_I)$.
\end{lemma}

\begin{proof}
  According to~\Cref{eq:in_general_position}, we have
   \begin{equation}\label{eq:lemlem}
     \bigcap_{i \in I} \widetilde{X}_i \setminus \widetilde{X}_{[n] \setminus I} = \{p_I\}.
   \end{equation}
   Thus, the lemma follows from Theorem~\ref{thm:hu_kuo_ting_generalized}.
\end{proof}

Thus, Theorem~\ref{thm:hu_kuo_ting_generalized} can be visualized as follows:
For each element $X_1, \dots, X_n$, draw a disk $\widetilde{X}_i$ such that they intersect ``in general position'', meaning that all intersections of (part of) the disks are present.
Assign the function $\eta_I$ to each atom $p_I$, as in the preceding lemma.
Furthermore, assign subsets of $\widetilde{X}$ to information functions according to Equation~\eqref{eq:hu_kuo_ting_equation}.
See Figures~\ref{fig:hu_kuo_ting_simple_visualization} and~\ref{fig:hu_kuo_ting_quadruple_visualization} for examples.
In Figure~\ref{fig:hu_kuo_ting_triple_visualization}, we exemplify how to use these diagrams to visually represent and prove identities of information functions.
Note that in all figures, we write sets $I = \{i_1, \dots, i_k\}$ for simplicity just as the sequence $i_1i_2\dots i_k$.

\subsection{Explicit Construction of the \texorpdfstring{$G$}{G}-Valued Measure \texorpdfstring{$\mu$}{mu}}\label{sec:explicit_construction}

Assume all notation as in part 1 of Theorem~\ref{thm:hu_kuo_ting_generalized}.
In this subsection, we explain how one could ``guess'' Equation~\eqref{eq:measure_pointwise_definition} without knowledge of Möbius inversion theory.
This section is meant as motivation, and other sections do not depend on it.

The high-level idea is the following: 
we have the sequence of functions $F_1, F_2, \dots, $ as our data to work with.
We also know that $F_q$ is constructed from $F_{q-1}$ for all $q \geq 2$, which means that we should be able to express the measure $\mu$ in terms of $F_1$ alone.
Additionally, we must have $F_1(X_K) = \mu(\widetilde{X}_K)$ in the end.
Thus, our aim is to explain how, for arbitrary $\emptyset \neq I \subseteq [n]$, we can express $\mu(p_I)$ using only terms $\mu(\widetilde{X}_K)$ with $K \subseteq [n]$.
This idea, while carried out differently, is also at the heart of the proof of the existence of information diagrams given in~\citep{Yeung1991,Yeung2002}.

We now look at some examples for $n$ and $I$ and derive $\mu(p_I)$ from the $\mu(\widetilde{X}_K)$.
In the following visual computations, each Venn diagram always depicts the measure of the grey area.
We frequently make use of the fact that $\mu$ is a $G$-valued measure.
For $n = 1$ and $I = \{1\} = 1$,\footnote{For simplicity, we write sets as a sequence of their elements.} we obtain:

\begin{figure}[H]
 \centering
 \includegraphics{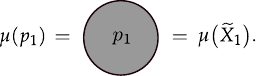}
\end{figure}

For $n = 2$ and $I = \{1\} = 1$, we have:

\begin{figure}[H]
 \centering
 \includegraphics{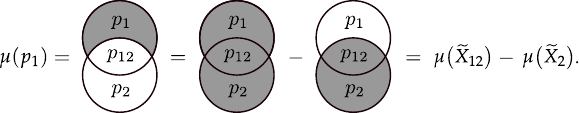}
\end{figure}

For $n = 2$ and $I = \{2\} = 2$, we get the same situation with $1$ and $2$ exchanged:

\begin{figure}[H]
 \centering
 \includegraphics{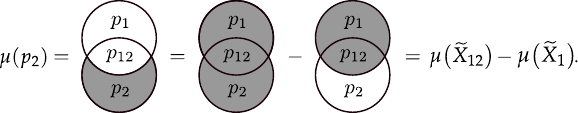}
\end{figure}

Next, we look at the case $n = 2$, $I = \{1, 2\} = 12$:

\begin{figure}[H]
 \centering
 \includegraphics{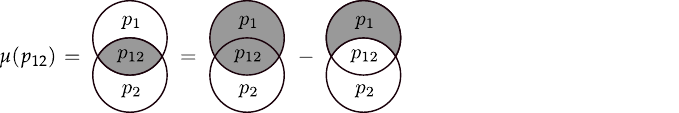}
\end{figure}

\begin{figure}[H]
 \centering
 \includegraphics{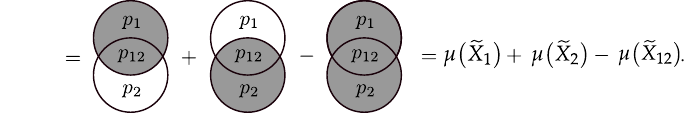}
\end{figure}

Finally, for $n = 3$ and $I = \{1, 2\} = 12$, we obtain:

\begin{figure}[H]
 \centering
 \includegraphics{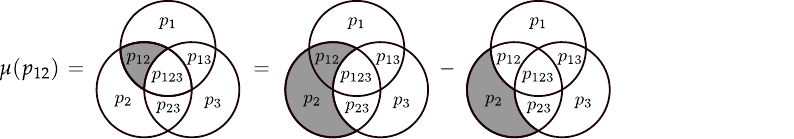}
\end{figure}

\begin{figure}[H]
 \centering
 \includegraphics{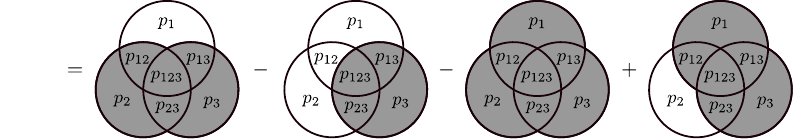}
\end{figure}

\begin{figure}[H]
 \centering
 \includegraphics{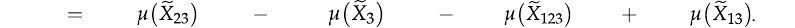}
\end{figure}

In all cases, we managed to achieve our goal to only use terms of the form $\mu(\widetilde{X}_K)$.
Additionally, a close look at the coefficients shows that these examples obey Equation~\eqref{eq:measure_pointwise_definition}, as desired.

\subsection{General Consequences of the Explicit Construction of \texorpdfstring{$\mu$}{mu}}\label{sec:general_consequences}

Assume the setting as in part 1 of Theorem~\ref{thm:hu_kuo_ting_generalized}, which we now consider proven.
In this section, we consider general consequences of Hu's theorem that specifically use the explicit construction, Equation~\eqref{eq:measure_pointwise_definition}, of the $G$-valued measure $\mu: 2^{\widetilde{X}} \to G$.
Corollary~\ref{cor:general_consequences} explains how three different information functions can be expressed with respect to each other.

\begin{corollary}[See Proof~\ref{prf:cor:general_consequences}]\label{cor:general_consequences}
  Recall the functions $\eta_I$ from Equation~\eqref{eq:eta_I_def}.
    We obtain the following identities:
    \begin{enumerate}
      \item Let $1 \leq q \leq n$ and $\emptyset \neq I = \{i_1, \dots, i_q\} \subseteq [n]$. 
	Then 
	\begin{equation*}
	  \eta_I = \sum_{\emptyset \neq K \supseteq I^c} (-1)^{|K| + |I| + 1 - n} \cdot F_1(X_K).
	\end{equation*}
      \item Let $K \subseteq [n]$ arbitrary. Then 
	\begin{equation*}
	  F_1(X_K) = \sum_{\substack{I \subseteq [n] \\ I \cap K \neq \emptyset}} \eta_I.
	\end{equation*}
      \item Let $1 \leq q \leq n$ and $\emptyset \neq J = \{j_1, \dots, j_q\} \subseteq [n]$ be arbitrary. 
	Then 
	\begin{equation*}
	  F_q(X_{j_1}; \dots ; X_{j_q}) = \sum_{I \supseteq J} \eta_I.
	\end{equation*}
      \item For $\emptyset \neq I \subseteq [n]$, we have 
	\begin{equation*}
	  \eta_I = \sum_{J \supseteq I}(-1)^{|J| - |I|} \cdot F_{|J|}(X_{j_1}; \dots ; X_{j_{|J|}}).
	\end{equation*}
      \item Let $K \subseteq [n]$ arbitrary.
	Then one has 
	\begin{equation*}
	  F_1(X_K) = \sum_{\emptyset \neq J \subseteq K} (-1)^{|J| + 1} \cdot F_{|J|}(X_{j_1}; \dots ; X_{j_{|J|}}).
	\end{equation*}
      \item Let $1 \leq q \leq n$ and $\emptyset \neq J = \{j_1, \dots, j_q\} \subseteq [n]$. 
	Then one has 
	\begin{equation*}
	  F_q(X_{j_1}; \dots; X_{j_q}) = \sum_{\emptyset \neq K \subseteq J}  (-1)^{|K| + 1} \cdot F_1(X_K).
	\end{equation*}
      \end{enumerate}
\end{corollary}

\section{Hu's Theorem for Kolmogorov Complexity}\label{sec:kolmogorov_complexity}

In this section, we establish the generalization of Hu's theorem for two-argument functions, Corollary~\ref{cor:corollary_for_kolmogorovy}, for different versions of Kolmogorov complexity.
All of these versions satisfy a chain rule up to certain error terms.
These can all be handled in our framework, but the most exact chain rule holds for \emph{Chaitin's prefix-free Kolmogorov complexity}, on which we therefore focus our attention.
Our main references are~\cite{chaitin1987,li1997,grunwald2008}.
In this whole section, we work with the binary logarithm, which we denote by $\log$, instead of the natural logarithm $\ln$. 

This section is written with minimal prerequisites on the reader. We proceed as follows: in Section~\ref{sec:preliminaries}, we explain the preliminaries of prefix-free Kolmogorov complexity.
Then in Section~\ref{sec:chain_rule_chaitin}, we state the chain rule of Chaitin's prefix-free Kolmogorov complexity, which holds up to an additive constant. 
We reformulate this chain rule in Section~\ref{sec:hu_kuo_ting_chaitin} to satisfy the general assumptions of Corollary~\ref{cor:corollary_for_kolmogorovy} for two-argument functions.
In Section~\ref{sec:HKT_theorem_chaitin_interpreted}, we then define interaction complexity analogously to interaction information, and make the resulting Hu theorem explicit.

Then in Section~\ref{sec:interaction_inf_is_expected_complexity}, we combine the two Hu theorems for interaction complexity and Shannon interaction information and show that expected interaction complexity is up to an error term equal to interaction information.
This leads to the remarkable result that in all degrees, the ``per-bit'' expected interaction complexity equals interaction information for sequences of well-behaved probability measures on increasing sequence lengths.

Finally, the Sections~\ref{sec:hkt_prefix_free} and~\ref{sec:hkt_plain_complexity} then summarize the resulting chain rules for standard prefix-free Kolmogorov complexity and plain Kolmogorov complexity, leaving more concrete interpretations of the resulting Hu theorems to future work.

Most proofs for this section can be found in Appendix~\ref{sec:proofs_section_6}.

\subsection{Preliminaries on Prefix-Free Kolmogorov Complexity}\label{sec:preliminaries}

Let the \emph{alphabet} be given by $\{0, 1\}$.
The set of \emph{binary strings} is given by
\begin{equation*}
  \bins \coloneqq \{\epsilon, 0, 1, 00, 01, 10, 11, 000, \dots\},
\end{equation*} 
where $\epsilon$ is the empty string.
The above lexicographical ordering defines a bijection $\N \to \bins$ that we use to identify natural numbers with binary strings.
Concretely, this identification maps
\begin{equation}\label{eq:identification_binary}
  0 \mapsto \epsilon, \ \ \ \ 1 \mapsto 0, \ \ \ \ 2 \mapsto 1, \ \ \ \  3 \mapsto 00, \ \ \ \  4 \mapsto 01, \ \ \ \ 5 \mapsto 10, \dots
\end{equation}
We silently switch between viewing natural numbers as ``just numbers'' and viewing them as binary strings and vice versa. 

If $x, y \in \bins$ are two binary strings, then we can concatenate them to obtain a new binary string $xy \in \bins$.
A string $x \in \bins$ is a proper prefix of the string $y \in \bins$ if there is a string $z \in \bins$ with $z \neq \epsilon$ such that $y = xz$. 
A set $\mathcal{A} \subseteq \bins$ is called \emph{prefix-free} if no element in $\mathcal{A}$ is a proper prefix of any other element in $\mathcal{A}$.

Let $\mathcal{X}$ and $\mathcal{Y}$ be sets. A \emph{partial function} $f: \mathcal{X} \dashrightarrow \mathcal{Y}$ is a function $f: \mathcal{A} \to \mathcal{Y}$ defined on a subset $\mathcal{A} \subseteq \mathcal{X}$.
A \emph{decoder} for a set $\mathcal{X}$ is a partial function $D: \bins \dashrightarrow \mathcal{X}$.\footnote{Often, the word \emph{code} is used instead of \emph{decoder}.
We find ``decoder'' less confusing.}
A decoder can be thought of as \emph{decoding} the \emph{code words} in $\bins$ into \emph{source words} in $\mathcal{X}$.
A decoder $D: \bins \dashrightarrow \mathcal{X}$ is called a \emph{prefix-free decoder} if its domain $\mathcal{A} \subseteq \bins$ is prefix-free.\footnote{In the literature, this is often called a \emph{prefix code}. 
We choose the name ``prefix-free'' as it avoids possible confusions.}

For a binary string $x$, $l(x)$ is defined to be its \emph{length}, meaning the number of its symbols. 
Thus, for example, we have $l(\epsilon) = 0$ and $l(01) = 2$.
Let $D: \bins \dashrightarrow \mathcal{X}$ be a decoder.
We define the length function $L_D: \mathcal{X} \to \N \cup \{\infty\}$ via
\begin{equation*}
  L_D(x) \coloneqq \min\big\lbrace l(y) \mid y \in \bins, \ D(y) = x\big\rbrace,
\end{equation*}
which is $\infty$ if $D^{-1}(x) = \emptyset$.

In the following, we make use of the notion of a \emph{Turing machine}.
This can be imagined as a machine with very simple rules that implements an algorithm. 
We will not actually work with concrete definitions of Turing machines;
instead, we let Church's Thesis~\ref{thesis:churchs_thesis} do the work, which we describe below --- it will guarantee that any function that \emph{intuitively} resembles an algorithm could equivalently be described by a Turing machine.
Concrete definitions can be found in Chapter 1.7 of~\cite{li1997}.

A \emph{partial computable function} is any partial function $T: \bins \dashrightarrow \bins$ that can be computed by a Turing machine. 
The Turing machine \emph{halts} on precisely the inputs on which $T$ is defined.
We do not distinguish between Turing machines and the corresponding partial computable functions:
If $T$ is a partial computable function, then we say that $T$ \emph{is} a Turing machine.
If $x \in \bins$ is in the domain of the Turing machine $T$, we say that $T$ \emph{halts} on $x$ and write $T(x) < \infty$.
If $T$ does not halt on $x$, we sometimes write $T(x) = \infty$.

By the Church-Turing thesis, partial computable functions are precisely the partial functions for which there is an ``algorithm in the intuitive sense'' that computes the output for each input.
We reproduce the formulation from~\cite{li1997}:

\begin{thesis}[Church's Thesis]\label{thesis:churchs_thesis}
  The class of algorithmically computable partial functions (in the intuitive sense) coincides with the class of partial computable functions.
\end{thesis}

We now define two prefix-free decoders for binary sequences.
To do that, we first define the corresponding \emph{encoders}:
define the encoder $(\cdot)': \bins \to \bins$ by
\begin{equation}\label{eq:integer_coding_formula}
  x' \coloneqq 1^{l(l(x))}0l(x)x.\footnotemark
\end{equation}
\footnotetext{In the literature, this is viewed as a code for the \emph{natural numbers} instead of $\bins$. But both viewpoints are equivalent due to the bijection $\N \cong \bins$.}Note that the natural number $l(x)$ is viewed as a binary string using the identification in Equation~\eqref{eq:identification_binary}.

The decoder corresponding to $(\cdot)'$ is a partial computable function $D': \bins \dashrightarrow \bins$ that is only defined on inputs of the form $x'$. 
The underlying algorithm reads until the first $0$ to know the length of the bitstring representing $l(x)$. 
Then it reads until the end of $l(x)$ to know the length of $x$.
Subsequently, it can read until the end of $x$ to know $x$ itself, which it then outputs. 
This decoder is prefix-free:
if $x'$ is a prefix of $y'$, then $l(x) = l(y)$ and $x$ is a prefix of $y$, from which $x = y$ and thus $x' = y'$ follows.

Let a pairing function $\bins \times \bins \to \bins$ be given by
\begin{equation*}
  (x, y) \mapsto  x'y.
\end{equation*}
Note that we can algorithmically recover both $x$ and $y$ from $x'y$:
reading the string $x'y$ from the left, the algorithm first recovers $l(x)$ and then $x$, after which the rest of the string automatically is $y$.

A Turing machine $T: \bins \dashrightarrow \bins$ is called a \emph{prefix-free machine} if it is a prefix-free decoder.
The input is then imagined to be a code word encoding the output string.
There is a bijective, computable enumeration, called \emph{standard enumeration}, $T_1, T_2, T_3, \dots$, of all prefix-free machines
(\cite{li1997}, Section 3.1).
Computable here means the following: 
if we would encode the set of \emph{rules} of any Turing machine as a binary sequence, then the map from natural numbers to binary sequences corresponding to the standard enumeration is itself computable.

A Turing machine $T: \bins \dashrightarrow \bins$ is called a \emph{conditional Turing machine} if for all $x$ such that $T$ halts on $x$ we have $x = y'p$ for some elements $y, p \in \bins$; 
$p$ is then called the \emph{program}, and $y$ the \emph{input}.
A \emph{universal} conditional prefix-free machine is a conditional prefix-free machine $U: \bins \dashrightarrow \bins$ such that for all $i \in \N$ and $y, p \in \bins$, we have $U(y'i'p) = T_i(y'p)$, and $U$ does not halt on inputs of any other form.
Here, again, $i$ is viewed as a binary string via Equation~\eqref{eq:identification_binary}.
One can show that such universal conditional prefix-free machines indeed do exist (\cite{li1997}, Theorem 3.1.1).

For the rest of this article, let $U$ be a fixed universal conditional prefix-free machine.

\begin{definition}[Prefix-Free Kolmogorov Complexity]\label{def:kolmogorov_complexity}
  The \emph{conditional prefix-free Kolmogorov complexity} is the function $K: \bins \times \bins \to \N$ given by
  \begin{align*}
    K(x \mid y) & \coloneqq \min \Big\lbrace l(p) \ \ \big| \ \  p \in \bins, \  U(y'p) = x \Big\rbrace \\
    & = \min \Big\lbrace l(i') + l(q) \ \ \big| \ \  i \in \N, \  q \in \bins, \  U(y'i'q) = x \Big\rbrace \\
    & = \min \Big\lbrace l(i') + l(q) \ \ \big| \ \  i \in \N, \  q \in \bins, \  T_i(y'q) = x \Big\rbrace \\
    & < \infty.
  \end{align*}
  We define the \emph{non-conditional} prefix-free Kolmogorov complexity by $K: \bins \to \N$, $K(x) \coloneqq K(x \mid \epsilon)$.
  As $\epsilon' = 1^{l(l(\epsilon))}0 l(\epsilon) = 0$,\footnote{Here, we used $l(\epsilon) = 0$, which is a \emph{natural number} corresponding to the \emph{string} $\epsilon$ that is plucked back into the formula.} we obtain
  \begin{equation*}
    K(x) = \min \big\lbrace l(p) \mid U(0p) = x\big\rbrace.
  \end{equation*}
  Here, the $0$ can be thought of as simply signaling that there is no input, while each ``actual'' input starts with a $1$ due to the definition of $y'$.
\end{definition}

\begin{definition}[Joint Conditional Prefix-Free Kolmogorov Complexity]
  \label{def:joint_kolmogorov_complexity}
  Define $\Concat: (\bins)^n \to \bins$ by $\Concat(x_1, \dots, x_n) \coloneqq x_1' \cdots x_{n-1}'x_n$.
  For $x_1, \dots, x_n \in \bins$ and $y_1, \dots, y_m \in \bins$, we define the \emph{(joint conditional) prefix-free Kolmogorov complexity} by
  \begin{equation*}
    K\big(x_1, \dots, x_n \mid y_1, \dots, y_m\big) \coloneqq K\big(\Concat(x_1, \dots, x_n) \mid \Concat(Y_1, \dots, y_m)\big).
  \end{equation*}
  We then simply set $K\big(x_1, \dots, x_n\big) \coloneqq K\big(x_1, \dots, x_n \mid \epsilon\big)$.
\end{definition}

\subsection{The Chain Rule for Chaitin's Prefix-Free Kolmogorov Complexity}\label{sec:chain_rule_chaitin}

Let $f, g: \mathcal{X} \to \R$ be two functions on a set $\mathcal{X}$.
We adopt the following notation from~\cite{grunwald2008}: 
$f \plle g$ means that there is a constant $c \geq 0$ such that $f(x)  < g(x) + c$ for all $x \in \mathcal{X}$.
We write $f \plge g$ if $g \plle f$.
Finally, we write $f \ple g$ if $f \plle g$ and $f \plge g$, which means that there is a constant $c \geq 0$ such that $\big|f(x) - g(x)\big| < c$ for all $x \in \mathcal{X}$.
If we want to emphasize the inputs, we may, for example, also write $f(x) \ple g(x)$.

Let $x \in \bins$ be arbitrary and $K(x)$ its prefix-free Kolmogorov complexity.
Let $x^* \in \bins$ be chosen as follows:
we look at all $y \in \bins$ of length $l(y) = K(x)$ such that $U(0y) = x$.
Among those, we look at all $y$ such that $U$ computes $x$ on input $0y$ with the smallest number of computation steps.
And finally, among those, we define $x^*$ to be the lexicographically first string.
Based on this, Chaitin's prefix-free Kolmogorov complexity is given by 
\begin{equation*}
  \Kc: \bins \times \bins \to \R, \quad \Kc(x \mid y) \coloneqq K(x \mid y^*)
\end{equation*}
and $\Kc(x) \coloneqq \Kc(x \mid \epsilon)$.

Clearly, there is a program that, on input $x'K(x)$, outputs $x^*$ --- we simply run $U(0y)$ for all programs $y$ of length $K(x)$ in parallel, and the one that outputs $x$ the fastest and is lexicographically first among those is the output $x^*$.
Vice versa, given $x^*$, one can compute $x'K(x)$ by simply computing $U(0x^*)'l(x^*)$.
In this sense, $x^*$ and $x'K(x)$ can be said to ``contain the same information''.
In the literature, Chaitin's prefix-free Kolmogorov complexity is, for this reason, also often defined by $\Kc(x \mid y) \coloneqq K(x \mid y, K(y))$.

The following result might have for the first time been written down in~\cite{Gacs1974}, and was attributed therein to Leonid Levin.

\begin{theorem}[Chain Rule for Chaitin's Prefix-Free Kolmogorov Complexity]
  \label{thm:chain_rule_kolmogorov_compl}
  The following identity holds:
  \begin{equation}\label{eq:chain_rule_for_chaitins_prefix_kolm}
    \Kc(x, y) \ple \Kc(x) + \Kc(y \mid x).
  \end{equation} 
  Here, both sides are viewed as functions $(\bins)^2 \to \R$ that map inputs of the form $(x, y)$.
\end{theorem}

\begin{proof}
  See~\cite{li1997}, Theorem 3.8.1 for the proof of the inequality $\Kc(x, y) \plle \Kc(x) + \Kc(y \mid x)$. 
  The proof of the other direction, namely $\Kc(y \mid x) \plle K(x, y) - K(x)$, in \cite{li1997} seems incorrect to us, as it only seems to show that the constant is independent of $x$ and not of $y$.
  See the proof in \cite{chaitin1987} for that direction.
\end{proof}

\subsection{A Reformulation of the Chain Rule in Terms of Our General Framework}\label{sec:hu_kuo_ting_chaitin}

Our goal is to express the result, Equation~\eqref{eq:chain_rule_for_chaitins_prefix_kolm}, in terms of the assumptions of Corollary~\ref{cor:corollary_for_kolmogorovy}.
To do this, we need to find a framework under which the chain rule becomes \emph{exact} instead of correct up to a constant, and in which the inputs come from a monoid.
We will solve this by identifying functions whose difference is bounded by a constant.

For $n \geq 0$ any fixed natural number, we define $\Maps\big((\bins)^n, \R\big)$ as the abelian group of functions from $(\bins)^n$ to $\R$.
We define the equivalence relation $\sim_{\Kc}$ on $\Maps\big( (\bins)^n, \R\big)$ by
\begin{equation*}
  F \sim_{\Kc} H \ \ \ : \Longleftrightarrow F \ple H.
\end{equation*}
The reason we put $\Kc$ in the subscript of $\sim_{\Kc}$ is that later, we will investigate different equivalence relations $\sim_{K}$ and $\sim_C$ for prefix-free and plain Kolmogorov complexity.
Note that the functions $F$ with $F \sim_{\Kc} 0$, i.e., $F \ple 0$, form a subgroup of $\Maps\big((\bins)^n, \R\big)$.
Consequently, we obtain an abelian group $\Maps\big((\bins)^n, \R\big)/ \sim_{\Kc}$ with elements written as $[F]_{\Kc}$.

Now, let the variables $X_1, \dots, X_n$ be defined as the following projections:
\begin{equation*}
  X_i: (\bins)^n \to \bins, \ \ \ \boldsymbol{x} = (x_1, \dots, x_n) \mapsto x_i.
\end{equation*}
Then, for any $i_1, \dots, i_k \in [n]$, we can form the product variable $X_{i_1}\cdots X_{i_k}$:
\begin{equation*}
  X_{i_1} \cdots X_{i_k}: (\bins)^n \to (\bins)^k, \ \ \ \boldsymbol{x} = (x_1, \dots, x_n) \mapsto (x_{i_1}, \dots, x_{i_k}).
\end{equation*}
These strings of projections form the elements of the monoid $\widetilde{M} = \{X_1, \dots, X_n\}^*$, with multiplication simply given by concatenation.
Then from $\Kc: \bins \times \bins \to \R$, we can define the function
\begin{align*}
  [\Kc]_{\Kc}: \ \ & \widetilde{M} \times \widetilde{M} \to \Maps\big( (\bins)^n, \R\big)/\sim_{\Kc}, \\
  & (Y, Z) \mapsto [\Kc(Y \mid Z)]_{\Kc},
\end{align*}
with $\Kc(Y \mid Z)$ simply being the function that inserts tuples from $(\bins)^n$ into the variables $Y$ and $Z$:
\begin{equation*}
  \Kc(Y \mid Z): (\bins)^n \to \R, \ \ \ \boldsymbol{x} \mapsto \Kc(Y(\boldsymbol{x}) \mid Z(\boldsymbol{x})).
\end{equation*}
Similarly as before, one can then define $\Kc(Y): (\bins)^n \to \R$ by $\Kc(Y) \coloneqq \Kc(Y \mid \epsilon)$ with $\epsilon \in \widetilde{M}$ being the empty string of variables.
In the same way, $[\Kc]_{\Kc}(Y) \coloneqq [\Kc]_{\Kc}(Y \mid \epsilon) = [\Kc(Y)]_{\Kc}$.
Since $\epsilon(\boldsymbol{x}) = \epsilon$ for all $\boldsymbol{x} \in (\bins)^n$, these definitions are compatible with the earlier definition $\Kc(x) \coloneqq \Kc(x \mid \epsilon)$ for $x \in \bins$:
we have $\big(\Kc(Y)\big)(\boldsymbol{x}) = \Kc\big(Y(\boldsymbol{x})\big)$.

\begin{proposition}[See Proof~\ref{prf:pro:exact_equality}]
  \label{pro:exact_equality}
  For arbitrary $Y, Z \in \widetilde{M}$, we have the \emph{exact} equality
  \begin{equation}\label{eq:exact_equality}
    [\Kc]_{\Kc}(YZ) = [\Kc]_{\Kc}(Y) + [\Kc]_{\Kc}(Z \mid Y)
  \end{equation}
  of elements in $\Maps\big((\bins)^n, \R\big) / \sim_{\Kc}$.
\end{proposition}

To obtain a commutative, idempotent monoid, we show that we can permute and ``reduce'' the elements in $\widetilde{M}$ without affecting the resulting functions in $\Maps\big((\bins)^n, \R\big)/\sim_{\Kc}$:
for arbitrary $Y = X_{i_1}\cdots X_{i_k} \in \widetilde{M}$ we define the reduction $\overline{Y} \in \widetilde{M}$ by
\begin{align}\label{eq:definition_reduction}
  \overline{Y} \coloneqq X_I \coloneqq \prod_{i \in I} X_i, \quad \text{with } I \coloneqq \Big\lbrace i \in [n] \ \ \big| \ \  \exists s \in [k]: \ i_s = i \Big\rbrace.
\end{align}
Here, the factors $X_i$ with $i \in I$ are assumed to appear in increasing order of the index $i$.

\begin{lemma}[See Proof~\ref{prf:lem:reduction_step}]
  \label{lem:reduction_step}
  For all $Y, Z \in \widetilde{M}$, we have the equality
  \begin{equation*}
    [\Kc]_{\Kc}\big(Y \mid Z\big) = [\Kc]_{\Kc}\big(\overline{Y} \mid \overline{Z}\big)
  \end{equation*}
  in $\Maps\big( (\bins)^n, \R\big) / \sim_{\Kc}$.
\end{lemma}

Now, define the equivalence relation $\sim$ on $\widetilde{M}$ by $Y \sim Z$ if $\overline{Y} = \overline{Z}$, with $\overline{(\cdot)}: \widetilde{M} \to \widetilde{M}$ defined as in Equation~\eqref{eq:definition_reduction}.
We define $M \coloneqq \widetilde{M}/\sim$.
Each element $[Y] \in M$ is then represented by $\overline{Y}$ since $\overline{\overline{Y}} = \overline{Y}$;
it is of the form $\overline{Y} = X_I$ for some $I \subseteq [n]$.
Additionally, if $I \neq J$, then obviously we have $X_I \nsim X_J$, and consequently, there is a one-to-one correspondence between representatives of the form $X_I$ and elements in $M$.
Therefore, we can write elements in $M$ for convenience, and by abuse of notation, simply as $[Y] = X_I$.
We then define the multiplication in $M$ by $[Y] \cdot [Z] \coloneqq [YZ]$, which in the new notation can be written as $X_I \cdot X_J = X_{I \cup J}$ and thus makes $M$ a well-defined commutative, idempotent monoid generated by $X_1, \dots, X_n$.
We define, by abuse of notation, $[\Kc]_{\Kc}: M \times M \to \Maps \big( (\bins)^n, \R\big)/\sim_{\Kc}$ in the obvious way on representatives,
which is well-defined by Lemma~\ref{lem:reduction_step}.
Overall, we obtain by Corollary~\ref{cor:corollary_for_kolmogorovy} a Hu theorem for Chaitin's prefix-free Kolmogorov complexity, which we next explain in more detail.

\subsection{Hu's Theorem for Chaitin's Prefix-Free Kolmogorov Complexity}\label{sec:HKT_theorem_chaitin_interpreted}

We now deduce a Hu theorem for Chaitin's prefix-free Kolmogorov complexity.
We formulate it without the abstraction of equivalence classes from the previous subsection (which is mainly important for the proof), with the goal to obtain an intrinsically more interesting version.
For formulating the result, we first name the higher-degree terms analogously to the interaction information from Definition~\ref{def:interaction_information}:

\begin{definition}
  [Interaction Complexity]
  \label{def:interaction_complexity}
  Define $\Kc_1 \coloneqq \Kc: \bins \times \bins \to \R$ and $\Kc_q: (\bins)^q \times \bins \to \R$ inductively by
  \begin{equation*}
    \Kc_q(y_1; \dots ; y_q \mid z) \coloneqq \Kc_{q-1}(y_1; \dots ; y_{q-1} | z) - \Kc_{q-1}(y_1; \dots ; y_{q-1} \mid y_q, z).
  \end{equation*}
  We call $\Kc_q$ the \emph{interaction complexity of degree $q$}.
\end{definition}

For example, $\Kc_2(x;y) = \Kc_1(x) - \Kc_1(x \mid y)$ measures the reduction of the encoding length of $x$ when having access to $y$. 
E.g., if $x$ is thought of as ``data'' and $y$ thought of as a ``theory'', then $\Kc_2(x; y)$ measures the extent to which $y$ helps in compressing $x$.
See also the last paragraph in~\Cref{sec:unanswered_questions} for more interpretation of the potential meaning of these quantities.
The interpretation of higher-order terms is future work.

For $Y_1, \dots, Y_q, Z \in \widetilde{M} = \{X_1, \dots, X_n\}^*$, define $\Kc_q(Y_1; \dots ; Y_q \mid Z) \in \Maps\big( (\bins)^n, \R\big)$ by
\begin{align*}
  \Kc_q(Y_1; \dots ; Y_q \mid Z): \ \ \boldsymbol{x} \mapsto \Kc_q\big( Y_1(\boldsymbol{x}); \dots ; Y_q(\boldsymbol{x}) \mid Z(\boldsymbol{x})\big).
\end{align*}
One can easily inductively show that
\begin{equation}\label{eq:easy_induction}
  \Kc_q(Y_1; \dots ; Y_q \mid Z) \ple \Kc_{q-1}(Y_1; \dots ; Y_{q-1} \mid Z) - \Kc_{q-1}(Y_1; \dots ; Y_{q-1} \mid Y_q Z).
\end{equation}

The full proof of the following theorem can be found in Appendix~\ref{sec:proofs_section_6}, Proof~\ref{prf:thm:hu_kuo_ting_chaitin_finalized}.
The main ingredient is the chain rule, Proposition~\ref{pro:exact_equality}, together with Corollary~\ref{cor:corollary_for_kolmogorovy}.

\begin{theorem}[See Proof~\ref{prf:thm:hu_kuo_ting_chaitin_finalized}]
  \label{thm:hu_kuo_ting_chaitin_finalized}
  Let $\widetilde{X} = \widetilde{X}(n)$.
  There exists a measure $\mu: 2^{\widetilde{X}} \to \Maps \big( (\bins)^n, \R\big)$ such that for all $L_1, \dots, L_q, J \subseteq [n]$, the relation
  \begin{equation}\label{eq:equality_hkt_chaitin_kolm}
     \Kc_q \big( X_{L_1} ; \dots ; X_{L_q} \mid X_J\big)  \ple  \mu \Bigg( \bigcap_{k = 1}^q \widetilde{X}_{L_k} \setminus \widetilde{X}_J \Bigg)
  \end{equation}
  of functions $(\bins)^n \to \R$ holds.
  Concretely, $\mu$ can be defined as the unique measure that is on individual atoms $p_I \in \widetilde{X}$ defined by
  \begin{equation}
    \mu(p_I) \coloneqq \sum_{\emptyset \neq K \supseteq I^c} (-1)^{|K| + |I| + 1 - n} \cdot \Kc_1(X_K),
    \label{eq:def_mu_other_special_case}
  \end{equation}
  where $I^c = [n] \setminus I$ is the complement of $I$ in $[n]$.
\end{theorem}

\begin{remark}
  \label{rem:write_down_dependences}
  In Theorem~\ref{thm:hu_kuo_ting_chaitin_finalized}, the equality holds up to a constant independent of the input in $(\bins)^n$.
  However, there \emph{is} a dependence on $q$, the degree, and $n$, the number of generating variables.
  We now briefly analyze this.
  
  For analyzing the dependence on $q$, we note that the inductive step of the proof of the generalized Hu Theorem~\ref{thm:hu_kuo_ting_generalized} uses the theorem for degree $q-1$ \emph{twice}.
  That means that the number of comparisons doubles in each degree, leading to a dependence of $q$ of the form $O(2^q)$.
  Can one do better than this?
  One idea might be to not define $\Kc_q$ inductively, but with an inclusion-exclusion--type formula motivated by Corollary~\ref{cor:general_consequences}, part 6.
  One sensible definition is the following:
  \begin{equation*}
    \Kc_q(y_1; \dots ; y_q \mid z) \coloneqq \sum_{K \subseteq [q] }(-1)^{|K|+1} \cdot \Kc_1(\boldsymbol{y}_Kz)
  \end{equation*}
  with 
  \begin{equation}\label{eq:inclusion-exclusion-type}
    \boldsymbol{y}_K \coloneqq \prod_{k \in K} y_k'.
  \end{equation}
  However, this now leads to $2^q$ \emph{summands}, which one would, for a proof of Hu's theorem, individually compare with the evaluation of $\mu$ on a ``disk'' in $\widetilde{X} = \widetilde{X}(n)$.
  As in the general definition Equation~\eqref{eq:inclusion-exclusion-type}, the order of the factors in $\boldsymbol{y}_K$ does not follow the ordering of the generators $x_1, \dots, x_n$, we expect there a reordering of the factors to be necessary for the comparison. 
  This has each time a cost of $O(1)$, thus again leading to a dependence of the form $O(2^q)$.
  We currently do not see a way to improve this.

  Now, for each of the $2^q$ comparisons, we would like to know the dependence on $n$.
    One possible algorithm for bringing $\boldsymbol{y}_K z$ ``in order'' works as follows:
    assuming that all of $y_k$, $k \in K$, and $z$ are given by a permutation (with omissions) of $x_1, \dots, x_n$, then we have to specify $q+1$ permutations, which each involves to specify the position of $n$ elements.
    The position is one of $1, \dots, n$ plus ``omission'', which together has a cost of $\log(n + 1)$.
    Overall, this leads to a dependence on $n$ of $O\big( (q+1) \cdot n \cdot \log (n + 1)\big)$.
 
    Overall, the dependence on $q$ and $n$ together is thus $O\big( 2^q \cdot (q+1) \cdot n \cdot \log (n+1)\big)$.
  \end{remark}

\begin{figure}[H]
  \centering
  \includegraphics[width=\textwidth]{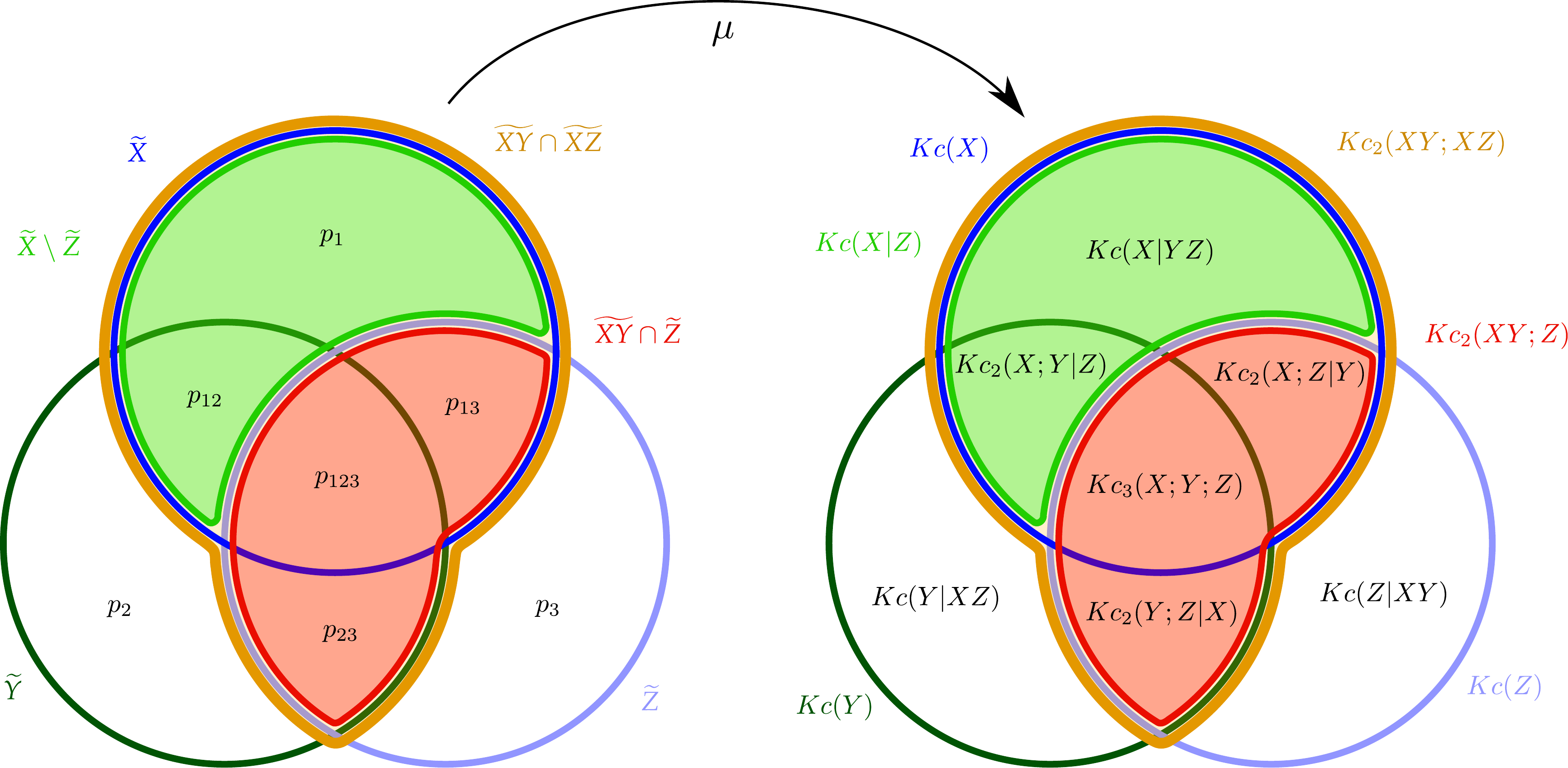}
  \caption{A visualization of Hu's theorem for Kolmogorov complexity for three variables $X, Y, Z$.
	  On the left-hand-side, three subsets of the abstract set $\widetilde{XYZ}$ are emphasized, namely $\widetilde{XY} \cap \widetilde{XZ}$, $\widetilde{X} \setminus \widetilde{Z}$, and $\widetilde{XY} \cap \widetilde{Z}$.
	  On the right-hand-side, Equation~\eqref{eq:equality_hkt_chaitin_kolm} turns them up to a constant error into the Kolmogorov complexity terms $\Kc_2(XY; XZ)$, $\Kc(X \mid Z)$, and $\Kc_2(XY; Z)$, respectively.
	Many decompositions of complexity terms into sums directly follow from the theorem by using that $\mu$ turns disjoint unions into sums, as exemplified by the equation $\Kc_2(XY;XZ) \ple  \Kc(X \mid Z) + \Kc_2(XY; Z)$.}
  \label{fig:triple_visualization_kolmogorov}
\end{figure}

As an Example, we recreate Figure~\ref{fig:hu_kuo_ting_triple_visualization} for the case of Kolmogorov complexity in Figure~\ref{fig:triple_visualization_kolmogorov}.
We can also translate back from the notation with variables to the more familiar notation in which elements of $\bins$ are inserted in the formulas. 
If we do this, then the example equation from Figure~\ref{fig:triple_visualization_kolmogorov} becomes
\begin{equation*}
  \Kc_2(x,y ; x,z) \ple \Kc(x \mid z) + \Kc_2(x, y ; z),
\end{equation*}
where both sides are viewed as functions $(\bins)^3 \to \R$.

\subsection{Expected Interaction Complexity is Interaction Information}\label{sec:interaction_inf_is_expected_complexity}

Recall Definition~\ref{def:interaction_information} of the interaction information of $q$ discrete random variables $Y_1, \dots, Y_q$, denoted $I_q(Y_1; \dots ; Y_q)$.
Additionally, recall that for another discrete random variable $Z$ defined on the same sample space, we can define the averaged conditioning $Z.I_q(Y_1; \dots ; Y_q)$, see Definition~\ref{def:averaged_conditioning}, which is again an information function. 
Its evaluation on a probability measure $P$ on the sample space is denoted $Z.I_q(Y_1; \dots ; Y_q ; P)$.

In this section, we want to establish a relationship between interaction information of random variables defined on $(\bins)^n$ with values in $(\bins)^k$ for some $k$ on the one hand, and the expectation of interaction complexity as defined in Definition~\ref{def:interaction_complexity} on the other hand.
The deviation from an equality between interaction information and interaction complexity will be quantified by the Kolmogorov complexity of probability mass functions.

For this aim, we first need to interpret outputs of Turing machines as rational numbers:
If $T$ is a Turing machine and $T(x) = m'n$ for some $m, n \in \bins$, then interpret $m, n$ as natural numbers via the identification map in~\Cref{eq:identification_binary}, and consequently $m'n$ as the rational number $m/n$, see also~\citet{li1997}, Section 1.7.3.
Interpret the output as $0$ if it is not of the form $m'n$.

\begin{definition}[Kolmogorov Complexity of Probability Mass Functions]
  \label{def:complexity_of_mass_function}
  Let $P: (\bins)^n \to \R$ be a probability mass function.
  Its \emph{Kolmogorov complexity} is defined by
  \begin{equation*}
    K(P) \coloneqq \min_{p \in \bins} \Big\lbrace l(p) \ \ \big| \ \  \forall q \in \N, \ \  \forall \boldsymbol{x} \in (\bins)^n:  \  \ \big| T_p(\boldsymbol{x}'q) - P(\boldsymbol{x}) \big| \leq 1/q\Big\rbrace,
  \end{equation*}
  where $T_p$ is the p'th prefix-free Turing machine.
\end{definition}

\begin{definition}
  [Computability of Probability Mass Functions]
  \label{def:computable_mass_function}
  A probability mass function $P: (\bins)^n \to \R$ is called \emph{computable} if $K(P) < \infty$.
\end{definition}

In other words, a probability mass function $P$ is computable if there exists a prefix-free Turing machine $T_p$ that can, for all natural numbers $q$, approximate $P$ up to precision $1/q$.

We now unify the viewpoint of the variables $X_i$ as ``placeholders'' with the viewpoint that they are random variables:
remember that the $X_i: (\bins)^n \to \bins$ are given by projections:
$X_i(\boldsymbol{x}) = x_i$.
They form the monoid $\widetilde{M} = \{X_1, \dots, X_n\}^*$, with multiplication given by concatenation.
Furthermore, we defined an equivalence relation $\sim$ with $Y \sim Z$ if $\overline{Y} = \overline{Z}$.

Now, interpret $(\bins)^n$ as a discrete sample space.
Then the strings in $Y \in \widetilde{M}$ can be interpreted as random variables on $(\bins)^n$ with values in $(\bins)^k$ for some $k$. 
The concatenation of these strings is identical to the product of random variables defined in Equation~\eqref{eq:def_product_random_variable}.
Now, remember that in Section~\ref{sec:equivalence_classes} we also defined an equivalence relation for random variables, which we now call $\sim_r$ to distinguish it from $\sim$.
For $Y: (\bins)^n \to (\bins)^{k_y}$ and $Z: (\bins)^n \to (\bins)^{k_z}$, we have $Y \sim_r Z$ if there exist functions $f_{ZY}: (\bins)^{k_y} \to (\bins)^{k_z}$ and $f_{YZ}: (\bins)^{k_z} \to (\bins)^{k_y}$ such that $f_{ZY} \circ Y = Z$ and $f_{YZ} \circ Z = Y$.

\begin{lemma}[See Proof~\ref{prf:lem:equivalence_relations_identical}]
  \label{lem:equivalence_relations_identical}
  For all $Y, Z \in \widetilde{M}$, we have
  \begin{equation*}
    Y \sim Z \quad \Longleftrightarrow \quad Y \sim_r Z.
  \end{equation*}
  That is, the equivalence relations $\sim$ and $\sim_r$ are identical.
\end{lemma}

This shows that the commutative, idempotent monoids $M = \{X_1, \dots, X_n\}^*/\sim$ and $\M(X_1, \dots, X_n)$ from Definition~\ref{def:simplicial_monoid} are the same. 
The only difference is simply that the neutral element in $\{X_1, \dots, X_n\}^*/\sim$ was denoted $\epsilon$, whereas the one of $\M(X_1, \dots, X_n)$ was denoted $\one$.
We denote both monoids simply by $M$ from now on.
For the following theorems, recall that a probability measure $P \in \Delta(\Omega)$ has a Shannon entropy $I_1(P)$ which equals $I_1(\id_{\Omega}; P)$, see Definitions~\ref{def:shannon_entropy},~\ref{def:entropy_function}.
Our aim is to generalize the following theorem:

\begin{theorem}[\cite{li1997}, Theorem 8.1.1]
  \label{thm:kolm_is_entropy_original_result}
  We have
  \begin{equation*}
    0 \leq \Bigg( \sum_{\boldsymbol{x} \in (\bins)^n}P(\boldsymbol{x})\Kc(\boldsymbol{x}) - I_1(P)\Bigg) \plle K(P),
  \end{equation*}
  where both sides are viewed as functions in computable probability measures $P: (\bins)^n \to \R$ with finite entropy $I_1(P) < \infty$.
  That is, up to $K(P) + c$ for some constant $c$ independent of $P$, entropy equals expected Kolmogorov complexity.
\end{theorem}

In the following theorem, if we write $f = g + O(h)$ for functions $f, g, h: \mathcal{X} \to \R$, we mean that there exists a $c \geq 0$ such that $|f(x) - g(x)| < c \cdot h(x)$ for all $x \in \mathcal{X}$.
This is in contrast to our use of that notation in the following parts, Sections~\ref{sec:hkt_prefix_free} and~\ref{sec:hkt_plain_complexity}, where the inequality only needs to hold starting from some threshold value $x_0 \in \mathcal{X}$.

We prove the result in Appendix~\ref{sec:proofs_section_6}, Proof~\ref{prf:thm:averaged_kolm_is_interaction}, with the main ingredients being Hu's theorems for both Shannon entropy --- which follows using Summary~\ref{sum:summary} from Theorem~\ref{thm:hu_kuo_ting_generalized} --- and Chaitin's prefix-free Kolmogorov complexity (Theorem~\ref{thm:hu_kuo_ting_chaitin_finalized}).
Both together allow a reduction to the well-known special case, Theorem~\ref{thm:kolm_is_entropy_original_result}.

\begin{theorem}[See Proof~\ref{prf:thm:averaged_kolm_is_interaction}]
  \label{thm:averaged_kolm_is_interaction}
  Let $X_1, \dots, X_n: (\bins)^n \to \bins$ be the (random) variables given by $X_i(\boldsymbol{x}) = x_i$.
  Let $M = \{X_1, \dots, X_n\}^*/\sim \ \ = \  \M(X_1, \dots, X_n)$ be the idempotent, commutative monoid generated by $X_1, \dots, X_n$, with elements written as $X_I$ for $I \subseteq [n]$.
  Then for all $q \geq 1$ and $Y_1, \dots, Y_q, Z \in M$, the following relation holds:
  \begin{equation}
    \sum_{\boldsymbol{x} \in (\bins)^n} P(\boldsymbol{x}) \cdot \big( \Kc_q(Y_1; \dots; Y_q \mid Z)\big)(\boldsymbol{x}) = Z.I_q(Y_1; \dots ; Y_q ; P) + O\big(K(P)\big),
    \label{eq:averaged_asymptotics}
  \end{equation}
  where both sides are viewed as functions in computable probability mass functions $P: (\bins)^n \to \R$ with finite entropy $I_1(P) < \infty$.
\end{theorem}

\begin{remark}
  \label{rem:another_analysis_of_constants}
  Similar to Remark~\ref{rem:write_down_dependences}, one can also for this theorem wonder about the dependence on $n$ and $q$.
  A similar analysis shows that our techniques lead to a dependence of the form 
  \begin{equation*}
    O\Big( 2^q \big( (q+1)n\log (n+1) + K(P)\big)\Big).
  \end{equation*}
\end{remark}

\begin{corollary}
  \label{cor:dependence_on_m}
  Assume that $(P_m)_{m \in \N}$ is a sequence of computable probability mass functions $P_m: (\bins)^n \to \R$ with finite entropy.
  Additionally, we make the following two assumptions:
  \begin{itemize}
    \item $P_m$ has all its probability mass on elements $\boldsymbol{x} = (x_1, \dots, x_n) \in (\bins)^n$ with sequence lengths $l(x_i) = m$ for all $i \in [n]$;
    \item $K(P_m)$ grows sublinearly with $m$, i.e.,
    \begin{equation*}
      \lim_{m \to \infty} \frac{ K(P_m)}{ m} = 0.
    \end{equation*} 
  \end{itemize}
  Let $q \geq 1$ and $Y_1, \dots, Y_q, Z \in M$ be arbitrary.
  Then the ``per-bit'' difference between expected interaction complexity and interaction information goes to zero for increasing sequence length:
  \begin{equation*}
    \lim_{m \to \infty} \frac{\sum_{\boldsymbol{x} \in (\{0,1\}^m)^n} P_m(\boldsymbol{x}) \cdot \big( \Kc_q(Y_1; \dots; Y_q \mid Z)\big)(\boldsymbol{x}) -  Z.I_q(Y_1; \dots ; Y_q ; P_m)}{m}  = 0.
  \end{equation*}
\end{corollary}

\begin{proof}
  This follows immediately from Theorem~\ref{thm:averaged_kolm_is_interaction}.
\end{proof}

\begin{example}
  \label{exa:example_per_bit_error}
  As an example to Corollary~\ref{cor:dependence_on_m}, consider the case that we have $n$ parameters $p_1, \dots, p_n \in (0,1)$ for Bernoulli distributions.
  Let $P_m$ be the probability mass function given on $\boldsymbol{x} \in (\{0,1\}^m)^n$ by
  \begin{equation*}
    P_m(\boldsymbol{x}) \coloneqq \prod_{i = 1}^n P_m^{p_i}(x_i) \coloneqq \prod_{i = 1}^n \prod_{k = 1}^m p_i^{x_i^{(k)}} \cdot (1 - p_i)^{1 - x_i^{(k)}}.
  \end{equation*}
  That is, $P_m$ consists of $n$ independent probability mass functions $P_m^{p_i}$ that correspond to $m$ independent Bernoulli distributions with parameter $p_i$. 
  We have $K(P_m) = O(\log m)$ since $m$ is the only moving part in the preceding description for $P_m$, with $p_1, \dots, p_n$ being independent of $m$.
  Consequently, Corollary~\ref{cor:dependence_on_m} can be applied, meaning that the per-bit difference between an expected interaction complexity term and the corresponding interaction information goes to zero.
  This generalizes the observation after~\cite{grunwald2008}, Theorem 10, to $n > 1$ and more complicated interaction terms.
\end{example}

\subsection{Hu's Theorem for Prefix-Free Kolmogorov Complexity}\label{sec:hkt_prefix_free}

We now argue that there is also a Hu theorem for prefix-free Kolmogorov complexity.
It requires a logarithmic error term and is therefore less strong than the corresponding theorem for Chaitin's prefix-free Kolmogorov complexity.
Additionally, we need to now use $O$-notation, since the equalities only hold for \emph{almost all} inputs:
for three functions $f, g, h: (\bins)^n \to \R$, different from Section~\ref{sec:interaction_inf_is_expected_complexity}, we now write $f = g + O(h)$ if there is a constant $c \geq 0$ and a threshold $\boldsymbol{x}_0 \in (\bins)^n$ such that 
\begin{equation*}
  \big|f(\boldsymbol{x}) - g(\boldsymbol{x})\big| \leq c \cdot h(\boldsymbol{x})
\end{equation*}
for all $\boldsymbol{x} \geq \boldsymbol{x}_0$.
The latter condition means that $\boldsymbol{x}$ is greater than or equal to $\boldsymbol{x}_0$ in at least one entry, where $\bins$ is ordered lexicographically.

\cite{li1997}, Exercise 3.9.6, shows the following relation:
\begin{equation}\label{eq:comparison_chaitin_plainy}
  K(y \mid x^*) = K(y \mid x) + O\big(\log K(x) + \log K(y)\big).
\end{equation}
Overall, this results in the following chain rule for prefix-free Kolmogorov complexity:

\begin{theorem}[Chain Rule for Prefix-Free Kolmogorov Complexity]
  \label{thm:chain_rule_prefixxx}
  The following identity holds:
  \begin{equation}
     K(x, y) = K(x) + K(y \mid x) + O\big(\log K(x) + \log K(y)\big).
     \label{eq:chain_rule_plain_prefix_K}
  \end{equation}
  Here, both sides are viewed as functions $\bins \times \bins \to \R$ that map inputs of the form $(x, y)$.
\end{theorem}

\begin{proof}
  Combine Theorem~\ref{thm:chain_rule_kolmogorov_compl} with Equation~\eqref{eq:comparison_chaitin_plainy}.
\end{proof}

To get a precise chain rule, we can, similarly to the case of Chaitin's prefix-free Kolmogorov complexity and motivated by Equation~\eqref{eq:comparison_chaitin_plainy}, define a new equivalence relation $\sim_K$ on $\Maps\big( (\bins)^n, \R \big)$ by
\begin{equation*}
  F \sim_K H \ \ \ : \Longleftrightarrow \ \ \ F(\boldsymbol{x}) = H(\boldsymbol{x}) + O\Bigg( \sum_{i = 1}^n \log K(x_i)\Bigg), \ \ \ \text{where } \boldsymbol{x} = (x_1, \dots, x_n) \in (\bins)^n.
\end{equation*}
We denote the equivalence class of a function $F$ by $[F]_K \in \Maps\big( (\bins)^n, \R\big)/\sim_K$.
Then, we again use the monoid $M = \{X_1, \dots, X_n\}^*/\sim$ and define
\begin{align*}
  [K]_K:  M \times M & \to \Maps\big( (\bins)^n, \R\big)/\sim_K, \\
   (Y, Z) & \mapsto [K(Y \mid Z)]_K
\end{align*}
with 
\begin{equation*}
  K(Y \mid Z): \boldsymbol{x} \mapsto K\big( Y(\boldsymbol{x}) \mid Z(\boldsymbol{x})\big).
\end{equation*}
Again, this is well-defined by the same arguments as in Lemma~\ref{lem:reduction_step}, only that this time, we don't need to use the chain rule in the proof.
Furthermore, we can prove an analog of the chain rule given in Proposition~\ref{pro:exact_equality}.

\begin{proposition}[See Proof~\ref{prf:pro:exact_chain_rule_prefix_kolmogorov}]
  \label{pro:exact_chain_rule_prefix_kolmogorov}
  For arbitrary $Y, Z \in M$, the following equality
  \begin{equation*}
    [K]_K(YZ) = [K]_K(Y) + [K]_K(Z \mid Y)
  \end{equation*}
  of elements in $\Maps\big( (\bins)^n, \R\big)/\sim_K$ holds.
\end{proposition}

Thus, $[K]_K: M \times M \to \Maps\big( (\bins)^n, \R\big)/\sim_K$ satisfies all conditions of Corollary~\ref{cor:corollary_for_kolmogorovy} and we obtain a corresponding Hu theorem for prefix-free Kolmogogorov complexity.
This could be worked out similarly to Theorem~\ref{thm:hu_kuo_ting_chaitin_finalized}, which we leave to the interested reader.

\subsection{Hu's Theorem for Plain Kolmogorov Complexity}\label{sec:hkt_plain_complexity}

Here, we briefly consider Hu's theorems for plain Kolmogorov complexity $C: \bins \times \bins \to \R$.
Recall the $O$-notation from Section~\ref{sec:hkt_prefix_free}.

The plain Kolmogorov complexity $C: \bins \times \bins \to \R$ is defined in the same way as prefix-free Kolmogorov complexity, but it allows the set of halting programs to not form a prefix-free set, see~\cite{li1997}, Chapter 2.
This version satisfies the following chain rule:

\begin{theorem}
  [Chain Rule for Plain Kolmogorov Complexity]
  \label{thm:plain_kolm_complexity_chain_rule}
  The following identity holds:
  \begin{equation}
    C(x, y) = C(x) + C(y \mid x) + O\big(\log C(x, y)\big).
    \label{eq:plain_kolm_complexity_chain_rule}
  \end{equation}
  Here, both sides are viewed as functions $\bins \times \bins \to \R$ that are defined on inputs of the form $(x, y)$.
\end{theorem}

\begin{proof}
  This is proved in~\cite{li1997}, Theorem 2.8.
\end{proof}

To get a precise chain rule, we can, similarly as for (Chaitin's) prefix-free Kolmogorov complexity, define a new equivalence relation $\sim_C$ on $\Maps\big( (\bins)^n, \R \big)$ by
\begin{equation*}
  F \sim_C H \ \ \ : \Longleftrightarrow \ \ \ F(\boldsymbol{x}) = H(\boldsymbol{x}) + O \big( \log C(\boldsymbol{x})\big), \ \ \ \text{where } \boldsymbol{x} = (x_1, \dots, x_n) \in (\bins)^n.
\end{equation*}
We denote the equivalence class of a function $F$ by $[F]_C \in \Maps\big( (\bins)^n, \R\big)/\sim_C$.
Using again the monoid $M = \{X_1, \dots, X_n\}^*/\sim$, one can define
\begin{align*}
  [C]_C: M \times M & \to \Maps\big( (\bins)^n, \R\big)/\sim_C \\
  (Y, Z) & \mapsto [C(Y \mid Z)]_C
\end{align*}
with 
\begin{equation*}
  C(Y \mid Z): \boldsymbol{x} \mapsto C\big( Y(\boldsymbol{x}) \mid Z(\boldsymbol{x})\big).
\end{equation*}
Again, this is well-defined by the same arguments as in Lemma~\ref{lem:reduction_step}, and as for prefix-free Kolmogorov complexity, we do not need to use the chain rule in the proof.
Furthermore, we can prove an analog of the chain rules given in Proposition~\ref{pro:exact_equality} and Proposition~\ref{pro:exact_chain_rule_prefix_kolmogorov}:

\begin{proposition}[See Proof~\ref{prf:pro:exact_chain_rule_plain_kolmogorov}]
  \label{pro:exact_chain_rule_plain_kolmogorov}
  For arbitrary $Y, Z \in M$, the equality
  \begin{equation*}
    [C]_C(YZ) = [C]_C(Y) + [C]_C(Z \mid Y)
  \end{equation*}
  of elements in $\Maps\big( (\bins)^n, \R\big)/\sim_C$ holds.
\end{proposition}

Thus, $[C]_C: M \times M \to \Maps\big( (\bins)^n, \R\big)/\sim_C$ satisfies all conditions of Corollary~\ref{cor:corollary_for_kolmogorovy}, and we obtain a corresponding Hu theorem for plain Kolmogogorov complexity.
This could again be worked out similarly to Theorem~\ref{thm:hu_kuo_ting_chaitin_finalized}.

\section{Further Examples of the Generalized Hu Theorem}\label{sec:example_applications}

In this section, we establish further examples of the premises of Theorem~\ref{thm:hu_kuo_ting_generalized} and Corollary~\ref{cor:corollary_for_kolmogorovy}, which essentially boils down to finding a chain rule for a function with the correct type signature.
For the case of Shannon entropy, the premises were summarized in Summary~\ref{sum:summary}.
We mostly leave investigations of the specific \emph{meaning} of the resulting higher-order terms to future work, though we do briefly look at the second degree terms for both Kullback-Leibler divergence and the generalization error in machine learning.
To keep things simple, we diverge from Sections~\ref{sec:technical_introduction} by only working with \emph{finite} discrete random variables, in the cases where the monoid is based on random variables.
As a result, we do not have to worry about questions of convergence and can replace $\Delta_f(\Omega)$ by $\Delta(\Omega)$ and $\Meas_{\con}$ by $\Meas$ everywhere.

Concretely, we investigate Tsallis $\q$-entropy (Section~\ref{sec:alpha_entropy}), Kullback-Leibler divergence (Section~\ref{sec:KL-div}), $\q$--Kullback-Leibler divergence (Section~\ref{sec:alpha_KL}), and cross-entropy (Section~\ref{sec:cross_entropy}). 
We also study arbitrary functions on commutative, idempotent monoids (Section~\ref{sec:arbitrary_functions}), the special case of submodular information functions (Section~\ref{sec:submodular_functions}), and the generalization error from machine learning (Section~\ref{sec:generalization_error}).
Some of the proofs for chain rules are found in Appendix~\ref{sec:proofs_section_7}.
The whole section is written in a self-contained way that requires minimal knowledge from the reader. 

\subsection{Tsallis \texorpdfstring{$\q$}{a}-Entropy}\label{sec:alpha_entropy}

We now investigate the Tsallis $\q$-entropy, which was introduced in~\cite{tsallis1988}.
We follow the investigations in~\cite{vigneaux2019a} and translate them into our framework.

That is, assume a finite, discrete sample space $\Omega$, $n$ finite, discrete random variables $X_1, \dots, X_n$ on $\Omega$, and the monoid $\M(X_1, \dots, X_n)$ generated by equivalence classes of these random variables, see Definition~\ref{def:simplicial_monoid}.
Now, fix an arbitrary number $\q \in \R \setminus \{1\}$.
Then we define the monoid action
\begin{equation*}
  ._{\q}: \M(X_1, \dots, X_n) \times \Meas\big( \Delta(\Omega), \R\big) \to \Meas\big( \Delta(\Omega), \R\big),
\end{equation*}
which we define for $X \in \M(X_1, \dots, X_n)$, $F \in \Meas\big( \Delta(\Omega), \R\big)$, and $P \in \Delta(\Omega)$ by
\begin{equation*}
  (X._{\q}F)(P) \coloneqq \sum_{x \in \vs{X}} P_X(x)^{\q} \cdot F(P|_{X = x}).
\end{equation*}
This is well-defined --- meaning that equivalent random variables act in the same way --- by the same arguments as in Proposition~\ref{pro:averaged_conditioning_equivalence}.
That it is a monoid action can be proved as in Proposition~\ref{pro:conditioning_rules}.
Now, define for arbitrary $\q \in \R \setminus \{1\}$ the $\q$-logarithm by
\begin{equation*}
  \ln_{\q}: (0, \infty) \to \R, \ \ \ \ln_{\q}(p) \coloneqq \frac{p^{\q - 1} - 1}{\q - 1}.
\end{equation*}
We have $\lim_{\q \to 1} \ln_{\q}(p) = \ln(p)$, as can be seen using l'Hospital's rule.
Finally, we can define the Tsallis $\q$-entropy $I_1^{\q}: \M(X_1, \dots, X_n) \to \Meas\big( \Delta(\Omega), \R\big)$ by 
\begin{equation*}
  \big[ I_1^{\q}(X) \big](P) \coloneqq - \sum_{x \in \vs{X}} P_X(x)\ln_{\q}P_X(x) =  \frac{\sum_{x \in \vs{X}} P_X(x)^{\q} - 1}{1 - \q}.
\end{equation*}
This can be shown to be well-defined similarly as in Proposition~\ref{pro:inequality_of_rvs_preserved}.
Since $\lim_{\q \to 1} \ln_{\q}p = \ln p$, we consequently also have $\lim_{\q \to 1} I_1^{\q}(X; P) = I_1(X;P)$.
That is, the $\q$-entropy generalizes the Shannon entropy.

The following chain rule guarantees the existence of a corresponding Hu theorem.

\begin{proposition}[See Proof~\ref{prf:pro:chain_rule_tsallis_entropy}]
  \label{pro:chain_rule_tsallis_entropy}
  $I_1^{\q}: \M(X_1, \dots, X_n) \to \Meas\big(\Delta(\Omega), \R\big)$ satisfies the chain rule 
  \begin{equation*}
    I_1^{\q}(XY) = I_1^{\q}(X) + X._{\q}I_1^{\q}(Y)
  \end{equation*}
  for all $X, Y \in \M(X_1, \dots, X_n)$.
\end{proposition}

\subsection{Kullback-Leibler Divergence}\label{sec:KL-div}

In this section, we study the chain rule of Kullback-Leibler divergence.
It resembles the one described in~\cite{vigneaux2019a}, chapter 3.7, in the language of information cohomology.
A more elementary formulation of the chain rule can also be found in~\cite{Cover2005}, Theorem 2.5.3, which is applied in their Section 4.4 to prove a version of the second law of thermodynamics.
In the end, we will also briefly study and interpret \emph{KL divergence of degree 2}, in analogy to mutual information $I_2$, in Example~\ref{exa:binary_symmetric_channel}.

Let again the monoid $\M(X_1, \dots, X_n)$ of $n$ discrete random variables on $\Omega$ be given.
For $P, Q \in \Delta(\Omega)$, we write $P \ll Q$ if for all $\omega \in \Omega$, the following implication is true: $Q(\omega) = 0 \Longrightarrow  P(\omega) = 0.$
In the literature, $P$ is then called \emph{absolutely continuous} with respect to the measure $Q$.
We set 
\begin{equation*}
  \widetilde{\Delta(\Omega)^2} \coloneqq \Big\lbrace (P, Q) \in \Delta(\Omega)^2 \ \ \ \big| \ \ \  P \ll Q \Big\rbrace.
\end{equation*}
We will silently make use of the fact that $P \ll Q$ implies $P_X \ll Q_X$ and $P|_{X = x} \ll Q|_{X = x}$ for all discrete random variables $X: \Omega \to E_X$ and $x \in E_X$.

We now define $G \coloneqq \Meas\Big(\widetilde{\Delta(\Omega)^2}, \R\Big)$.
We write elements $F \in G$ applied to inputs $(P, Q)$ as $F(P \| Q)$.
Define for $X \in \M(X_1, \dots, X_n)$ and $F \in \Meas\Big(\widetilde{\Delta(\Omega)^2}, \R\Big)$, and $P \ll Q \in \Delta(\Omega)$ the monoid action by:
\begin{equation*}
  (X.F)(P \| Q ) \coloneqq \sum_{x \in \vs{X}} P_X(x) F\big(P|_{X = x} \| Q|_{X = x}\big). 
\end{equation*}
Similarly as before, this is a well-defined, additive monoid action.
In the following, we use the convention that $0 \cdot x = 0$ for $x \in \R \cup \{\pm \infty\}$ and $\ln(0) = - \infty$. 
Finally, we define the function $D_1: \M(X_1, \dots, X_n) \to \Meas\Big(\widetilde{\Delta(\Omega)^2}, \R\Big)$ as the Kullback-Leibler divergence, given for all $X \in \M(X_1, \dots, X_n)$ and $P \ll Q \in \Delta(\Omega)$ by
\begin{equation*}
  \big[D_1(X)\big] (P \| Q) \coloneqq D_1\big(X ; P \| Q\big) \coloneqq - \sum_{x \in \vs{X}} P_X(x) \ln \frac{Q_X(x)}{P_X(x)}.
\end{equation*}
This is well-defined, and we obtain:
\begin{proposition}[See Proof~\ref{prf:pro:cocycle_condition_kl_divergence}]\label{pro:cocycle_condition_kl_divergence}
  $D_1: \M(X_1, \dots, X_n) \to \Meas\Big(\widetilde{\Delta(\Omega)^2}, \R\Big)$ satisfies the chain rule for all $X, Y \in \M(X_1, \dots, X_n)$:
      \begin{equation*}
	D_1(XY) = D_1(X) + X.D_1(Y) .
      \end{equation*}
\end{proposition}

\begin{example}\label{exa:binary_symmetric_channel}
  In~\cite{Fullwood2021}, the following situation is discussed:
  $\mathcal{X}$ and $\mathcal{Y}$ are finite sets, and $\Omega = \mathcal{X} \times \mathcal{Y}$.
  One can consider the two marginal variables
  \begin{align*}
    X: \mathcal{X} \times \mathcal{Y} & \to \mathcal{X}, \quad (x, y) \mapsto x, \\
    Y: \mathcal{X} \times \mathcal{Y} & \to \mathcal{Y}, \quad (x, y) \mapsto y.
  \end{align*}
  A \emph{channel} from $X$ to $Y$ is a conditional distribution $P(Y \mid X)$.
  Together with a prior distribution $P(X)$, it forms a joint $P(X, Y)$ over $\mathcal{X} \times \mathcal{Y}$.
  Now, take two distributions $P \ll Q \in \Delta(\mathcal{X} \times \mathcal{Y})$.
  Then, as noted in~\cite{Fullwood2021}, the chain rule Proposition~\ref{pro:cocycle_condition_kl_divergence} shows the following:
  \begin{equation*}
    D_1\big(P \| Q\big) = D_1\big(P(X) \| Q(X)\big) + \sum_{x \in \mathcal{X}} P(x) \cdot  D_1\big(P(Y \mid x) \| Q(Y \mid x)\big).
  \end{equation*}
  Note that for ease of notation, we write $P(X)$ for $P_X$, $D_1\big(P(X) \| Q(X)\big)$ for $\big[ D_1(X)\big](P \| Q)$, $P(x)$ for $P_X(x)$, $P(Y \mid x)$ for $(P|_{X = x})_Y$, etc.
  
  In our context, the ``mutual Kullback-Leibler divergence'' $D_2(X ; Y)$ is of interest. 
  With respect to $P$ and $Q$, it is given according to Equation~\eqref{eq:inductive_definition} and using symmetry of $D_2$ (which follows from Theorem~\ref{thm:hu_kuo_ting_generalized} due to set operations being symmetric) as follows:
  \begin{equation*}
    \big[D_2(X; Y)\big] (P \| Q) = D_1\big(P(Y) \| Q(Y)\big) - \sum_{x \in \mathcal{X}} P(x) \cdot D_1\big( P(Y \mid x) \| Q(Y \mid x)\big).
  \end{equation*}
  It is well-known that a simple use of Jensen's inequality proves the non-negativity of the Kullback-Leibler divergence $D_1$.
  We also know that mutual information $I_2$ is non-negative.
  Can the same be said about the mutual Kullback-Leibler divergence $D_2$?
  
  The answer is \emph{no}.
  Consider the case $\mathcal{X} = \mathcal{Y} = \{0, 1\}$, and let the prior distributions $P(X) = Q(X)$ both be uniform.
  Furthermore, let $P(Y \mid X)$ and $Q(Y \mid X)$ be \emph{binary symmetric channels} (~\cite{Cover2005}, Section 7.1.4), given as in Figure~\ref{fig:channels_P_Q}.
  \begin{figure}
    \centering
    \begin{tikzcd}
      0 \ar[rrrr, "1/2"] \ar[ddrrrr, "1/2", near end] & & & &   0     &  &     0 \ar[rrrr, "1 - \epsilon"] \ar[ddrrrr, "\epsilon", near end] & & & & 0 \\
      \\
       1 \ar[rrrr, "1/2"'] \ar[uurrrr, "1/2", near start] & &  & &  1     & &      1 \ar[rrrr, "1 - \epsilon"'] \ar[uurrrr, "\epsilon", near start] & & & & 1 \\
       \mathcal{X} \ar[rrrr, rightsquigarrow, "P(Y \mid X)"] & & & & \mathcal{Y} & & \mathcal{X} \ar[rrrr, rightsquigarrow, "Q(Y \mid X)"] & & & & \mathcal{Y}  \\
    \end{tikzcd}
    \caption{Binary symmetric channels for the joint distributions $P$ and $Q$ in Example~\ref{exa:binary_symmetric_channel}.
    For a uniform prior $P(X) = Q(X)$, $P$ and $Q$ have the same marginals $P(Y) = Q(Y)$, but differ in their conditionals $P(Y \mid X)$ and $Q(Y \mid X)$. This leads for small $\epsilon > 0$ to an arbitrarily large negative mutual Kullback-Leibler divergence $\big[D_2(X;Y)\big](P \| Q)$.}
    \label{fig:channels_P_Q}
  \end{figure}
  Note that the marginal distributions $P(Y)$ and $Q(Y)$ are identical, and so
  \begin{equation*}
    D_1\big(P(Y) \| Q(Y)\big) = 0.
  \end{equation*}
  We now work with binary logarithms $\log$.
  For the second term, we then obtain
  \begin{align*}
    \sum_{x \in \{0, 1\}} P(x) \cdot D_1\big( P(Y \mid x) & \| Q(Y \mid x)\big) = \sum_{x \in \{0, 1\}} P(x) \sum_{y \in \{0, 1\}} P(y \mid x) \log \frac{P(y \mid x)}{Q(y \mid x)} \\
    & = \frac{1}{4}  \cdot \Bigg[ \log\frac{P(0 \mid 0)}{Q(0 \mid 0)} + \log \frac{P(1 \mid 0)}{Q(1 \mid 0)} + \log \frac{P(0 \mid 1)}{Q(0 \mid 1)} + \log \frac{P(1 \mid 1)}{Q(1 \mid 1)}\Bigg] \\
    & = \frac{1}{4} \cdot \Big[ -4 - 2 \log (1 - \epsilon) - 2 \log (\epsilon)\Big] \\
    & = -1 - \frac{1}{2}  \cdot \Big[ \log(1 - \epsilon) + \log(\epsilon)\Big]
  \end{align*}
  Note that for very small $\epsilon$, $\log(1 - \epsilon)$ becomes negligible and $\log(\epsilon)$ approaches $- \infty$, and so the term above approaches $+ \infty$.
  Overall, this means that
  \begin{equation*}
    \big[D_2(X ; Y) \big](P \| Q) = -\sum_{x \in \{0, 1\}} P(x) \cdot  D_1\big( P(Y \mid x)  \| Q(Y \mid x)\big) < 0
  \end{equation*}
  is negative, and even unbounded, reaching $- \infty$ as $Q$ becomes deterministic.
  We can compare this conceptually to mutual information as follows:
  $I_2(X ; Y)$ is the average reduction of uncertainty in $Y$ when learning about $X$.
  Similarly, we can interpret $D_2(X; Y)$ as the average reduction of Kullback-Leibler divergence between two marginal distributions in $Y$ when learning about $X$.
  However, in this case, the divergence \emph{only becomes visible} when the evaluation of $X$ is known, since there is no difference in the marginals $P(Y)$ and $Q(Y)$.
  Thus, the ``reduction'' is actually negative.
\end{example}

\subsection{\texorpdfstring{$\q$}{a}--Kullback-Leibler Divergence}\label{sec:alpha_KL}

Similarly to the Tsallis $\q$-entropy from Section~\ref{sec:alpha_entropy}, one can also define a $\q$--Kullback-Leibler divergence, as is done in~\cite{vigneaux2019a}, Chapter 3.7.\footnote{Our definition differs from the one given in~\cite{vigneaux2019a} by using a slightly different definition of the $\q$-logarithm.
We did this to be consistent with the definition of the Tsallis $\q$-entropy above.}
The monoid action $._{\q}: \M(X_1, \dots, X_n) \times \Meas\Big( \widetilde{\Delta(\Omega)^2}, \R\Big) \to \Meas\Big( \widetilde{\Delta(\Omega)^2}, \R\Big)$ is now given by
\begin{equation*}
  (X._{\q}F)(P \| Q) \coloneqq \sum_{x \in \vs{X}} P_X(x)^{\q} Q_X(x)^{1 - \q} \cdot F\big(P|_{X = x} \| Q|_{X = x}\big).
\end{equation*}
Now, we define the $\q$--Kullback-Leibler divergence $D_1^{\q}: \M(X_1, \dots, X_n) \to \Meas\Big(\widetilde{\Delta(\Omega)^2}, \R\Big)$ for all $X \in \M(X_1, \dots, X_n)$ and $P \ll Q \in \Delta(\Omega)$ as the following generalization of standard Kullback-Leibler divergence:
\begin{equation*}
  \big[ D_1^{\q}(X) \big](P \| Q) \coloneqq \sum_{x \in \vs{X}} P_X(x) \ln_{\q} \frac{P_X(x)}{Q_X(x)} = \frac{\sum_{x \in \vs{X}} P_X(x)^{\q} Q_X(x)^{1 - \q} - 1}{\q - 1}.
\end{equation*}

\begin{proposition}[See Proof~\ref{prf:pro:chain_rule_alpha_deformed_KL}]
  \label{pro:chain_rule_alpha_deformed_KL}
  $D_1^{\q}: \M(X_1, \dots, X_n) \to \Meas\Big(\widetilde{\Delta(\Omega)^2}, \R\Big)$ satisfies the chain rule
  \begin{equation*}
    D_1^{\q}(XY) = D_1^{\q}(X) + X._{\q}D_1^{\q}(Y)
  \end{equation*}
  for all $X, Y \in \M(X_1, \dots, X_n)$.
\end{proposition}

\subsection{Cross-Entropy}\label{sec:cross_entropy}

We choose the same monoid action $. : \M(X_1, \dots, X_n) \times \Meas\Big(\widetilde{\Delta(\Omega)^2}, \R\Big) \to \Meas\Big(\widetilde{\Delta(\Omega)^2}, \R\Big)$ as for the Kullback-Leibler divergence.
The cross-entropy $C_1: \M(X_1, \dots, X_n) \to \Meas\Big(\widetilde{\Delta(\Omega)^2}, \R\Big)$ is given by:
\begin{equation*}
  \big[ C_1(X) \big](P \| Q) \coloneqq C_1\big(X; P \| Q\big) \coloneqq - \sum_{x \in \vs{X}} P_X(x) \ln Q_X(x).
\end{equation*}

\begin{proposition}
  $C_1$ satisfies the chain rule for all $X, Y \in \M(X_1, \dots, X_n)$:
  \begin{equation*}
    C_1(XY) = C_1(X) + X.C_1(Y).
  \end{equation*}
\end{proposition}

\begin{proof}
  This follows with the same arguments as Proposition~\ref{pro:cocycle_condition_kl_divergence}.
\end{proof}

\begin{remark}
  \label{rem:simple_observation}
  One can easily show the following well-known relation between cross-entropy $C_1$, Shannon entropy $I_1$, and Kullback-Leibler divergence $D_1$:
  \begin{equation*}
    \big[ C_1(X)\big](P \| Q) = \big[ I_1(X)\big](P) +  \big[ D_1(X)\big] (P \| Q).
  \end{equation*}
  This means that the study of $C_q$ is entirely subsumed by that of $I_q$ and $D_q$.
  Since we already looked at $D_2$ in Example~\ref{exa:binary_symmetric_channel}, we omit looking at $C_2$ here.
\end{remark}

\subsection{Arbitrary Functions on Commutative, Idempotent Monoids}\label{sec:arbitrary_functions}

Let $M$ be any commutative monoid, 
and $R: M \to G$ be \emph{any} function into an abelian group $G$.
Define the two-argument function $R_1: M \times M \to G$ by $R_1(A \mid B) \coloneqq R(AB) - R(B)$.
Set $R_1(A) \coloneqq R_1(A \mid \one) = R(A) - R(\one)$, where $\one \in M$ is the neutral element.
These definitions mean that the chain rule is satisfied \emph{by definition}, making Hu's theorem a purely combinatorial fact.
The reader can verify the following proposition:

\begin{proposition}
  \label{pro:chain_rule_arbitrary_function}
  $R_1: M \times M \to G$ satisfies the chain rule
  \begin{equation*}
    R_1(AB) = R_1(A) + R_1(B \mid A)
  \end{equation*}
  for all $A, B \in M$.
\end{proposition}

Therefore, if $M$ is also idempotent and finitely generated, then $R_1: M \times M \to G$ satisfies all conditions of Corollary~\ref{cor:corollary_for_kolmogorovy}, and one obtains a corresponding Hu theorem.

\subsection{Submodular Information Functions}\label{sec:submodular_functions}

Using the framework of Section~\ref{sec:arbitrary_functions}, we can study the submodular information functions from~\cite{Edmonds2003,Nemhauser1978,steudel2010}, which they use to formulate generalizations of conditional independence and the causal Markov condition.
Alternatively, we could also analyze general submodular set functions~\citep{schrijver2003}, but decided to restrict to submodular information functions since they are closer to our interests.

Recall that a lattice is a tuple $L = (L, \vee, \wedge)$ consisting of a set $L$ together with commutative, associative, and idempotent operations $\vee, \wedge: L \times L \to L$ that satisfy the absorption rules $a \vee (a \wedge b) = a$ and $a \wedge (a \vee b) = a$.
Given a lattice $L$, one can define a corresponding partial order on $L$ by $a \leq b$ if $a = a \wedge b$.

From now on, let $(L, \vee, \wedge)$ be a finite lattice, meaning that $L$ is a finite set.
One can define $\zero \coloneqq \bigwedge_{a \in L}a$, the meet of the finitely many elements in $L$.
By the axioms above, this is neutral with respect to the join operation, that is $b \vee \zero = b$.
Note that $\zero \wedge b = \zero$ for all $b \in L$ due to the second absorption rule above. 
Consequently, $\zero \leq b$ for all $b \in L$.
\cite{Edmonds2003,Nemhauser1978,steudel2010} then study the concept of a submodular (information) function.
We follow the version outlined in~\cite{steudel2010}:
\begin{definition}
  [Submodular Information Function]
  \label{def:submodular_information_function}
  Let $L$ be a finite lattice. 
  Then a function $R: L \to \R$ is called a \emph{submodular information function} if all of the following conditions hold for all $a, b \in L$:
  \begin{enumerate}
    \item normalization: $R(\zero) = 0$;
    \item monotonicity: $a \leq b$ implies $R(a) \leq R(b)$;
    \item submodularity: $R(a) + R(b) \geq R(a \vee b) + R(a \wedge b)$.
  \end{enumerate}
  In particular, the second property implies $R(b) \geq R(\zero) = 0$, meaning $R$ is non-negative.
\end{definition}

They then define the conditional $R_1: L \times L \to \R$ by $R_1(a \mid b) \coloneqq R(a \vee b) - R(a)$.
Furthermore, to define conditional independence and obtain a generalized causal Markov condition, they define the conditional mutual information $I: L^2 \times L \to \R$ by
\begin{equation*}
  I(a;b \mid c) \coloneqq R(a \vee c) + R(b \vee c) - R(a \vee b \vee c) - R(c).
\end{equation*}
Now, note that $(L, \vee, \zero)$ is a finitely generated, commutative, idempotent monoid.
Thus, Proposition~\ref{pro:chain_rule_arbitrary_function} shows that $R_1$ gives rise to Hu's theorem for higher-order functions $R_2, R_3, \dots$, as defined in Corollary~\ref{cor:corollary_for_kolmogorovy}.
We can easily see that $R_2$ agrees with the definition of $I$ from above:
\begin{align*}
  R_2(a; b \mid c) & \coloneqq R_1(a \mid c) - R_1(a \mid b \vee c) \\
  & = R(a \vee c) - R(c) - R(a \vee b \vee c) + R(b \vee c) \\
  & = I(a;b \mid c).
\end{align*}
As special cases of submodular information functions,~\cite{steudel2010} consider Shannon entropy on sets of random variables, Chaitin's prefix-free Kolmogorov complexity, other compression-based information functions, period lengths of time series, and the size of a vocabulary in a text.

\subsection{Generalization Error}\label{sec:generalization_error}

Before coming to the generalization error, we briefly consider the situation dual to that of Section~\ref{sec:arbitrary_functions}. 
Let $M$ be a commutative, idempotent monoid.
Let $G$ be an abelian group and $\mathcal{E}: M \to G$ be any function.
Define $\Ad: M \times M \to G$ by $\Ad(A \mid B) \coloneqq \mathcal{E}(B) - \mathcal{E}(AB)$
Here, $\Ad$ stands intuitively for ``advantage'', a terminology that becomes clear in the machine learning example below.
Similarly as in the case of Kolmogorov complexity, define $\Ad(A) \coloneqq \Ad(A \mid \one) = \mathcal{E}(\one) - \mathcal{E}(A)$.
The reader can easily verify the chain rule:

\begin{proposition}
  \label{pro:chain_rule_trivial}
  $\Ad: M \times M \to G$ satisfies the chain rule:
   one has
  \begin{equation*}
    \Ad(A B) = \Ad(A) + \Ad(B \mid A)
  \end{equation*}
  for all $A, B \in M$.
\end{proposition}

Consequently, $\Ad: M \times M \to G$ satisfies the assumptions of Corollary~\ref{cor:corollary_for_kolmogorovy}.
One then obtains a corresponding Hu theorem.

We now specialize this investigation to the \emph{generalization error} from machine learning~\citep{mohri2018,shalev2014}.
In this case, let $J = [n]$ be a finite set and the monoid be given by $2^J = (2^J, \cup, \emptyset)$, i.e., $\cup$ is the operation and $\emptyset$ the neutral element.
This monoid is idempotent, commutative, and finitely generated by $\{1\}, \dots, \{n\}$.

For all $j \in J$, let $\mathcal{X}_j$ be a measurable space. 
Let $(X_j)_{j \in J}$ be the random variable of feature tuples with values in $\prod_{j \in J} \mathcal{X}_j$.
Similarly, let $\mathcal{Y}$ be another measurable space and $Y$ the random variable of labels in $\mathcal{Y}$.
A typical assumption is that there exists a joint distribution $P \coloneqq P\big((X_j)_{j \in J}, Y\big)$ from which ``the world samples the data''.
Additionally, let $\Delta(\mathcal{Y})$ be the space of probability measures on $\mathcal{Y}$, and $L: \Delta(\mathcal{Y}) \times \mathcal{Y} \to \overline{\R} \coloneqq \R \cup \{+ \infty\}$ a loss function that compares a model distribution over labels to the true label.

For all $A \subseteq J$, assume that $\mathcal{F}(A) \subseteq \Maps\big(\prod_{a \in A} \mathcal{X}_a, \Delta(\mathcal{Y})\big)$ is a class of functions\footnote{Further below, we will make the assumption that $A \subseteq B$ implies $\mathcal{F}(A) \subseteq \mathcal{F}(B)$, in a suitable sense. We make no other assumptions on the collection of $\mathcal{F}(A)$ for $A \subseteq J$.} that, given a feature tuple with indices in $A$, predicts a distribution over $\mathcal{Y}$.
We call this the set of \emph{hypotheses} for predicting the labels given features in $A$. 
For a hypothesis $q \in \mathcal{F}(A)$ and $x_A \in \prod_{a \in A} \mathcal{X}_a$, we denote the output by $q(Y \mid x_A) \coloneqq q(x_A) \in \Delta(\mathcal{Y})$. 
A learning algorithm with access to features in $A$ is supposed to find a hypothesis $q \in \mathcal{F}(A)$ that minimizes the \emph{generalization error}:
\begin{equation*}
  \mathcal{E}(A) \coloneqq \inf_{q \in \mathcal{F}(A)} \Exp_{(\hat{x}, \hat{y}) \sim P}\Big[L\big(q(Y \mid \hat{x}_A) \ \  \| \ \ \hat{y}\big)\Big].
\end{equation*}
Then, as above, define $\Ad_Y: 2^J \times 2^J \to \R$ by 
\begin{equation*}
  \Ad_Y\big(X_A \mid X_B\big) \coloneqq \mathcal{E}(B) - \mathcal{E}(A \cup B).\footnotemark 
\end{equation*}
\footnotetext{There is a one-to-one correspondence between all $A \in 2^J$ and all variables $X_A$ with $A \in 2^J$.
We simply denote the monoid of all $X_A$ again by $2^J$, with the multiplication rule becoming $X_A X_B = X_{A \cup B}$.
}
From Proposition~\ref{pro:chain_rule_trivial}, we obtain the following chain rule:
\begin{equation}\label{eq:empirical_risk_chain_rule}
  \Ad_Y(X_{A \cup B}) = \Ad_{Y}(X_A) + \Ad_Y(X_B \mid X_A).
\end{equation}
To interpret this chain rule sensibly, we make one further assumption:
namely that, when having access to \emph{more features}, the learning algorithm can still use all hypotheses that simply \emph{ignore these additional features}.
More precisely, for $B \subseteq C \subseteq J$,
let us interpret each map $q_B \in \mathcal{F}(B)$ as a function $\widetilde{q_B}: \prod_{c \in C}\mathcal{X}_c \to \Delta(\mathcal{Y})$ by
\begin{equation*}
  \widetilde{q_B}\big( (x_c)_{c \in C}\big) \coloneqq q_B\big( (x_b)_{b \in B}\big).
\end{equation*}
The assumption is that $\widetilde{q_B} \in \mathcal{F}(C)$, for all $B \subseteq C \subseteq J$ and $q_B \in \mathcal{F}(B)$.
Overall, we can interpret this as $\mathcal{F}(B) \subseteq \mathcal{F}(C)$.
It follows that $\mathcal{E}(B) \geq \mathcal{E}(C)$.
Consequently, for all $A, B \subseteq J$ (without any inclusion imposed), it follows
\begin{equation}\label{eq:positive_advantage}
  \Ad_Y(X_A \mid X_B) = \mathcal{E}(B) - \mathcal{E}(A \cup B) \geq 0. 
\end{equation}
The meaning of this is straightforward:
$\Ad_Y(X_A \mid X_B)$ measures what a perfect learning algorithm can gain from knowing all the features in $A$ if it already has access to all the features in $B$ --- the \emph{advantage} motivating the notation $\Ad_Y(X_A \mid X_B)$.
The chain rule, Equation~\eqref{eq:empirical_risk_chain_rule}, thus says the following:
for a perfect learning algorithm, the advantage from getting access to features in $A \cup B$ equals the advantage it receives from the features in $A$, plus the advantage it receives from $B$ when it already has access to $A$.

We can then ask: is then the ``mutual advantage'', as defined from Equation~\eqref{eq:inductive_definition_K} by
\begin{equation*}
  \Ad_Y^2(X_A; X_B) \coloneqq \Ad_Y(X_A) - \Ad_Y(X_A \mid X_B),
\end{equation*}
necessarily also positive, as we expect from the case of entropy and mutual information? 
The answer is \emph{no}, as the following simple example shows:

\begin{example}\label{exa:synergy_in_machine_learning}
  Let $J = \{1, 2\}$, $\mathcal{X}_1 = \mathcal{X}_2 = \mathcal{Y} = \{0, 1\}$, $X_1, X_2$ two independent Bernoulli distributed random variables, and $Y$ be the result of applying a XOR gate to $X_1$ and $X_2$.
  In other words, the joint distribution $P(X_1, X_2, Y) \in \Delta\big(\{0, 1\}^3\big)$ is the unique distribution with
  \begin{align*}
    P\big(X_1 = 0, X_2 = 0, Y = 0\big) & = 1/4, \\
    P\big(X_1 = 0, X_2 = 1, Y = 1\big) & = 1/4, \\
    P\big(X_1 = 1, X_2 = 0, Y = 1\big) & = 1/4, \\
    P\big(X_1 = 1, X_2 = 1, Y = 0\big) & = 1/4.
  \end{align*}
  We define the loss function $L: \Delta\big( \{0, 1\}\big) \times \{0, 1\} \to \overline{\R}$ as the cross-entropy loss: $L\big(q(Y) \  \| \   y\big) \coloneqq - \log q(y),$
  where $\log$ is the binary logarithm.
  Furthermore, we define $\mathcal{F}(A) \coloneqq \big\lbrace q: \mathcal{X}_A \to \Delta(\{0, 1\})\big\rbrace$ as the space of \emph{all} possible prediction functions with access to features in $A \subseteq J = \{1, 2\}$.
  Now, note that if one does not have access to both features, i.e. $A \neq \{1, 2\}$, then it is impossible to do better than random, since $X_1 \indep Y$ and $X_2 \indep Y$.
  Thus, in that case, the best prediction is $q(\hat{y} \mid \hat{x}_A) = 1/2$, irrespective of $\hat{x}$ and $\hat{y}$.
  If, however, one has access to both features, then perfect prediction is possible, since $Y$ is a deterministic function of $(X_1, X_2)$.
  Using $- \log (1/2) = 1$ and $- \log(1) = 0$, this leads to the following generalization errors:
  \begin{equation*}
    \mathcal{E}\big(\emptyset\big) = 1, \ \ \ \mathcal{E}\big(\{1\}\big) = 1, \ \ \ \mathcal{E}\big(\{2\}\big) = 1, \ \ \ \mathcal{E}\big(\{1, 2\}\big) = 0.
  \end{equation*}
  Consequently, the mutual advantage of $X_1$ with $X_2$ is given by
  \begin{align*}
    \Ad_Y^2(X_1; X_2) & = \Ad_Y(X_1) - \Ad_Y(X_1 \mid X_2) \\
    & = \mathcal{E}\big(\emptyset\big) - \mathcal{E}\big(\{1\}\big) - \mathcal{E}\big(\{2\}\big) + \mathcal{E}\big(\{1, 2\}\big) \\
    & = -1 \\
    & < 0.
  \end{align*}
  Thus, in this example, the mutual advantage is negative.
  Rearranging the inequality, we can read this as $\Ad_Y(X_1) < \Ad_Y(X_1 \mid X_2)$.
  In general, beyond the specifics of this example, the inequality
  \begin{equation*}
    \Ad_Y(X_A) < \Ad_Y(X_A \mid X_B)
  \end{equation*}
  means that features in $A \subseteq J$ are more predictive of $Y$ if we already have access to features in $B$.
  This indicates a case of \emph{feature interaction} or \emph{synergy}:
  the contribution of a set of features in predicting $Y$ is greater than the individual contribution of each single feature.
  Intuitively, we expect such situations in many machine learning applications, and think it might be worthwhile to investigate the meaning of the higher degree interaction terms $\Ad_Y^q$ appearing in Hu's theorem as in Corollary~\ref{cor:corollary_for_kolmogorovy}.
\end{example}

\section{Discussion}\label{sec:discussion}

\subsection{Major Findings: a Generalization of Hu's Theorem and its Applications}\label{sec:main_findings}

In this work, we have systematically abstracted away from the details of Shannon's information theory~\citep{Shannon1948, Shannon1964} to generalize Hu's theorem~\citep{Hu1962} to new situations.
To obtain information diagrams, one simply needs a finitely generated commutative, idempotent monoid $M$ --- also known under the name of a join-semilattice --- acting additively on an abelian group $G$, and a function $F_1: M \to G$ satisfying the chain rule of information: $F_1(XY) = F_1(X) + X.F_1(Y)$.
Alternatively, with $M$ and $G$ being as above, the additive monoid action and $F_1$ together can be replaced by a two-argument function $K_1: M \times M \to G$ satisfying the chain rule: $K_1(XY) = K_1(X) + K_1(Y \mid X).$
The proof of the main result --- Theorem~\ref{thm:hu_kuo_ting_generalized} together with Corollary~\ref{cor:corollary_for_kolmogorovy} --- is very similar to the one given in~\cite{Yeung1991} for the case of Shannon entropy;
the main insight is that it is possible to express the basic \emph{atoms} of an information diagram with an inclusion-exclusion type expression over ``unions of disks'':
\begin{equation*}
  \mu(p_I) = \sum_{\emptyset \neq K \supseteq I^c} (-1)^{|K| + |I| + 1 - n} \cdot F_1(X_K) = \sum_{K \subseteq I} (-1)^{|K| + 1} \cdot F_1(X_K X_{I^c}).
\end{equation*}
This formula is visually motivated in Section~\ref{sec:explicit_construction}.
Relations to different interaction terms are explored in Section~\ref{sec:general_consequences}.

With the monoid given by equivalence classes of (countably infinite) discrete random variables, the abelian group by measurable functions on probability measures, and the additive monoid action by the conditioning of information functions, we recover information diagrams for Shannon entropy, see Summary~\ref{sum:summary}.
Beyond this classical case, we obtained Hu's theorems for several versions of Kolmogorov complexity~\citep{li1997} (Section~\ref{sec:kolmogorov_complexity}), Tsallis $\q$-entropy~\citep{tsallis1988}, Kullback-Leibler divergence, $\q$--Kullback-Leibler divergence, cross-entropy~\citep{vigneaux2019a}, general functions on commutative, idempotent monoids, submodular information functions~\citep{steudel2010}, and the generalization error from machine learning~\citep{shalev2014,mohri2018} (all in Section~\ref{sec:example_applications}).
For Kolmogorov complexity, we generalized the well-known theme that ``expected Kolmogorov complexity is close to Shannon entropy'':
\begin{equation*}
  \text{``expected interaction complexity''}  \quad \approx \quad \text{``interaction information''}.
\end{equation*}
For well-behaved probability distributions, this results in the limit of infinite sequence length in an actual \emph{equality} of the per-bit quantities for the two concepts (Section~\ref{sec:interaction_inf_is_expected_complexity}).

\subsection{The Cohomological Context of this Work}\label{sec:context}

The main context in which our ideas developed is information cohomology~\citep{baudot2015a,vigneaux2019a,vigneaux2020,bennequin2020}.
The setup of that work mainly differs by using partition lattices instead of equivalence classes of random variables and generalizing this further to so-called \emph{information structures}.
The functions satisfying the chain rule are reformulated as so-called ``cocycles'' in that cohomology theory, which are ``cochains'' whose ``coboundary'' vanishes:
\begin{equation*}
  (\delta F_1)(X;Y) \coloneqq X.F_1(Y) - F_1(XY) + F_1(X) = 0.
\end{equation*}
That gives these functions a context in the realm of many cohomology theories that were successfully developed in mathematics.
The one defined by Gerhard Hochschild for associative algebras is maybe most closely related~\citep{hochschild1945}.
For the special case of probabilistic information cohomology,~\cite{baudot2015a,vigneaux2019a} were able to show that Shannon entropy is not only \emph{a} cocycle, but is in some precise sense the \emph{unique} cocycle generating all others of degree 1.
Thus, Shannon entropy finds a fully cohomological interpretation.
Arguably, without the abstract nature of that work and the consistent emphasis on abstract structures like monoids and monoid actions, our work would not have been possible.

There is one way in which information cohomology tries to go beyond Shannon information theory:
it tries to find higher degree cocycles that \emph{differ} from the interaction terms $F_q$.
This largely unsolved task has preliminary investigations in~\cite{vigneaux2019a}, Section 3.6, and~\cite{Dube2023}.
In that sense, information cohomology can be viewed as a generalization of Hu's theorem.
Since some limitations in the expressiveness of interaction information are well-known~\citep{James2017},
we welcome any effort to make progress on that task.

\subsection{Unanswered Questions and Future Directions}\label{sec:unanswered_questions}

\paragraph{Further generalizations}
On the theoretical front, it should be possible to generalize Hu's theorem further from commutative, idempotent monoids to what~\cite{vigneaux2019a} calls \emph{conditional meet semi-lattices}.
As these \emph{locally} are commutative, idempotent monoids, the generalization can probably directly use our result.

\paragraph{A transport of ideas}
More practically, we hope that the generalization of Hu's theorem leads to a transport of ideas from the theory of Shannon entropy to other functions satisfying the chain rule.
There are many works that study information-theoretic concepts based on the interaction information functions and thus ultimately Shannon entropy, for example O-information~\citep{rosas2019,gatica2021}, total correlation~\citep{watanabe1960}, dual total correlation~\citep{han1978}, and information paths~\citep{baudot2019,baudot2019b}.
All of these can trivially be defined for functions satisfying the chain rule that go beyond Shannon entropy, and can thus be generalized to all the example applications in Sections~\ref{sec:kolmogorov_complexity} and~\ref{sec:example_applications}.
Most of the basic algebraic properties should carry over since they often follow from Hu's theorem itself.
It is our hope that studying such quantities in greater generality may lead to new insights into the newly established application areas of Hu's theorem.

Additionally, it should not be forgotten that even Shannon interaction information \emph{itself} deserves to be better understood.
Understanding these interaction terms in a more general context could help for resolving some of the persisting confusions about the topic.
One of them surrounds the possible negativity of interaction information $I_3(X; Y; Z)$ of three (and more) random variables~\citep{bell2003,baudot2021a}, which is sometimes understood as meaning that there is more synergy than redundancy present~\citep{Williams2010public, Williams2010private}.
Similarly, we saw in Example~\ref{exa:synergy_in_machine_learning} that the mutual feature advantage $I_Y^2(X_A; X_B)$ can be negative as well, which has a clear interpretation in terms of synergy.
Example~\ref{exa:binary_symmetric_channel} shows that the mutual Kullback-Leibler divergence $D_2(X; Y)$ of two distributions $P \ll Q$ can be negative if knowing $X$ ``reveals'' the divergence of $P$ and $Q$ in $Y$.
We would welcome more analysis in this direction, ideally in a way that transcends any particular applications and could thus shed new light on the meaning of classical interaction information.

\paragraph{Further chain rules}
It goes without saying that we were likely not successful in finding \emph{all} functions satisfying a chain rule.
One interesting candidate seems to be differential entropy $h$ (\cite{Cover2005}, Theorem 8.6.2):
\begin{equation*}
  h(X, Y) = h(X) + h(Y \mid X).
\end{equation*}
However, it seems to us that differential entropy is not well-behaved.
For example, if $X$ is a random variable with values in $\R$, then even if $h(X)$ exists, the differential entropy of the joint variable $(X, X)$ with values in $\R^2$ is negative infinity:
\begin{equation*}
  h(X, X) = - \infty.
\end{equation*}
In particular, we have $h(X) \neq h(X, X)$, and so Hu's theorem cannot hold.

As clarified, for example, in~\cite{Vigneaux2021}, differential entropy is measured \emph{relative to a given base measure}.
Given that $(X, X)$ takes values only in the diagonal of $\R^2$, which has measure $0$, explains why the differential entropy degenerates.
To remedy this, one would need to change the base measure to also live on the diagonal;
it is unclear to us how to interpret this, or if a resulting Hu theorem could indeed be deduced.

Another possible candidate is quantum entropy, also called von Neumann entropy, which also allows for a conditional version that satisfies a chain rule (\cite{Cerf1999}, Theorem 1).
Interestingly, conditional quantum entropy, also called partial quantum information, can be negative~\citep{Cerf1997,Horodecki2005}, which contrasts it from classical Shannon entropy.

In analogy to the Kullback-Leibler divergence (Section~\ref{sec:KL-div}), also quantum entropy admits a relative version, which has many applications in quantum information theory~\citep{Vedral2002}.
In~\cite{Fang2020}, a chain rule for quantum relative entropy was proven, which, however, is an \emph{inequality}.
In~\cite{Parzygnat2021}, Proposition 1 and Example 1, one can find a chain rule--type statement for quantum relative entropy that generalizes the one for non-relative quantum conditional entropy.
We leave the precise meaning or interpretation of these results in the context of our work to future investigations.

\paragraph{Kolmogorov complexity and information decompositions}
In the context of Kolmogorov complexity, we would welcome a more thorough analysis of the size of the constants involved in Theorems~\ref{thm:hu_kuo_ting_chaitin_finalized} and~\ref{thm:averaged_kolm_is_interaction}, potentially similar to~\cite{Zvonkin1970}.
More precisely, it would be worthwhile to improve on the dependence on $q$ or $n$ that we explain in Remarks~\ref{rem:write_down_dependences} and~\ref{rem:another_analysis_of_constants}.

More broadly, one could try to understand complex interactions that go beyond interaction information in the context of Kolmogorov complexity.\footnote{Or in the context of any other of the application areas in Section~\ref{sec:example_applications} of our generalized Hu theorem.}
For example, partial information decomposition (PID)~\citep{Williams2010public,Williams2010private}\footnote{The only privately communicated version,~\cite{Williams2010private}, of~\cite{Williams2010public}, has a stronger emphasis on the axiomatic framework and is more up to date.} aims to complement the usual information functions with unique information, shared information, and complementary information.
It argues that the mutual information of a random variable $Z$ with a joint variable $(X, Y)$ can be decomposed as follows:
\begin{equation*}
  I_2\big( (X, Y); Z\big)   = \underbrace{UI(X \setminus Y ; Z)}_{\text{unique}} + \underbrace{UI(Y \setminus X ; Z)}_{\text{unique}} + \underbrace{SI(X, Y ; Z)}_{\text{shared}} + \underbrace{CI(X, Y; Z)}_{\text{complementary}}.
\end{equation*}
Here, $UI(X \setminus Y; Z)$ is the information that $X$ provides about $Z$ that is not also contained in $Y$; $SI(X, Y; Z)$ is the information that $X$ and $Y$ both share about $Z$; and finally, $CI(X, Y; Z)$ is the information that $X$ and $Y$ can \emph{only together} provide about $Z$, but neither on its own.
$SI$ is also called ``redundant information'', and $CI$ ``synergistic information''.
This then leads to an interpretation of interaction information as a difference of shared and complementary information:
\begin{equation*}
  I_3(X, Y, Z) = \underbrace{SI(X, Y ; Z)}_{\text{shared}} - \underbrace{CI(X, Y; Z)}_{\text{complementary}}.
\end{equation*}
While it is known that such functions exist, no proposals have yet satisfied all axioms that are considered desirable. 
In this sense, the search for shared, redundant, and synergistic information in the framework of PID is still ongoing~\citep{Lizier2018a}.
See also~\cite{Bertschinger2014,Quax2017,Finn2018} for related work.

We could imagine that attempting a similar decomposition for Kolmogorov complexity could provide new insights.
To argue that this might be possible, we can look, for example, at the thought experiment of $x$ and $y$ being binary strings encoding physical theories, and $z$ being a binary string containing data about a physical phenomenon.
Then a hypothesized ``algorithmic complementary information'' $CI(x, y; z)$ would intuitively be high if the theories $x$ and $y$ \emph{only together} allow explaining (parts of) the data $z$; 
a high shared information $SI(x, y; z)$ would mean that $x$ and $y$ are theories that are \emph{equally} able to explain (parts of) the data in $z$.
One hope is that averaging such quantities leads to a partial information decomposition in the usual information-theoretic sense, thus providing a new bridge that helps with the transport of ideas between fields:
\begin{equation*}
  \text{``expected algorithmic PID''} \quad   \overset{?}{\approx}   \quad  \text{``PID''}.
\end{equation*}

\subsection{Conclusion}

To restate our main finding, we can say:
whenever you find a chain rule
\begin{equation*}
  F_1(XY) = F_1(X) + X.F_1(Y),
\end{equation*}
you will under mild conditions obtain information diagrams.
Most of their implications are yet to be understood.

\newpage

\appendix

\section{Measure Theory for Countable Discrete Spaces}\label{sec:measure_theory_technicalities}

In this section, we investigate some technical details related to the measurability of certain functions.
For more background on measure theory, any book on the topic suffices, for example~\cite{Tao2013} and~\cite{Schilling2017}.
As the results are elementary, we leave most of them to the reader to prove.

Recall that for a measurable space $\mathcal{Z}$, the space of probability measures $\Delta(\mathcal{Z})$ on $\mathcal{Z}$ carries the smallest $\sigma$-algebra that makes all evaluation maps 
  \begin{equation*}  
    \ev_A: \Delta(\mathcal{Z}) \to [0,1],  \ \ \ P \mapsto P(A)
  \end{equation*}
for measurable $A \subseteq \mathcal{Z}$ measurable.
Also recall that discrete random variables are functions $X: \Omega \to \vs{X}$ such that both $\Omega$ and $\vs{X}$ are discrete, meaning they are countable and all of their subsets are measurable.
Finally, recall that for a discrete sample space $\Omega$, $\Delta_f(\Omega)$ is the measurable subspace of probability measures $P \in \Delta(\Omega)$ with finite Shannon entropy $H(P)$.

\begin{lemma}
  \label{lem:measurability_of_push_forward}
  Let $X: \Omega \to \vs{X}$ be a random variable. 
  Then the function
  \begin{equation*}
    X_*: \Delta(\Omega) \to \Delta(\vs{X}), \quad P \mapsto \Big( P_X: A \mapsto P \big( X^{-1}(A)\big)\Big)
  \end{equation*}
  is measurable.
\end{lemma}

\begin{proof}
  This is elementary and left to the reader to prove.
\end{proof}

To investigate the measurability of the Shannon entropy function and ``conditioned'' information functions, we need the result that pointwise limits of measurable functions are again measurable:

\begin{lemma}
  \label{lem:pointwise_limit_of_measurable_is_measurable}
  Let $(f_n)_{n \in \N}$ be a sequence of measurable functions $f_n: \mathcal{X} \to \R$ from a measurable space $\mathcal{X}$ to the real numbers $\R$.
  Assume that the pointwise limit function
  \begin{equation*}
    f: \mathcal{X} \to \R, \quad x \mapsto \lim_{n \to \infty}f_n(x)
  \end{equation*}
  exists.
  Then $f$ is also measurable.
\end{lemma}

\begin{proof}
  See \cite{Schilling2017}, Corollary 8.10.
\end{proof}

\begin{corollary}
  \label{cor:measurability_of_entropy_function}
  Let $X: \Omega \to \vs{X}$ be a discrete random variable.
  Then the corresponding Shannon entropy function
  \begin{equation*}
    H(X): \Delta_f(\Omega) \to \R, \ \ \ P \mapsto H(X; P) \coloneqq - \sum_{x \in \vs{X}} P_X(x)  \ln P_X(x)
  \end{equation*}
  is measurable.
\end{corollary}

\begin{proof}
  We already know from Lemma~\ref{lem:measurability_of_push_forward} that the function $P \mapsto P_X$ is measurable.
  Therefore, we can reduce to the case $X = \id_{\Omega}$, i.e.:
  we need to show that the function
  \begin{equation*}
    H: \Delta_f(\Omega) \to \R, \ \ \ P \mapsto - \sum_{\omega \in \Omega} P(\omega) \ln P(\omega)
  \end{equation*}
  is measurable.
  Note that $P(\omega) = \ev_{\omega}(P)$.
  $\ev_{\omega}$ is measurable by definition of the $\sigma$-algebra on $\Delta_f(\Omega)$.
  Also, $\ln: \R_{> 0} \to \R$ is known to be measurable. 
  Since also limits of measurable functions are measurable by Lemma~\ref{lem:pointwise_limit_of_measurable_is_measurable}, the result follows.
\end{proof}

\begin{lemma}
  \label{lem:measurability_of_conditional}
  Let $X: \Omega \to \vs{X}$ be a discrete random variable and $x \in \vs{X}$ any element. 
  Then the function
  \begin{equation*}
    (\cdot)|_{X = x}: \Delta(\Omega) \to \Delta(\Omega), \ \ \ P \mapsto P|_{X = x},
  \end{equation*}
  with $P|_{X = x}$ defined as in Equation~\eqref{eq:conditional_distribution_def}, is measurable.
\end{lemma}

\begin{proof}
  This is elementary and left to the reader to prove.
\end{proof}

\begin{corollary}
  \label{cor:result_after_pliability_measurable}
  Let $\Omega$ be a discrete measurable space and $F: \Delta_f(\Omega) \to \R$ a \emph{conditionable} measurable function, meaning that for all discrete random variables $X: \Omega \to \vs{X}$ and all $P \in \Delta_f(\Omega)$, the series
  \begin{equation*}
    (X.F)(P) = \sum_{x \in \vs{X}} P_X(x) \cdot F\big(P|_{X = x}\big)
  \end{equation*}
  converges unconditionally.
  Then the function $X.F: \Delta_f(\Omega) \to \R$ is also measurable.
\end{corollary}

\begin{proof}
  We have
  \begin{equation*}
    (X.F)(P) = \sum_{x \in \vs{X}} (\ev_x \circ X_*)(P) \cdot \big(F \circ (\cdot)|_{X = x}\big) (P).
  \end{equation*}
  The result follows from the measurability of $\ev_x: \Delta(\vs{X}) \to \R$, $X_*$ as stated in Corollary~\ref{lem:measurability_of_push_forward}, $F$, $(\cdot)_{X = x}: \Delta(\Omega) \to \Delta(\Omega)$ as proven in Lemma~\ref{lem:measurability_of_conditional}, and finally the fact that limits of measurable functions are measurable, see Lemma~\ref{lem:pointwise_limit_of_measurable_is_measurable}.
\end{proof}

\section{Proofs for Section~\ref{sec:technical_introduction}}

\begin{Prf}{pro:inequality_of_rvs_preserved}
  Let $P: \Omega \to [0,1]$ be any probability measure with finite entropy.
  Since $Y \precsim X$, there is a function $f_{YX}: \vs{X} \to \vs{Y}$ such that $f_{YX} \circ X = Y$.
  We obtain
  \begin{align*}
    I_1(Y; P)
    & = - \sum_{y \in \vs{Y}} P\big(Y^{-1}(y)\big) \ln P\big(Y^{-1}(y)\big) \\
    & = - \sum_{y \in \vs{Y}} P_X\big(f_{YX}^{-1}(y)\big) \ln P_X\big(f_{YX}^{-1}(y)\big) \\
    & = - \sum_{y \in \vs{Y}} \sum_{x \in f_{YX}^{-1}(y)} P_{X}(x) \ln \sum_{x' \in f_{YX}^{-1}(y)} P_X(x') \\
    & \overset{(1)}{\leq} - \sum_{y \in \vs{Y}} \sum_{x \in f_{YX}^{-1}(y)} P_X(x) \ln P_X(x) \\
    & = I_1(X; P).
  \end{align*}
  In step (1) we use that $- \ln$ is a monotonically decreasing function and $\sum_{x' \in f_{YX}^{-1}(y)} P_X(x') \geq P_X(x)$ for each $x \in f_{YX}^{-1}(y)$.
\end{Prf}

\begin{Prf}{pro:averaged_conditioning_equivalence}
  We have $f_{XY}$ and $f_{YX}$ with $f_{XY} \circ Y = X$ and $f_{YX} \circ X = Y$.
  For every conditionable measurable function $F: \Delta_f(\Omega) \to \R$ and probability measure $P: \Omega \to \R$, we obtain
  \begin{align*}
    (X.F)(P)     & = \sum_{x \in \im{X}} P_X(x) F(P|_{X = x}) \\
    & \overset{(1)}{=} \sum_{y \in \im{Y}} P_X\big(f_{XY}(y)\big) F\big(P|_{X = f_{XY}(y)}\big) \\
    & \overset{(2)}{=} \sum_{y \in \im{Y}} P_Y(y) F(P|_{Y = y}) \\
    & = (Y.F)(P).
  \end{align*}
  In step $(1)$, we use that $f_{XY}: \im Y \to \im X$ is a bijection.  Step $(2)$ can easily be verified.
\end{Prf}

\begin{Prf}{pro:monoid_of_rvs}
	All required properties follow from Lemma~\ref{lem:joint_and_equivalence}:
	first of all, the multiplication $\cdot: M \times M \to M$ is well-defined, i.e., does not depend on the representatives of the factors $\ec{X}$, $\ec{Y}$ by property $0$.
	We get $\ec{\one} \cdot \ec{X} = \ec{X} = \ec{X} \cdot \ec{\one}$ from property $1$.
	$\ec{X} \cdot \ec{Y} = \ec{Y} \cdot \ec{X}$ follows from property $3$.
	We have $\ec{X} \cdot \ec{X} = \ec{X}$ due to property $4$.

	We now prove the rule $(\ec{X} \cdot \ec{Y}) \cdot \ec{Z} = \ec{X} \cdot (\ec{Y} \cdot \ec{Z})$.
	For any two random variables $U, V \in \widehat{M}$, we write $Z_{UV} \in \widehat{M}$ for a chosen random variable with $UV \sim Z_{UV}$.
	Then, we obtain:
	\begin{align*}
	  \big(\ec{X} \cdot \ec{Y}\big) \cdot \ec{Z} = \ec{Z_{Z_{XY}Z}}
	  \overset{(\star)}{=} \ec{Z_{XZ_{YZ}}} = \ec{X} \cdot \big(\ec{Y} \cdot \ec{Z}\big).
	\end{align*}
	For step ($\star$), one uses the equivalence $Z_{Z_{XY}Z} \sim Z_{XZ_{YZ}}$ that follows from Lemma~\ref{lem:joint_and_equivalence}.
\end{Prf}

\section{Proofs for Section~\ref{sec:generalized_hu_kuo_ting}}\label{sec:proofs_section_5}

\subsection{Proof of the Generalized Hu Theorem~\ref{thm:hu_kuo_ting_generalized} and Corollary~\ref{cor:corollary_for_kolmogorovy}}\label{sec:the_proof}

All notation and assumptions are as in Theorem~\ref{thm:hu_kuo_ting_generalized}.
\begin{lemma}
  \label{lem:inclusion-exclusion-application}
  Let $\mu$ be the measure given on atoms by Equation~\eqref{eq:measure_pointwise_definition}.
  For all $I \subseteq [n]$, we have $F_1(X_I) = \mu(\widetilde{X}_I)$.
\end{lemma}

\begin{proof}
  This is an application of a version of the inclusion-exclusion principle~\citep{beeler2015}, a special case of the Möbius inversion formula on a poset~\citep[3.7.1 Proposition]{Stanley_2011}.
  It states the following:
  For any two functions $f, g: 2^{[n]} \to G$, the following implication holds:
  \begin{equation*}
    g(I) = \sum_{K \supseteq I} f(K) \quad \Longrightarrow \quad f(I) = \sum_{K \supseteq I} (-1)^{|K| - |I|} g(K).
  \end{equation*}
  Set $\mu(p_{\emptyset}) \coloneqq - F_1(X_{[n]})$.
  We apply the principle to the functions $g(I) \coloneqq (-1)^{|I|} \mu(p_{I^c})$ and $f(K) \coloneqq (-1)^{|K|+1} F_1(X_K)$.
  Then Equation~\eqref{eq:measure_pointwise_definition} implies the premise of the inclusion-exclusion principle.
  From the conclusion, we obtain:
  \begin{equation*}
    (-1)^{|I| + 1} F_1(X_I) = \sum_{K \supseteq I} (-1)^{|K| - |I|} \cdot(-1)^{|K|} \mu(p_{K^c}),
  \end{equation*}
  which implies
  \begin{equation*}
    F_1(X_I) = - \sum_{K \supseteq I}\mu(p_{K^c}) = - \sum_{K \colon K \cap I = \emptyset} \mu(p_K) = F_1(X_{[n]}) - \sum_{\emptyset \neq K \colon K \cap I = \emptyset} \mu(p_K). 
  \end{equation*}
  In the last step we used $\mu(p_{\emptyset}) = - F_1(X_{[n]})$.
  Thus, showing that $F_1(X_I) = \mu(\widetilde{X}_I) = \sum_{\emptyset \neq K \colon K \cap I \neq \emptyset} \mu(p_K)$ reduces to the following special case for $I = [n]$:
  \begin{equation*}
    F_1(X_{[n]}) = \mu\big(\widetilde{X}_{[n]}\big).
  \end{equation*}
  To show this, note that~\Cref{eq:measure_pointwise_definition} implies
  \begin{align}\label{eq:expansion_of_measure}
    \begin{split}
      \mu\big( \widetilde{X}_{[n]} \big) = \sum_{K} (-1)^{|K| + 1 - n} \Bigg[ \sum_{\emptyset \neq I: \ I \supseteq K^c} (-1)^{|I|} \Bigg] F_1(X_K).
    \end{split}
  \end{align}
  Ignoring that $\emptyset \neq I$, the inner coefficient is given by
  \begin{align*}
    \sum_{I \supseteq K^c} (-1)^{|I|} 
     = (-1)^{n - |K|} \sum_{i = 0}^{|K|} (-1)^{i} \binom{|K|}{i} 
     = 
    \begin{cases}
      0, \ \ \ \ \ \ \ \ \  K \neq \emptyset \\
      (-1)^n, \text{ else.}
    \end{cases}
  \end{align*}
  Note that $F_1(X_{\emptyset}) = 0$, so the last case is irrelevant.
  Also, note that the condition $I \neq \emptyset$ only restricts the inner sum in~\Cref{eq:expansion_of_measure} when $K = [n]$.
  Thus, in that case, we need to subtract $1$ from the result just computed and obtain:
  \begin{equation*}
    \mu \big( \widetilde{X}_{[n]} \big) = (-1)^{n + 1 - n} \cdot (-1) \cdot F_1(X_{[n]}) = F_1(X_{[n]}),
  \end{equation*}
  proving the claim.
\end{proof}

\begin{proposition}\label{pro:entropy_satisfy_hu_kuo_ting}
  For all $n \in \N_{\geq 0}$, for $\mu$ being the $G$-valued measure constructed from $F_1$ as in Equation~\eqref{eq:measure_pointwise_definition}, for all $L_1, J \subseteq [n]$, the following identity holds:
  \begin{equation*}
    X_J.F_1(X_{L_1}) = \mu(\widetilde{X}_{L_1} \setminus \widetilde{X}_J)
  \end{equation*}
\end{proposition}

\begin{proof}
  This follows immediately from Lemma~\ref{lem:inclusion-exclusion-application} and the chain rule, Equation~\eqref{eq:cocycle_equationn}, together with the fact that $\mu$ is a measure.
\end{proof}

We have now done all the hard work for finishing the proof of Theorem~\ref{thm:hu_kuo_ting_generalized}:

\begin{Prf}{thm:hu_kuo_ting_generalized}
  \textbf{Part 1. \ \ \ \ }
  This is a simple inductive argument, using Proposition~\ref{pro:entropy_satisfy_hu_kuo_ting} for $q = 1$, and Equation~\eqref{eq:inductive_definition} for showing the step from $q-1$ to $q$.

  \textbf{Part 2. \ \ \ \ }
  For part 2, using Equation~\eqref{eq:hu_kuo_ting_equation}, we observe
  \begin{equation*}
    X_J.F_1(X_I) - F_1(X_{J \cup I}) + F_1(X_J) = \mu(\widetilde{X}_{I} \setminus \widetilde{X}_J) - \mu(\widetilde{X}_{J} \cup \widetilde{X}_I) + \mu(\widetilde{X}_J) = 0.
  \end{equation*}
  Thus, $F_1$ satisfies Equation~\eqref{eq:cocycle_equationn}.
  For $q \geq 2$, using Equation~\eqref{eq:hu_kuo_ting_equation} again, we similarly observe
 \begin{align*}
    F_{q-1}(X_{L_1}; \dots ; X_{L_{q-1}}) - X_{L_q}.F_{q-1}(X_{L_1}; \dots ; X_{L_{q-1}}) 
    & = F_q(X_{L_1}; \dots ; X_{L_q}).
     \end{align*}
     That finishes the proof.
\end{Prf}

\begin{Prf}{cor:corollary_for_kolmogorovy}
  Define $\widetilde{G} \coloneqq \Maps(M, G)$ as the group of functions from $M$ to $G$.
  Define, using \emph{currying}, the function $\widetilde{K}_1: M \to \widetilde{G}$ by
  \begin{equation*}
    \big[\widetilde{K}_1(X)\big](Y) \coloneqq K_1(X \mid Y).
  \end{equation*}
  Define the additive monoid action $.: M \times \widetilde{G} \to \widetilde{G}$ by
  \begin{equation*}
    (X.F)(Y) \coloneqq F(XY)
  \end{equation*}
  for all $X, Y \in M$.
  Note that we need $M$ to be commutative to show that this is indeed a monoid action.
  Then clearly, $\widetilde{K_1}$ satisfies the chain rule. 
  
  Define $\widetilde{K}_q: M^q \to \widetilde{G}$ as in Theorem~\ref{thm:hu_kuo_ting_generalized} inductively by
  \begin{equation*}
    \widetilde{K}_q(Y_1; \dots; Y_q) \coloneqq \widetilde{K}_{q-1}(Y_1; \dots ; Y_{q-1}) - Y_q.\widetilde{K}_{q-1}(Y_1; \dots ; Y_{q-1}).
  \end{equation*}
  By induction, one can show that $K_q(Y_1; \dots ; Y_q \mid Z) = \big[ \widetilde{K}_q(Y_1 ; \dots ; Y_q)\big](Z)$ for all $Y_1, \dots, Y_q, Z \in M$.
  
  By the conclusion of Theorem~\ref{thm:hu_kuo_ting_generalized}, we obtain a $\widetilde{G}$-valued measure $\widetilde{\mu}: 2^{\widetilde{X}} \to \widetilde{G}$ with
  \begin{equation*}
    \widetilde{\mu}\Bigg( \bigcap_{k = 1}^q \widetilde{X}_{L_k} \setminus \widetilde{X}_J\Bigg) = X_J.\widetilde{K}_q(X_{L_1}; \dots ; X_{L_q}).
  \end{equation*}
  Now, define $\mu: 2^{\widetilde{X}} \to G$ by $\mu(A) \coloneqq \big[ \widetilde{\mu}(A)\big](1)$ for all $A \subseteq \widetilde{X}$.
  Clearly, since $\widetilde{\mu}$ is a $\widetilde{G}$-valued measure, $\mu$ is a $G$-valued measure.
  The results immediately follow from these definitions and Hu's Theorem.
\end{Prf}

\subsection{Further Proofs for Section~\ref{sec:generalized_hu_kuo_ting}}\label{sec:proofs_section_5_2}

\begin{Prf}{cor:general_consequences}
  We proceed as follows:
  \begin{enumerate}
    \item This follows from Lemma~\ref{lem:p_I_charac} and Equation~\eqref{eq:measure_pointwise_definition}.
    \item By Lemma~\ref{lem:p_I_charac} and Theorem~\ref{thm:hu_kuo_ting_generalized}, we have
	      \begin{align*}
		\sum_{\substack{I \subseteq [n] \\ I \cap K \neq \emptyset}} \eta_I  = \mu\Big(  \Big\lbrace p_I \ \ \big| \ \  I \subseteq [n], \exists k \in K: k \in I\Big\rbrace \Big)
		= \mu(\widetilde{X}_K) = F_1(X_K).
	      \end{align*}
	    \item Using Lemma~\ref{lem:p_I_charac} and Theorem~\ref{thm:hu_kuo_ting_generalized} again, we obtain
	      \begin{align*}
		F_q(X_{j_1}; \dots ; X_{j_q}) = \mu \Bigg( \bigcap_{j \in J} \widetilde{X}_j\Bigg)
		= \sum_{I, \forall j \in J: j \in I} \eta_I  
		= \sum_{I \supseteq J} \eta_I.
	      \end{align*}
	    \item This is formally a consequence of $3$ and the inclusion-exclusion principle~\citep{beeler2015}.
	    \item This follows by combining results $2$ and $4$.
	    \item This follows by combining results $1$ and $3$, or by the inclusion-exclusion principle~\citep{beeler2015} applied to result $5$. 
    \end{enumerate}
\end{Prf}

\section{Proofs for Section~\ref{sec:kolmogorov_complexity}}\label{sec:proofs_section_6}

\begin{Prf}{pro:exact_equality}
  Let $Y, Z \in \widetilde{M}$ be arbitrary.
   In the following, for functions $f: (\bins)^n \to \R$, we write $f = f(\boldsymbol{x})$ for simplicity, and mean by it the function mapping $\boldsymbol{x}$ to $f(\boldsymbol{x})$.
   We obtain:
   \begin{align*}
     \Kc(YZ) & = \Kc\big( (YZ)(\boldsymbol{x})\big) \\
     & \ple \Kc \big( Y(\boldsymbol{x})'Z(\boldsymbol{x})\big) \\
     & \ple \Kc\big( Y(\boldsymbol{x})\big) + \Kc\big( Z(\boldsymbol{x}) \mid Y(\boldsymbol{x})\big) \\
     & \ple \Kc(Y) + \Kc(Z \mid Y).
   \end{align*}
   In the computation, the associativity rule in the second step holds as we can write a program of constant size that translates between the different nestings of the strings.\footnote{For this, we use that we can algorithmically extract all $x_i$ for indices appearing in $Y$ and $Z$ from the strings $(YZ)(\boldsymbol{x})$ and also $Y(\boldsymbol{x})'Z(\boldsymbol{x})$.
   This argument uses that the encoding $x \mapsto x'$ is prefix-free.}
   In the third step we use Theorem~\ref{thm:chain_rule_kolmogorov_compl}.
   The result follows.
\end{Prf}

\begin{Prf}{lem:reduction_step}
  We have
  \begin{align*}
    [\Kc]_{\Kc}(Y \mid Z) & \overset{(1)}{=} [\Kc]_{\Kc}(YZ) - [\Kc]_{\Kc}(Z) \\
    & \overset{(2)}{=} [\Kc]_{\Kc}(\overline{Y} \  \overline{Z}) - [\Kc]_{\Kc}(\overline{Z}) \\
    & \overset{(3)}{=} [\Kc]_{\Kc}(\overline{Y} \mid \overline{Z}). 
  \end{align*}
  In the computation, steps (1) and (3) follow from Proposition~\ref{pro:exact_equality}.
  For step (2) one can show that $\Kc(YZ) \ple \Kc(\overline{Y} \ \overline{Z})$ and $\Kc(Z) \ple \Kc(\overline{Z})$ in the same way as the associativity rule in Proposition~\ref{pro:exact_equality} was shown.
\end{Prf}

\begin{Prf}{thm:hu_kuo_ting_chaitin_finalized}
  Remember $M = \widetilde{M} / \sim$ and
  the function $[\Kc]_{\Kc}: M \times M \to \Maps \big( (\bins)^n, \R\big) / \sim_{\Kc}$, which we now denote by $[\Kc] = [\Kc]_1 \coloneqq [\Kc]_{\Kc}$. 
  From this, we can inductively define $[\Kc]_q: M^q \times M \to \Maps \big( (\bins)^n, \R\big) / \sim_{\Kc}$ as in Corollary~\ref{cor:corollary_for_kolmogorovy} by 
  \begin{equation*}
    [\Kc]_q(Y_1; \dots ; Y_q \mid Z) \coloneqq [\Kc]_{q-1}(Y_1; \dots ; Y_{q-1} \mid Z) - [\Kc]_{q-1}(Y_1; \dots ; Y_{q-1} \mid Y_q Z).
  \end{equation*}
  From Equation~\eqref{eq:easy_induction}, one can inductively show that
  \begin{equation}\label{eq:ingredient_1}
    [\Kc]_q(Y_1; \dots ; Y_q \mid Z) = [\Kc_q(Y_1; \dots ; Y_q \mid Z)]
  \end{equation}
  for all $Y_1, \dots, Y_q, Z \in M$.
  Note that $\Kc_q$ was defined on $\widetilde{M}$ and not $M$, which means that we plug in representatives of equivalence classes at the right-hand-side.
  Using Lemma~\ref{lem:reduction_step} and induction, one can show that this is well-defined.
  Then, construct $\mu: 2^{\widetilde{X}} \to \Maps \big( (\bins)^n, \R\big)$ explicitly as in Equation~\eqref{eq:def_mu_other_special_case}.
  Define, now, the measure $[\mu]: 2^{\widetilde{X}} \to \Maps \big( (\bins)^n, \R\big)/\sim_{\Kc}$ by
  \begin{equation}\label{eq:def_eq_mu}
    [\mu](A) \coloneqq [\mu(A)] \ \ \ \ \forall A \subseteq \widetilde{X}.
  \end{equation}
  Then Equation~\eqref{eq:ingredient_1} shows that
  \begin{equation}\label{eq:ingredient_2}
    [\mu](p_I) = \sum_{\emptyset \neq K \supseteq I^c} (-1)^{|K| + |I| + 1 - n} [\Kc]_1(X_K).
  \end{equation}
  Consequently, $[\mu]$ is the measure that results in Corollary~\ref{cor:corollary_for_kolmogorovy}, see Equation~\eqref{eq:explicit_construction_for_K}.
  We obtain for all $L_1, \dots, L_q, J \subseteq [n]$:
  \begin{align*}
    \Bigg[ \mu \Bigg( \bigcap_{k = 1}^q \widetilde{X}_{L_k} \setminus \widetilde{X}_J \Bigg)\Bigg] & = [\mu]\Bigg( \bigcap_{k = 1}^q \widetilde{X}_{L_k} \setminus \widetilde{X}_J \Bigg) & (\text{Equation~\eqref{eq:def_eq_mu}})\\
    & = [\Kc]_q(X_{L_1}; \dots ; X_{L_q} \mid X_J)  & (\text{Proposition~\ref{pro:exact_equality}, Corollary~\ref{cor:corollary_for_kolmogorovy}})\\
    & = \big[ \Kc_q(X_{L_1}; \dots ; X_{L_q} \mid X_J) \big] & (\text{Equation~\eqref{eq:ingredient_1}}).
  \end{align*}
  As two representatives of the same equivalence class in $\Maps\big( (\bins)^n, \R\big)$ differ by a constant, the result follows.
\end{Prf}

\begin{lemma}
  \label{lem:computability_of_marginals}
  Let $P: (\bins)^n \to \R$ be a computable probability mass function.
  Let $K \subseteq [n]$ a subset and $P_K$ the corresponding maginal distribution.
  Then $P_K$ is also computable, and the relation
  \begin{equation*}
    K(P_K) \plle K(P).
  \end{equation*}
  between their Kolmogorov complexities holds.
\end{lemma}

\begin{proof}
  We know that $P$ is computable, and so there exists a prefix-free Turing machine $T_p$ of length $l(p) = K(P)$ such that 
  \begin{equation*}
    \big|T_p(\boldsymbol{x}'q) - P(\boldsymbol{x})\big| \leq 1/q
  \end{equation*}
  for all $q \in \N$ and $\boldsymbol{x} \in (\bins)^n$.
  Now, fix $q \in \N$.
  Let $(\boldsymbol{x}^i)_{i \in \N}$ be a computable enumeration of $(\bins)^n$.
  Define the approximation $P_q: (\bins)^n \dashrightarrow \R$ of $P$ by 
  \begin{equation*}
    P_q(\boldsymbol{x}^i) \coloneqq T_p\big((\boldsymbol{x}^i)'(4q \cdot 2^i)\big).
  \end{equation*}
  Then for all subsets $I \subseteq \N$, we have
  \begin{align}\label{eq:important_inequality}
    \Bigg| \sum_{i \in I} P_q(\boldsymbol{x}^i) - \sum_{i \in I}P(\boldsymbol{x^i})\Bigg|    & \leq \sum_{i \in I} \Big|T_p\big((\boldsymbol{x}^i)'(4q \cdot 2^i)\big) - P(\boldsymbol{x}^i) \Big| \nonumber \\
    & \leq \sum_{i = 1}^{\infty} \frac{1}{4q \cdot 2^i} \\
    & = \frac{1}{4q}. \nonumber
  \end{align}
  Consequently, by setting $I = \N$ and using $\sum_{i \in \N}P(\boldsymbol{x}^i) = 1$, one can determine $i_q$ such that for the first time we have
  \begin{equation}\label{eq:another_important_inequality}
    \Bigg| \sum_{i = 1}^{i_q} P_q(\boldsymbol{x}^i) - 1\Bigg| \leq \frac{1}{2q}.
  \end{equation}
  Note that $i_q$ can be \emph{algorithmically} determined by computing one $P_q(\boldsymbol{x}^i)$ at a time and checking when the condition holds.
  Now, for arbitrary $\boldsymbol{x}_K \in (\bins)^{|K|}$ and $q \in \N$, we define
  \begin{equation*}
    T(\boldsymbol{x}_K'q) \coloneqq \sum_{\substack{i = 1 \\ (\boldsymbol{x}^i)_K = \boldsymbol{x}_K}}^{i_q} P_q(\boldsymbol{x}^i).
  \end{equation*}
  We now show that $T(\boldsymbol{x}_K'q)$ approximates $P_K(\boldsymbol{x}_K)$ up to an error of $1/q$:
  \begin{align*}
    \big| T(\boldsymbol{x}_K'q) - P_K(\boldsymbol{x}_K)\big| & =  \left| \sum_{\substack{i = 1 \\ (\boldsymbol{x}^i)_K = \boldsymbol{x}_K}}^{i_q} P_q(\boldsymbol{x}^i) - P_K(\boldsymbol{x}_K)\right| \\
    & \leq \left| \sum_{\substack{i = 1 \\ (\boldsymbol{x}^i)_K = \boldsymbol{x}_K}}^{i_q} P_q(\boldsymbol{x}^i) - \sum_{\substack{i = 1 \\ (\boldsymbol{x}^i)_K = \boldsymbol{x}_K}}^{i_q} P(\boldsymbol{x}^i) \right|
    + \left|\sum_{\substack{i = 1 \\ (\boldsymbol{x}^i)_K = \boldsymbol{x}_K}}^{i_q} P(\boldsymbol{x}^i) - P_K(\boldsymbol{x}_K) \right|  \\
    & \overset{\eqref{eq:important_inequality}}{\leq} \frac{1}{4q} + P_K(\boldsymbol{x}_K) - \sum_{\substack{i = 1 \\ (\boldsymbol{x}^i)_K = \boldsymbol{x}_K}}^{i_q} P(\boldsymbol{x}^i)  \\
    & =  \frac{1}{4q} + \Bigg| 1 - \sum_{i = 1}^{i_q} P(\boldsymbol{x}_i)\Bigg| \\
    & \leq \frac{1}{4q} + \Bigg| 1 -  \sum_{i = 1}^{i_q} P_q(\boldsymbol{x}^i) \Bigg| + \Bigg| \sum_{i = 1}^{i_q} P_q(\boldsymbol{x}^i) - \sum_{i = 1}^{i_q} P(\boldsymbol{x}^i)\Bigg| \\
    & \overset{\eqref{eq:another_important_inequality},\eqref{eq:important_inequality}}{\leq} \frac{1}{4q} + \frac{1}{2q} + \frac{1}{4q} = 1/q.
  \end{align*}
  Now, note that $T$ is computable, since it uses in its definition only the computable enumeration $(\boldsymbol{x}^i)_{i \in \N}$, the number $i_q$ for which we described an algorithm, and the Turing machine $T_p$ inside the definition of $P_q$. 
  Thus, $T$ is a prefix machine $T_{p_K}$ for a bitstring $p_K$ of length $l(p_K) \leq l(p) + c = K(P) + c$, where $c \geq 0$ is some constant.
  It follows $K(P_K) \leq l(p_K) \leq K(P) + c$, and we are done.
\end{proof}

\begin{Prf}{lem:equivalence_relations_identical}
  Assume that $Y \sim Z$.
  Then Lemma~\ref{lem:joint_and_equivalence} parts 3 and 4\footnote{What we denoted by $\sim$ in that lemma is denoted $\sim_r$ here.} show that $Y \sim_r \overline{Y} = \overline{Z} \sim_r Z$, and so $Y \sim_r Z$ by transitivity.

  On the other hand, if $Y \sim_r Z$, then also $X_I = \overline{Y} \sim_r \overline{Z} = X_J$ for some $I, J \subseteq [n]$, again by Lemma~\ref{lem:joint_and_equivalence} parts 3 and 4.
  Let $I = \big\lbrace i_1 < \dots < i_{|I|}\big\rbrace$ and $J = \big\lbrace j_1 < \dots < j_{|J|}\big\rbrace$.
  Then there exist functions $f_{JI}$ and $f_{IJ}$ such that $f_{JI} \circ X_I = X_J$ and $f_{IJ} \circ X_J = X_I$. 
  That is, for all $\boldsymbol{x} \in (\bins)^n$ we have
  \begin{align*}
    f_{JI}(x_{i_1}, \dots, x_{i_{|I|}}) = (x_{j_1}, \dots, x_{j_{|J|}}), \\
    f_{IJ}(x_{j_1}, \dots, x_{j_{|J|}}) = (x_{i_1}, \dots, x_{i_{|I|}}).
  \end{align*}
  The first equation shows $J \subseteq I$, as otherwise, changes in $\boldsymbol{x}_{J \setminus I}$ lead to changes in the right-hand-side, but not the left-hand-side.
  In the same way, the second equation shows $I \subseteq J$, and overall we obtain $I = J$.
  That shows $Y \sim \overline{Y} = X_I = X_J = \overline{Z} \sim Z$;
  due to transitivity, it follows $Y \sim Z$. 
\end{Prf}

\begin{Prf}{thm:averaged_kolm_is_interaction}
  We generalize the proof strategy that~\cite{li1997} use for their Lemma 8.1.1, which is a special case of our theorem for $n = 2$, $q = 2$, $Y_1 = X_1, Y_2 = X_2$, and $Z = \epsilon = \one$.

  We prove this in several steps by first handling convenient subcases.
  In the special case $q = 1$, $Z = \epsilon = \one$, and $Y_1 = X_K$ for some $K \subseteq [n]$, we can look at the marginal $P_K$ of $P$ and obtain 
  \begin{align*}
    \sum_{\boldsymbol{x} \in (\bins)^n}P(\boldsymbol{x}) \big(\Kc(X_K)\big)(\boldsymbol{x}) & = \sum_{\boldsymbol{x}_K \in (\bins)^{|K|}} P_K(\boldsymbol{x}_K) \big(\Kc(X_K)\big)(\boldsymbol{x}_K) \\
    & = I_1(P_K) + O\big( K(P_K)\big) & \big( \text{Theorem~\ref{thm:kolm_is_entropy_original_result}}\big)\\
    & =  I_1(X_K ; P) + O\big( K(P)\big),  & \big( \text{Lemma~\ref{lem:computability_of_marginals}} \big),
  \end{align*}
  which is the wished result.
  Now, let 
  \begin{align*}
    \mu: & 2^{\widetilde{X}} \to \Maps\big( (\bins)^n, \R\big), & \big( \text{Equation~\eqref{eq:def_mu_other_special_case}}\big)\\
    \mu^r: & 2^{\widetilde{X}} \to \Meas_{\con}\Big(\Delta_{f}\big((\bins)^n\big), \R\Big) & \big( \text{Equation~\ref{eq:measure_pointwise_definition}} \big)
  \end{align*}
  be the measures corresponding to Chaitin's prefix-free Kolmogorov complexity $\Kc: M \times M \to \Maps\big((\bins)^n, \R\big)$ and Shannon entropy $I_1: M \to \Meas_{\con}\Big( \Delta_{f}\big((\bins)^n\big), \R\Big)$, remembering that $\Delta_f\big( (\bins)^n\big)$ is the space of finite-entropy probability measures (or mass functions) on our \emph{countable}\footnote{The fact that $(\bins)^n$ is not finite but countably infinite is the main reason why we considered countable sample spaces in Summary~\ref{sum:summary}.} sample space $(\bins)^n$.\footnote{The superscript in $\mu^r$ is used to notationally distinguish it from $\mu$.
  $r$ can be thought of as meaning ``random''.}
  Let $I \subseteq [n]$ be any subset.
  Then we obtain
  \begin{align*}
    \sum_{\boldsymbol{x} \in (\bins)^n}P(\boldsymbol{x})\big(\mu(p_I)\big)(\boldsymbol{x}) & \overset{\eqref{eq:def_mu_other_special_case}}{=} \sum_{\boldsymbol{x} \in (\bins)^n} P(\boldsymbol{x}) \sum_{\emptyset \neq K \supseteq I^c}(-1)^{|K|+|I|+1-n} \big( \Kc(X_K) \big)(\boldsymbol{x}) \\
  & = \sum_{\emptyset \neq K \supseteq I^c} (-1)^{|K|+|I|+1-n} \sum_{\boldsymbol{x} \in (\bins)^n} P(\boldsymbol{x})\big( \Kc(X_K)\big)(\boldsymbol{x}) \\
  & \overset{(\star)}{=} \sum_{\emptyset \neq K \supseteq I^c} (-1)^{|K|+|I|+1-n} \Big( I_1(X_K; P) + O\big(K(P)\big)\Big) \\
  & = \Bigg( \sum_{\emptyset \neq K \supseteq I^c} (-1)^{|K|+|I|+1-n} I_1(X_K) \Bigg)(P) + O\big( K(P)\big) \\
  & =  \big(\mu^r(p_I)\big)(P) + O\big( K(P)\big),
  \end{align*}
  using our earlier result in step $(\star)$ and the definition of $\mu^r$.
  Now, using that $\mu$ and $\mu^r$ are additive over disjoint unions, we can deduce for all $A \subseteq \widetilde{X}$ the equality
  \begin{equation*}
    \sum_{\boldsymbol{x} \in (\bins)^n} P(\boldsymbol{x}) \big( \mu(A)\big)(\boldsymbol{x}) = \big( \mu^r(A)\big)(P) + O\big( K(P)\big).
  \end{equation*}
  Now, let $Y_1 = X_{L_1}, \dots, Y_q = X_{L_q}, Z = X_J$ for some $L_1, \dots, L_q, J \subseteq [n]$.
  Then, using Hu's theorems for interaction information (Theorem~\ref{thm:hu_kuo_ting_generalized} together with Summary~\ref{sum:summary}) and Kolmogorov complexity~\ref{thm:hu_kuo_ting_chaitin_finalized}, the result follows by setting $A \coloneqq \bigcap_{k = 1}^q \widetilde{X}_{L_k} \setminus \widetilde{X}_J$.
\end{Prf}

\begin{Prf}{pro:exact_chain_rule_prefix_kolmogorov}
  We have
  \begin{align*}
    K(YZ) & = K\big((YZ)(\boldsymbol{x})\big) \\
    & \overset{(1)}{=} K \big( Y(\boldsymbol{x})' Z(\boldsymbol{x})\big) + O(1)\\
    & \overset{(2)}{=} K \big( Y(\boldsymbol{x})\big) + K \big( Z(\boldsymbol{x}) \mid Y(\boldsymbol{x})\big) + O\Big( \log K\big( Y(\boldsymbol{x})\big) + \log K \big( Z(\boldsymbol{x})\big)\Big) \\
    & \overset{(3)}{=} K(Y) + K(Z \mid Y) + O\Bigg( \sum_{i = 1}^n \log K(x_i)\Bigg).
  \end{align*}
  where step (1) follows as in Proposition~\ref{pro:exact_equality}, step (2) uses Theorem~\ref{thm:chain_rule_prefixxx}, and step (3) follows from the subadditivity of $K$\footnote{The subadditivity property for $K$ says that $K(x, y) \leq K(x) + K(y) + O(1)$:
  one can construct a prefix-free Turing machine that extracts $x^*$ and $y^*$ from $x^* y^*$, which is of length $K(x) + K(y)$, and outputs $x'y$.
Note that since the set of halting programs of the universal Turing machine $U$ is \emph{prefix-free}, one does not need to indicate the place of separation between $x^*$ and $y^*$.}
and the logarithm, which holds for large enough inputs.
\end{Prf}

\begin{Prf}{pro:exact_chain_rule_plain_kolmogorov}
  Let $Y, Z \in M$ be arbitrary.
  Then, following the same arguments as in Proposition~\ref{pro:exact_equality} and Proposition~\ref{pro:exact_chain_rule_prefix_kolmogorov}, we are only left with showing the following:
  \begin{equation*}
    \log C\big(Y(\boldsymbol{x}), Z(\boldsymbol{x})\big) = O\big(\log C(\boldsymbol{x})\big),
  \end{equation*}
  where the left-hand-side is viewed as a function $(\bins)^n \to \R$.
  In fact, we even have
  \begin{equation*}
    \log C\big(Y(\boldsymbol{x}), Z(\boldsymbol{x})\big) \leq \log C(\boldsymbol{x}) + c
  \end{equation*}
  for some constant $c$ starting from some threshold $\boldsymbol{x}_0$: we can find a program in constant length that takes $\boldsymbol{x}$, extracts $x_1, \dots, x_n$ from it, and rearranges and concatenates them in such an order to obtain $Y(\boldsymbol{x})'Z(\boldsymbol{x})$, and the logarithm, being subadditive for large enough inputs, preserves the inequality.
\end{Prf}

\section{Proofs for Section~\ref{sec:example_applications}}\label{sec:proofs_section_7}

\begin{Prf}{pro:chain_rule_tsallis_entropy}
  For notational ease, we write $P(x) = P_X(x)$, $(P|_{X = x})_Y(y) = P(y \mid x)$ and $P(x, y) = P_{XY}(x, y)$ in this proof.
  We have
  \begin{align*}
    \big[I_1^{\q}(X) & + X._{\q}I_1^{\q}(Y) \big](P)  = \big[I_1^{\q}(X)\big](P) + \sum_{x \in \vs{X}} P(x)^{\q} \big[I_1^{\q}(Y)\big](P|_{X = x}) \\
    & = \frac{\sum_{x \in \vs{X}} P(x)^{\q} - 1}{ 1 - \q } + \sum_{x \in \vs{X}} P(x)^{\q} \frac{\sum_{y \in \vs{Y}} P(y \mid x)^{\q} - 1}{ 1 - \q } \\
    & = \frac{\sum_{x \in \vs{X}} P(x)^{\q} - 1 + \sum_{(x, y) \in \vs{X} \times \vs{Y}} \big( P(x) P(y \mid x)\big)^{\q} - \sum_{x \in \vs{X}} P(x)^{\q}}{ 1 - \q} \\
    & = \frac{\sum_{(x, y) \in \vs{X} \times \vs{Y}}P(x, y)^{\q} - 1}{ 1 - \q } \\
    & = \big[I_1^{\q}(XY) \big](P). & \qedhere
  \end{align*} 
\end{Prf}

\begin{Prf}{pro:cocycle_condition_kl_divergence}
  Let $X, Y \in \M(X_1, \dots, X_n)$ and $P \ll Q \in \Delta(\Omega)$.
  The following proof of the chain rule is similar to the one of Lemma~\ref{lem:converges_unconditionally} for Shannon entropy.
  For simplicity, we write $Q(x) = Q_X(x)$, $P(y \mid x) = (P|_{X = x})_Y(y)$ and $P(x, y) = P_{XY}(x, y)$ in this proof:
  \begin{align*}
    \big[X.D_1(Y) & + D_1(X) \big] (P \| Q) \\
    & = X.D_1(Y ; P \| Q) + D_1(X ; P \| Q) \\
    & = \sum_{x \in \vs{X}} P(x) D_1(Y ; P|_{X = x} \| Q |_{X = x}) - \sum_{x \in \vs{X}} P(x) \ln \frac{Q(x)}{P(x)} \\
    & \overset{(1)}{=} - \sum_{x \in \vs{X}} P(x) \sum_{y \in \vs{Y}}  P(y \mid x) \ln \frac{Q(y \mid x)}{P(y \mid x)} - \sum_{x \in \vs{X}} P(x) \Bigg( \sum_{y \in \vs{Y}} P(y \mid x) \Bigg) \ln \frac{Q(x)}{P(x)} \\
	& = - \sum_{(x, y) \in \vs{X} \times \vs{Y}} P(x) \cdot P(y \mid x) \cdot \Bigg[ \ln \frac{Q(y \mid x)}{ P(y \mid x)} + \ln \frac{Q(x)}{P(x)}\Bigg] \\
    & = - \sum_{(x, y) \in \vs{X} \times \vs{Y}} P(x, y) \ln \frac{Q(x, y)}{P(x, y)} \\
    & = \big[D_1(XY)\big](P \| Q).
  \end{align*}
  In step $(1)$, we used for the second sum that $P(y \mid x)$ is a probability measure in $y$ and thus sums to $1$. 
\end{Prf}

\begin{Prf}{pro:chain_rule_alpha_deformed_KL}
  Let $X, Y \in \M(X_1, \dots, X_n)$ and $P \ll Q \in \Delta(\Omega)$ be arbitrary.
  The following proof of the chain rule is similar to the one for the $\q$-entropy, Proposition~\ref{pro:chain_rule_tsallis_entropy}.
  For simplicity, we write $Q(x) = Q_X(x)$, $P(y \mid x) = (P|_{X = x})_Y(y)$ and $P(x, y) = P_{XY}(x, y)$ in this proof:
  \begin{align*}
    \big[ & D_1^{\q}(X)   + X._{\q}D_1^{\q}(Y)\big](P \| Q) = \big[D_1^{\q}(X)\big] (P \| Q) + \big[X._{\q}D_1^{\q}(Y)\big](P \| Q) \\
    & = \big[D_1^{\q}(X) \big](P \| Q) + \sum_{x \in \vs{X}} P(x)^{\q}Q(x)^{1 - \q} \big[D_1^{\q}(Y)\big]\big(P|_{X = x} \| Q|_{X = x}\big) \\
    & = \frac{\sum_{x \in \vs{X}} P(x)^{\q} Q(x)^{1 - \q} - 1}{\q - 1} + \sum_{x \in \vs{X}} P(x)^{\q}Q(x)^{1 - \q} \frac{\sum_{y \in \vs{Y}} P(y \mid x)^{\q}Q(y \mid x)^{1 - \q} - 1}{\q - 1} \\
    & = \frac{ - 1 + \sum_{(x, y) \in \vs{X} \times \vs{Y}}\big(P(x)P(y \mid x)\big)^{\q}\big(Q(x)Q(y \mid x)\big)^{1 - \q}}{\q - 1} \\
    & = \frac{\sum_{(x, y) \in \vs{X} \times \vs{Y}} P(x, y)^{\q}Q(x, y)^{1 - \q} - 1}{\q - 1}  \\
    & = \big[D_1^{\q}(XY) \big](P \| Q). & \qedhere
  \end{align*}
\end{Prf}

\phantomsection

\addcontentsline{toc}{section}{References}

\bibliographystyle{plainnat}
\bibliography{library}

\onecolumn\newpage

\end{document}